\documentclass[aps,prx,showpacs,twocolumn,amsmath,amssymb,superscriptaddress,letterpaper]{revtex4}
\usepackage{times}
\usepackage{amsfonts}
\usepackage{mathrsfs,ulem}
\usepackage{graphicx}
\usepackage{dcolumn}
\usepackage{bm}
\usepackage{color,xcolor}

\usepackage[colorlinks,bookmarks=false,citecolor=blue,linkcolor=red,urlcolor=blue]{hyperref}
\usepackage{MnSymbol}
\newcommand{\bra}[1]{\langle #1 |}
\newcommand{\ket}[1]{| #1 \rangle}
\newcommand{\braket}[2]{\left \langle #1 | #2 \right\rangle}

\newcommand{\bee}{\begin{equation}}
\newcommand{\ee}{\end{equation}}
\newcommand{\bma}{\begin{pmatrix}}
\newcommand{\ema}{\end{pmatrix}}
\newcommand{\balig}{\begin{align}}
\newcommand{\ealig}{\end{align}}

\newcommand{\clr}{\color{red}}

\newcommand{\clp}{\color{purple}}

\newcommand{\bk}{{\boldsymbol{k}}}
\newcommand{\bK}{{\boldsymbol{K}}}
\newcommand{\mH}{\mathcal{H}}
\newcommand{\br}{\boldsymbol{r}}
\newcommand{\bq}{\boldsymbol{q}}
\newcommand{\bp}{\boldsymbol{p}}
\newcommand{\bu}{\boldsymbol{u}}
\newcommand{\bR}{\boldsymbol{R}_n}

\newcommand{\bZ}{\mathbb{Z}}
\newcommand{\bI}{\mathbb{I}}

\def\be{\begin{equation}}       \def\ee{\end{equation}}
\def\bea{\begin{eqnarray}}      \def\eea{\end{eqnarray}}

\begin{document}

\title{Generalized Fermion Doubling Theorems: Classification of 2D Nodal Systems in Terms of Wallpaper Groups }

\author{Congcong Le}\thanks{These two authors contributed equally}
\affiliation{RIKEN Interdisciplinary Theoretical and Mathematical Sciences (iTHEMS), Wako, Saitama 351-0198, Japan}
\affiliation{Max Planck Institute for Chemical Physics of Solids, 01187 Dresden, Germany}
\affiliation{Kavli Institute for Theoretical Sciences, University of Chinese Academy of Sciences, Beijing 100190, China}

\author{Zhesen Yang}\thanks{These two authors contributed equally}
\affiliation{Beijing National Laboratory for Condensed Matter Physics, Institute of Physics, Chinese Academy of Sciences, Beijing 100190, China}
\affiliation{University of Chinese Academy of Sciences, Beijing 100049, China}
\affiliation{Kavli Institute for Theoretical Sciences, University of Chinese Academy of Sciences, Beijing 100190, China}

\author{Fan Cui}
\affiliation{Beijing National Laboratory for Condensed Matter Physics, Institute of Physics, Chinese Academy of Sciences, Beijing 100190, China}
\affiliation{University of Chinese Academy of Sciences, Beijing 100049, China}

\author{A. P. Schnyder}
\affiliation{Max-Planck-Institute for Solid State Research, Heisenbergstr. 1, D-70569 Stuttgart, Germany}

\author{Ching-Kai Chiu}\email{Corresponding: ching-kai.chiu@riken.jp}
\affiliation{Kavli Institute for Theoretical Sciences, University of Chinese Academy of Sciences, Beijing 100190, China}
\affiliation{RIKEN Interdisciplinary Theoretical and Mathematical Sciences (iTHEMS), Wako, Saitama 351-0198, Japan}

%
%
%
%
%
%
%
%
%
%
%

\date{\today}

\begin{abstract}
The Nielsen-Ninomiya Theorem has set up a ground rule for the minimal number of the topological points in a Brillouin zone.
Notably, in the 2D Brillouin zone, chiral symmetry and space-time inversion symmetry can properly define topological invariants as charges characterizing the stability of the nodal points so that the non-zero charges protect these points. Due to the charge neutralization, the Nielsen-Ninomiya Theorem requires at least two stable topological points in the entire Brillouin zone. However, additional crystalline symmetries might duplicate  the points. In this regard, for the wallpaper groups with crystalline symmetries, the minimal number of the nodal points in the Brillouin zone might be more than two. In this work, we determine the minimal numbers of the nodal points for the wallpaper groups in chiral-symmetric and space-time-inversion-symmetric systems separately and provide examples for new topological materials, such as topological nodal time-reversal-symmetric superconductors and Dirac semimetals. This generalized Nielsen-Ninomiya Theorem serves as a guide to search for 2D topological nodal materials and new platforms for twistronics. Furthermore, we show the Nielsen-Ninomiya Theorem can be extended to 2D non-Hermitian systems hosting topologically protected exceptional points and Fermi points for the 17 wallpaper groups and use the violation of the theorem on the surface to classify 3D Hermitian and non-Hermitian topological bulks.
\end{abstract}


\maketitle





\section{Introduction}













The study of various topologically protected points in lattice has deepened our understanding of profound knowledge in condensed matter physics~\cite{review_semimetal_a,Liu21022014,Xu_Weyl_2015_first,Weyl_discovery_TaAs,RevModPhys.88.035005,miriExceptionalPointsOptics2019,Lu_photonic_Weyl_2015}. In particular, the 2D nodal point, which is one type of these topological points, emerged in a variety of solid-state systems~\cite{Murakami2007,PhysRevLett.123.066405,Wieder246,ChiuSchnyder14,PhysRevLett.116.016401}. The manipulation of Dirac points in graphene has broadened our knowledge of strongly correlated systems~\cite{Cao:2018aa,Stepanov:2020aa,Arora:2020aa,Zondiner:2020aa} and topological phases~\cite{castroNetoRMP09,Kane:2005kx,Kane:2005vn,PhysRevX.1.021001}. Furthermore, nodal time-reversal symmetric superconductors host nodal points at zero energy revealing Majorana flat bands in the edges~\cite{RyuHatsugaiPRL02,matsuuraNJP13}. While the nodal points require symmetry for protection, exceptional points in non-Hermitian systems are robust even without symmetry protection~\cite{miriExceptionalPointsOptics2019a,PhysRevLett.120.146402,2019arXiv191202788Y}. All of these special topological points in 2D Hermitian and non-Hermitian lattices~\cite{PhysRevLett.127.196404} follow the Fermion doubling theorem (Nielsen-Ninomiya theorem)~\cite{NIELSEN1981173,NIELSEN198120,NIELSEN1981219} as a universal no-go theorem.

The Fermion doubling theorem in the literature is limited to the absence of additional global symmetries and crystalline symmetries. On the contrary, including these symmetries, the classification of topological phases of matter has rapidly grown since the discovery of the topological insulators~\cite{Chen:2009vn_short,Hsieh:2008fk,Xia:2009uq}. The ten-fold Hermitian classification in the AZ symmetry classes~\cite{Kitaev2009,altlandZirnbauerPRB10,Schnyder2008} and the thirty-eight-fold non-Hermitian classification~\cite{PhysRevX.9.041015} can provide a unified approach understand topological phases sorted by non-spatial (global) symmetries. In addition, the theory of the symmetry indicator~\cite{okumaTopologicalOriginNonHermitian2019,shenTopologicalBandTheory2018d,PhysRevLett.123.066405,2019arXiv191202788Y} is an established prominent approach classifying crystalline topological phases. However, a unified principle through Hermitian and non-Hermitian topological nodal systems regarding different types of symmetries is absent. This manuscript extends from the Fermion doubling theorem to a generalized no-go theorem and shows several distinct 2D topological nodal systems naturally following  this unified principle. 

In Hermitian systems, the Nielsen-Ninomiya Theorem in 2D is different from the one in 1D and 3D. That is, symmetries are required to protect 2D nodal points; 
either space-time inversion symmetry or chiral symmetry can protect 2D Dirac nodes with {\it 2-fold degeneracy} at any location in the Brillouin zone (BZ). Space-time inversion symmetry quantizes the $\bZ_2$ Berry phase in the 1D integral path~\cite{nodal_line_Yang}, whereas chiral symmetry leads to a well-defined integer winding number in the 1D integral path~\cite{ChiuSchnyder14,RevModPhys.88.035005}. When the 1D integral path encircles a Dirac node, the non-zero integral invariant stabilizes the Dirac node. The reason is that if the Dirac node is gapped, the integral path can become contractible and then vanishes. Since this contradicts the robustness of the non-zero invariant, the invariant quantized by the symmetries protects the Dirac node. This robustness manifests the emergence of multiple 2D topological Dirac semimetals~\cite{Novoselov:2005aa,PhysRevLett.102.236804,PhysRevLett.108.086804,PhysRevLett.119.226801}. 

	In chiral-symmetric systems, Dirac points with non-zero integer winding numbers are stable and fixed at zero energy, while in space-time inversion symmetric systems, Dirac points can be moved to any energy level and are the only stable nodal points due to the $Z_2$ invariant.  
	On the other hand, in 2D non-Hermitian systems, robust Fermi points~\cite{PhysRevLett.124.086801} and exceptional points~\cite{PhysRevLett.120.146402} are characterized by other non-zero integer winding numbers even in the absence of any symmetries. Those topological points (nodal points, Fermi points, exception points) in the Hermitian and non-Hermitian lattices always obey the Nielsen-Ninomiya Theorem as the ground rule~\cite{2019arXiv191202788Y}. In other words, in a 2D lattice, a topological point inevitably accompanies at least another topological point since the total charges characterizing the topological points in the entire BZ must be neutralized. However, in the presence of other symmetries, this no-go theorem might not be that simple. Generally, a 2D lattice system belongs to one of the 17 wallpaper groups (WGs) and preserves on-site symmetries, such as time-reversal symmetry. The additional symmetries limit the possible number of the topological points and confine the points in designated locations in the BZ. In other words, the minimal number of the topological points might be more than two; hence, the fermion ``doubling" theorem is not a proper name for the generalization of the theorem and we use Nielsen-Ninomiya theorem or no-go theorem for the name. In the literature, the generalized Nielsen-Ninomiya theorem has not been exhaustively studied for different wallpaper groups with or without time reversal symmetry. Due to the growing interest in topological materials, it is essential to set up the ground rule to understand the minimal multiplicity of the topological points for different crystalline symmetries and on-site symmetries.


	To investigate the generalized no-go theorem, we consider the three distinct 2D platforms --- nodal chiral-symmetric systems, space-time inversion symmetric semimetals, and non-Hermitian nodal systems. Our generalized no-go theorem exhaustively includes any 2D lattice possessing the topological points characterized by $\bZ$ and $\bZ_2$ invariants. For Hermitian lattices, we note that the fragile topology~\cite{PhysRevLett.121.126402,PhysRevB.100.205126,2019arXiv190503262S} or Dirac nodes protected by different crystalline symmetries~\cite{PhysRevLett.115.126803,2020arXiv200810175J} is not considered here, and our focus is on the stable nodal points protected by chiral symmetry or space-time inversion symmetry even in the presence of trivial bands.  
	First, we start with the minimal configurations and multiplicities of the nodal points at zero energy protected by chiral symmetry for all the 17 wallpaper groups in the BZ.  The chiral-symmetric systems without and with different types of time reversal symmetry and particle-hole symmetry correspond to five of the ten Altland-Zirnbauer (AZ) symmetry classes~\cite{Kitaev2009,altlandZirnbauerPRB10,Schnyder2008}. These five AZ classes (AIII, BDI, DIII, CII, CI) preserving chiral symmetry must be treated separately for the generalized no-go theorem since different time reversal symmetries lead to different charges and configurations for the nodal points. 
	Second, in space-time-reversal-symmetric semimetals the nodal points characterized by the $\bZ_2$ invariants might have different minimal configurations for the generalized no-go theorem with $\bZ$ invariants. The reason is that the two nodal points with non-zero $\bZ_2$ charges can be annihilated~\cite{PhysRevX.9.021013}, whereas the ones with non-zero integer charges together are always intact. However, space-time inversion symmetry can be realized in the wallpaper groups preserving inversion symmetry. Hence, only ten wallpaper groups, which possess inversion symmetry, can be studied for the generalized no-go theorem.
	Lastly, in 2D non-Hermitian systems without symmetries, exceptional points and Fermi points are characterized by integer winding numbers~\cite{okumaTopologicalOriginNonHermitian2019,shenTopologicalBandTheory2018d,PhysRevLett.123.066405,2019arXiv191202788Y}, whose mathematical structures are similar to the winding number quantized by chiral symmetry. 
	In this regard, we extend the generalized no-go theorem of the nodal chiral-symmetric systems in class AIII to the topological points in the non-Hermitian systems for all 17 wallpaper groups with small modifications. In addition, in non-Hermitian systems, the transpose operation and the complex-conjugation operation are not equivalent~\cite{PhysRevX.9.041015}. When those operations combine with crystalline operations, combined crystalline symmetries emerge. We also expand the generalized no-go theorem for the combined crystalline symmetries.  

		This generalized no-go theorem provides a new approach to understand the topological phases beyond the methodology of the symmetry indicators~\cite{Tang:2019aa,Zhang:2019aa,Vergniory:2019aa,2019arXiv190909634O,2019arXiv191011271G,Elcoro:2021uj,PhysRevB.104.085137,PhysRevX.7.041069,PhysRevResearch.3.023086}. In particular, the indicator approach can detect nodal points located at symmetric points and lines~\cite{2021arXiv210207676O,2021arXiv210611985T} but cannot sense most of the nodal points at general points in the BZ. Our generalization of the no-go theorem for the 17 wallpaper groups serves as a new paradigm to classify topological nodal platforms and thoroughly shows the minimal configurations for the topological points at all possible locations of the BZ, particularly including the general points. Furthermore, this generalized theorem can be applied for any topological points characterized by $\bZ$ and $\bZ_2$ invariants and governs various topological systems. For instance, spinless and spinful topological nodal time-reversal symmetric superconductors in class BDI and class DIII are two of the physical examples~\cite{beriPRB10,matsuuraNJP13}; the nodal points in the bulk Bogoliubov-de-Gennes Hamiltonian connect Majorana flat bands on the edges~\cite{RyuHatsugaiPRL02,BrydonSchnyderTimmFlat,SchnyderRyuFlat}. The generalized no-go theorem lists the minimal multiplicity and configurations of the nodal points for different symmetries. Furthermore, our theorem can be directly extended to 2D non-Hermitian systems and capture the minimal configuration of the exceptional points, which connect bulk Fermi arcs~\cite{Zhou1009}. In addition, Dirac semimetals are another promising example~\cite{PhysRevB.93.035401,Novoselov2005, 1Zhang2005, Wallace1947,Cahangirov2009,Malko2012,Huang2013,PhysRevLett.115.126803,C9NR00906J,Guo:2019aa,SciPostPhys.4.2.010,PhysRevLett.119.226801,PhysRevMaterials.1.054003}. For free fermion systems, space-time inversion symmetry protects Dirac points unfixed at any energy level. This theorem exhaustively provides the possible configurations of the Dirac points in 2D lattice and serves a guide to hunt new Dirac semimetals.  \\

\begin{figure*}[t!]
\centerline{\includegraphics[width=0.9\textwidth]{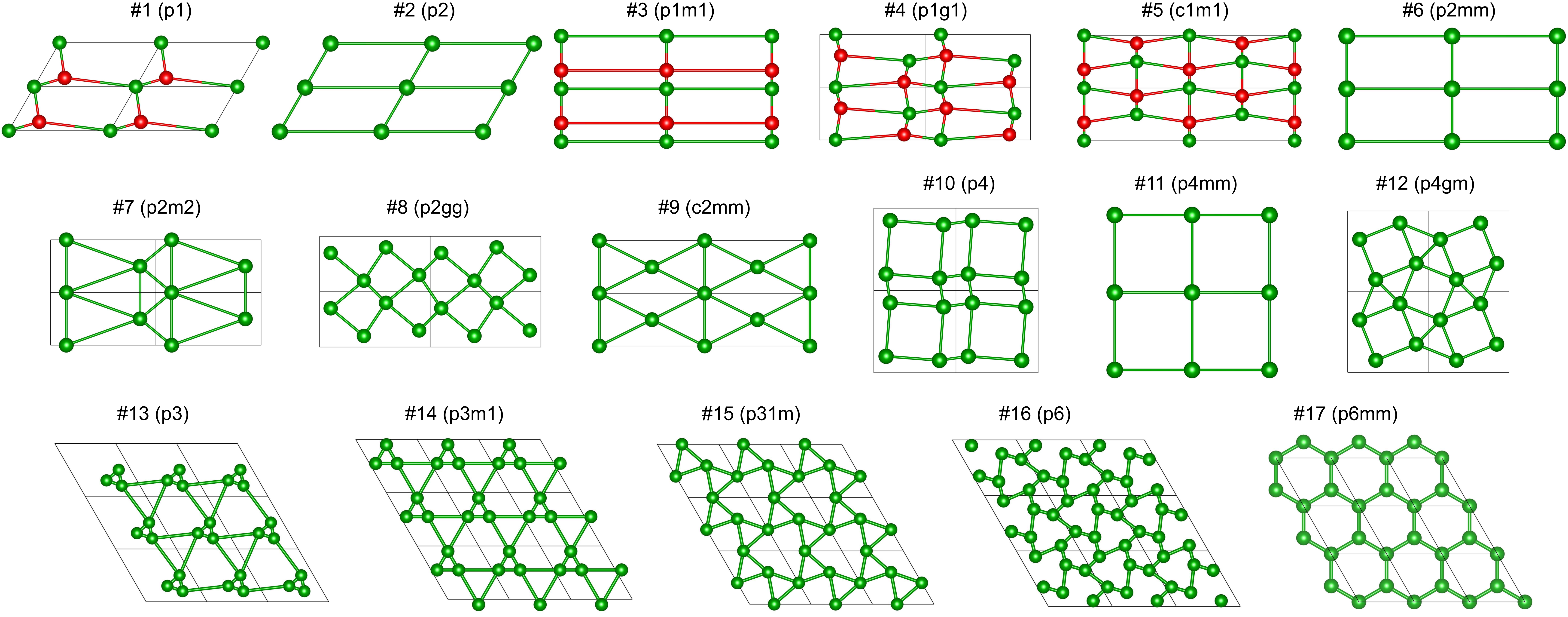}}
\caption{(color online) Seventeen 2D lattice structures illustrate the simplest examples for the seventeen wallpaper groups.  }
\label{wallpaper_structure}
\end{figure*}

		The remainder of this paper is organized as follows. In sec.~\ref{overview}, before going through the technicality, we provide an overview of the no-go theorem for the 17 WGs in topological points Hermitian and non-Hermitian lattices. In sec.~\ref{chiral section}, we thoroughly study the generalized no-go theorem for zero-energy nodal points protected by chiral symmetry. In sec.~\ref{time-space inversion section}, using a similar approach, we extend the no-go theorem to Dirac nodes protected by space-time inversion symmetry for the 10 WGs possessing inversion symmetry. In sec.~\ref{non-Hermitian section}, we show that the non-Hermitian no-go theorem for exceptional points and Fermi points inherits some features from the Hermitian chiral-symmetric one and distinctly have additional restrictions from rotation symmetries. In sec.~\ref{application}, we implement  the generalization no-go theorem for various condensed matter studies, such as layer groups, 3D bulk topology, and twistronics. Lastly, we conclude with a summary in Sec.~\ref{conclusion}. Some technical details have been relegated to appendices.

\section{ The overview of the no-go theorem for the 17 wallpaper groups} \label{overview}

	We begin with the overview of the generalized no-go theorem for 2D Hermitian and non-Hermitian lattices for the 17 wallpaper groups, and the following sections will serve all of the details. The 17 wallpaper groups exhaustively classify 2D lattices with crystalline symmetries and Fig.~\ref{wallpaper_structure} shows the simplest examples of the crystal structures for each wallpaper group. In general, in the 2D BZ, a stable topological point at $\bk_i$ is characterized by a non-zero charge $C(\bk_i)$, and the charge can be a $\bZ$ or $\bZ_2$ invariant. The key of the no-go theorem is that the summation of the charges carried by the topological points in the entire BZ must vanish as charge neutralization, i.e.,
\bee
\sum_{\bk_i \in \rm{BZ}} C(\bk_i)=0. \label{nogo}
\ee
This neutralization equation directly leads to the well-known statement of the Nielsen-Ninomiya theorem~\cite{NIELSEN1981173,NIELSEN198120,NIELSEN1981219} that a stable topological point must be accompanied by at least another topological point in the 2D BZ. To quantify the minimal multiplicity of the topological points, we definite the absolute charge number
\begin{equation}\begin{aligned}
\nu_{\rm{abs}}=\sum_{\bk_i \in \rm{BZ}} |C(\bk_i)|. \label{absolute number}
\end{aligned}\end{equation}
In chosen symmetry class, the minimum of this number corresponds to the minimal multiplicity and configuration of the topological points. The standard Nielsen-Ninomiya theorem~\cite{NIELSEN1981173,NIELSEN198120,NIELSEN1981219} shows $\nu_{\rm{abs}}\ge 2$ and $\nu_{\rm{abs}}$ is always an even number. We note that $\nu_{\rm{abs}}$ can be an integer greater than one, even though the charge $C(\bk_i)$ might have $\bZ_2$ property. The reason is that the topological points may be at different locations and cannot be annihilated. Beyond the Nielsen-Ninomiya theorem, our generalized no-go theorem shows that for some WGs, $\nu_{\rm{abs}}$ is always greater than 2 so that the minimal number of the topological points with charge weights is greater than 2. The main results of our generalized no-go theorem for Hermitian and non-Hermitian 2D lattices, which show the minimal number of $\nu_{\rm{abs}}$, are summarized in Table \ref{the number}. 

\renewcommand\arraystretch{1.5}
\begin{table*}
\caption{\label{the number} {\bf The minimum multiplicity list for topological points in Hermitian and non-Hermitian lattices.} The entire table exhaustively lists minimal $\nu_{\rm{abs}}$ describing the minimal multiplicities of the nodal points at zero energy for each WG and AZ symmetry class preserving chiral symmetry.
For each WG, one or two point group generators with signs indicate the algebra between the chiral symmetry operator and the generators. While `$0$' indicates the absence of nodal points protected by chiral symmetry, non-zero $\nu_{\rm{abs}}$ indicates the minimal number of Dirac points with $\pm1$ charges for most of the WG symmetry classes, with the exceptions labelled by $*$. That is, in the exceptions the minimal configurations require the presence of at least one high-ordered nodal point, of which the charge is not $\pm 1$. On the other hand, label $\diamond$ indicates the minimal numbers of the Dirac nodes for space-time-inversion-symmetric lattices for 10 of the WGs. For non-Hermitian systems, label $\bullet$ marks the minimal numbers of Fermi points and exceptional points for the 17 WGs.  }
\setlength{\tabcolsep}{7mm}{
\begin{tabular}{|c||c|c|c|c|c|}
  \hline
 WG & Generators & AIII & BDI/CII  & DIII & CI  \\\hline  \hline
$\#$1&  & 2$^\bullet$ & 2 & 4 & 4\\\hline
$\#$2& $C^{+}_2$  & 4$^\bullet$  & 0 & 4 & 4\\
    & $C^{-}_2$  & 2$^\diamond$ & 2 & 0 & 0  \\\hline
$\#$3,4,5& $M^{+}_{x}$ & 2$^\bullet$ & 2   & 4 & 4\\
     & $M^{-}_{x}$  & 2  & 2   & 4 & 4\\\hline
$\#$6,7 & $C^{+}_2$ and $M^{+}_{x}$ & 4$^\bullet$ & 0 & 4 & 4 \\
 ~~~8,9  &$C^{+}_2$ and $M^{-}_{x}$  & 4 & 0 & 4 & 4  \\
      & $C^{-}_2$ and $M^{\pm}_{x}$ & 2$^\diamond$ & 2  & 0 & 0 \\\hline
$\#$10& $C^{+}_4$ & 4, 8$^\bullet$ & 0 & 4 & 4$^*$  \\
     & $C^{-}_4$ & 4$^\diamond$ & 0 & 4 & 4 \\\hline
$\#$11,12 & $C^{+}_4$ and $M^{+}_{x}$ & 8$^\bullet$ & 0 & 8 & 8 \\
     & $C^{+}_4$ and $M^{-}_{x}$ & 4 & 0 & 4  & 4$^*$  \\
      & $C^{-}_4$ and $M^{\pm}_{x}$ & 4$^\diamond$ & 0  & 4 & 4 \\\hline
$\#$13   & $C^{+}_3$ & 2, 6$^\bullet$ & 2  & 4$^*$ & 4$^*$\\
     & $C^{-}_3$ & 0& 0  & 0 & 0\\\hline
$\#$14& $C^{+}_3$ and $M^{+}_{y}$ & 2, 6$^\bullet$ & 2  & 12 & 12\\
     & $C^{+}_3$ and $M^{-}_{y}$ & 6 & 6  & 4$^*$ & 4$^*$  \\
      & $C^{-}_3$ and $M^{\pm}_{y}$ & 0 & 0  & 0 & 0\\\hline
$\#$15& $C^{+}_3$ and $M^{+}_{x}$ & 6$^\bullet$  & 6  & 12  & 12\\
      & $C^{+}_3$ and $M^{-}_{x}$ & 2 & 2 & 4$^*$ & 4$^*$ \\
      & $C^{-}_3$ and $M^{\pm}_{x}$ & 0 & 0  & 0 & 0\\\hline
$\#$16& $C^{+}_6$ & 4$^{*}$, 12$^\bullet$ & 0  & 4$^*$ & 4$^*$  \\
     & $C^{-}_6$ & 2$^\diamond$ & 2  & 0 & 0\\\hline
$\#$17& $C^{+}_6$ and $M^{+}_{x}$ & 12$^\bullet$ & 0  & 12 & 12\\
     & $C^{+}_6$ and $M^{-}_{x}$ &  4$^*$ & 0 & 4$^*$ & 4$^*$ \\
      & $C^{-}_6$ and $M^{+}_{x}$ &  6  & 6   & 0 & 0\\
       & $C^{-}_6$ and $M^{-}_{x}$ & 2$^\diamond$ & 2  & 0 & 0\\\hline
\end{tabular}} \label{Summary table}
\end{table*}

	First, let us introduce the no-go theorem for 2D Hermitian lattices preserving chiral symmetry for the 17 WGs in Table \ref{the number}. Chiral symmetry makes the nodal points carry non-zero $\bZ$ winding numbers~\cite{ChiuSchnyder14,RevModPhys.88.035005} stable and locked at zero energy. The 17 WGs are simplified to 11 equivalent sets based on point groups, and 5 AZ symmetry classes (AIII, BDI, DIII, CI, and CII) can also be divided into four categories, since BDI and CII share the same no-go theorem. Hence, the number of symmetry combinations is 44 effective WG symmetry classes. With translation symmetry, each WG is formed by rotation symmetry operation $C_n$ or/and mirror symmetry operation $M_x$ as generators. Here we fix the direction of the mirror $M_x$ along $\overline{\rm \Gamma X}$ for the rectangle (square) BZ and $\overline{\rm \Gamma M}$ for the hexagonal BZs. The minimal configuration of the nodal points further depends on the algebra between the chiral symmetry operator $S$ and the two generators $C_n, M_x$. In Table \ref{the number}, $C_n^\pm$ represents the rotation symmetry operator commutes/anticommutes with $S$, while $M_x^\pm$ indicates the mirror symmetry operator commutes/anticommutes with $S$. (We note that WG$\#14$ without $M_x$, which is an exception, has mirror symmetry $M_y$ along $\overline{\Gamma K}$.) The numbers listed in the table represent the minimal absolute number $\nu_{\rm abs}$ for each WG symmetry class. Since some symmetries can trivialize the nodal points, `0' indicates the absence of the stable nodal points protected by chiral symmetry. Most of the remaining numbers represent the minimal numbers of the Dirac nodes with $\pm 1$ charge, with the exceptions labeled by $*$; due to the symmetry constraints, the minimal configurations of the exceptions require the presence of nodal points with high charges, whose absolute values are greater than one $|C|>1$. For example, in class CI in WG$\#10$ with $C_4^+$ the minimal configuration for $\nu_{\rm abs}=4$ only includes one nodal point with $+ 2$ winding number and the other with $- 2$ winding number.

	 For other Hermitian systems, stable nodal points protected by space-time inversion symmetry ($(C_{2}T)^2=1$) carry only $\pm 1$ charge, due to the $\bZ_2$ invariant of the Berry phase~\cite{nodal_line_Yang,PhysRevX.9.021013}. That is, there is only one type of stable nodal points. Being different from topological points with $\bZ$ charges, two of the non-zero $\bZ_2$ nodal points together are annihilated. In this regard, the no-go theorem with $\bZ_2$ invariants is distinct from the theorem with $\bZ$ invariants. On the other hand, inversion symmetry is required in the WGs to form space-time inversion symmetry with time-reversal symmetry. Only 10 of the 17 WGs preserving inversion symmetry can possess the $\bZ_2$ nodal points. Furthermore, the (anti)commutation relation between $C_2T$ and the generators ($C_n,M_x$) does not affect the no-go theorem. In Table \ref{the number}, the numbers $\nu_{\rm abs}$ marked by $\diamond$ indicate the minimal numbers of the protected nodal points for the 10 WGs. We note that those numbers also represent the minimal absolute numbers $\nu_{\rm abs}$ for chiral symmetric lattices in class AIII.
	
	Lastly, for 2D non-Hermitian lattices, any symmetry is no longer required to protect topological points; robust Fermi points and exceptional points carrying non-zero charges naturally arise~\cite{PhysRevLett.120.146402,2019arXiv191202788Y,PhysRevLett.123.066405}. In the absence of the symmetry requirement for protection, the algebra between the symmetry operators is inapplicable to classify the minimal configuration of the non-Hermitian systems. Hence, the 17 WGs are the only symmetry classes in this classification. Since the charges of the Fermi points and exceptional points share the same $\bZ$ properties, these two distinct types of the topological points obey the same generalized rules of the no-go theorem. The non-Hermitian lattices and the Hermitian chiral-symmetric ones share $\bZ$ invariant, so the restriction of the nodal points in the chiral-symmetric systems with $C_n^+,M_x^+$ can be applied to the non-Hermitian for the 17 WGs. The only difference is that the charges of the non-Hermitian topological points at rotation centers, say $C_n$, must be the multiple of $n$
\bee	
C(\bk_{\rm rotation})=nj,\ j\in \bZ	
\ee	
Following this additional rule, we note that the non-Hermitian no-go theorem does not always share the same minimal absolute numbers $\nu_{\rm abs}$ with the chiral-symmetric lattices in class AIII. In Table \ref{the number}, the generalized no-go theorem for the non-Hermitian lattices is shown by the absolute charge numbers labeled by $\bullet$. 	
	

	Table \ref{the number} serves as a dictionary showing the minimum of the absolute charge numbers $\nu_{\rm{abs}}$ for the three distinct 2D lattice systems as the generalized no-go theorem. For most of the cases $\nu_{\rm{abs}}$ is greater than $2$ due to the symmetries; hence, the generalized no-go theorem goes beyond the {\it doubling} theorem. We note that for the minimal configurations, the topological points have to be located at specific momenta in the BZ. That is, the configuration of the topological points profoundly depends on their locations. The location-dependent multiplicities of the topological points are studied in great detail in the following sections. Furthermore, our generalized theorem lists all of the possible configurations of Dirac nodes; this prediction can lead to possible discoveries of new exotic Dirac materials. The following sections provide the tight-binding models for the potential realization and the detailed derivation of the generalized no-go theorem.

\section{2D Hermitian chiral-symmetric lattices } \label{chiral section}

	We start with 2D Hermitian lattices preserving chiral symmetry. Since there are 5 AZ symmetry classes with chiral symmetry, the generalized no-go theorem for chiral symmetry is much more complicated than space-time-inversion-symmetric systems and non-hermitian lattices. After going through the details of all the chiral symmetry classes, we directly extend these results of the no-go theorem to the two other cases. 

\subsection{Nodal points protected by chiral symmetry}
	We review the interplay of nodal points, chiral symmetry, and winding numbers by showing that the non-zero winding number quantized by chiral symmetry can protect the nodal point. Furthermore, we build the relation between winding numbers for two nodal points connected by crystalline symmetries or time reversal symmetry. This relation is an important building block for our no-go theorem.

	Chiral symmetry is our focus symmetry protecting nodal points with non-zero winding numbers in this section. We consider an (effective) non-interacting system and the Bloch Hamiltonian preserving chiral symmetry obeys
\bee
S \mathcal{H}(\boldsymbol{k}) S^{-1} =-  \mathcal{H}(\boldsymbol{k}),  \label{chiral symmetry}
\ee
where unitary matrix $S$ is defined as chiral symmetry operator.
One of the ideal platforms to realize chiral symmetry is time-reversal symmetric superconductors~\cite{beriPRB10,matsuuraNJP13,RyuHatsugaiPRL02,BrydonSchnyderTimmFlat,SchnyderRyuFlat}. Superconductor systems can be described by a Bogoliubov-de Gennes Hamiltonian~\cite{deGennesbook} obeying particle-hole symmetry
\bee
C \mathcal{H}(-\boldsymbol{k}) C^{-1} =-  \mathcal{H}(\boldsymbol{k}), \label{particle-hole symmetry}
\ee
where particle-hole symmetry operator $C$ is antiunitary. The time-reversal symmetric Hamiltonian also satisfies
\bee
T \mathcal{H}(-\boldsymbol{k}) T^{-1} =  \mathcal{H}(\boldsymbol{k}) \label{time reversal symmetry},
\ee
where time-reversal symmetry operator $T$ is also antiunitary.
Consequently, the combination of particle-hole symmetry and time-reversal symmetry becomes chiral symmetry with chiral symmetry operator $S=CT$ as a unitary one.

When an eigenstate $\ket{\Psi(\bk)}$ possesses energy $E$, the chiral symmetry equation (\ref{chiral symmetry}) leads to another eigenstate $S\ket{\Psi(\bk)}$ having energy $-E$; zero energy ($E=0$) is a special point. Furthermore, it will be shown later that stable nodal points are locked in zero energy so we choose zero energy as Fermi level.


Here we review the definition of the winding number quantized by chiral symmetry. By choosing a proper basis, the chiral symmetry operator is written as $S=\tau_z \otimes \bI_{n\times n}$ so the Hamiltonian is in the block off-diagonal form
\bee
\mathcal{H}(\bk)=
\bma
0 & h(\bk) \\
h^\dagger(\bk) & 0
\ema.
\ee
Assuming $h(\bk)$ is invertible ($\det(h(\bk))\neq0$), we can define the winding number in a closed loop integral path in the BZ
\begin{equation}\begin{aligned}
\nu&=\frac{i}{2\pi}\oint d \boldsymbol{k}\cdot Tr[h^{-1}(\bm{k})\partial_{\bm{k}}h(\bm{k})]\\
  &=\frac{i}{2\pi}\oint d \boldsymbol{k}\cdot \partial_{\bm{k}}Tr[\ln h(\bm{k})] \label{winding number eq}\\
  &=\frac{i}{2\pi}\oint d \big( \ln\det[h(\bm{k})] \big )
\end{aligned}\end{equation}
By rewriting $\det h(\bk)\equiv f(\bk) e^{i\alpha(\bk)}$, where $f(\bk)$ is a single-valued function and $\alpha(\bk)$ is a multivalued function, the winding number is given by
\bee
\nu =\frac{-1}{2\pi}\oint d \alpha(\bk)
\ee
Since in the loop the start and end points of the integral are identical, $\oint d \alpha(\bk)=2\pi l$ and then the winding number $\nu$ is quantized. We note that the definition of the winding number is equivalent to the one using the flattened Hamiltonian  (see Appendix \ref{winding connection} for the detail).

We demonstrate the interplay of nodal point protection, chiral symmetry, and winding number by using a simple $2\times 2$ Hamiltonian
\begin{align}
\mathcal{H}_n(\bk)=
\bma
0 & (\Delta k_x+i\Delta k_y )^n \\
(\Delta k_x-i \Delta k_y)^n & 0
\ema, \label{simple model}
\end{align}
where $n$ is a positive integer, $\Delta k_x=k_x-k_{x0}$, $\Delta k_y=k_y-k_{y0}$. The nodal point appears at $\boldsymbol{K}_{0}=(k_{x0},k_{y0})$ in the BZ with $E=0$. The energy dispersion $E=\pm \Delta k^{n}$ and $q=\Delta ke^{in\theta}$, where $\Delta k =\sqrt{\Delta k_x^2 +\Delta k_y^2}$. The meaning of the winding number is the change  of $U(1)$ phase of $\det{h(\bk)}$ along the close loop $\Gamma(K_0)$ encircling $\bK_0$ in the counterclockwise direction.
The chiral symmetry operator is given by $S=\tau_z$ and $\mH(\bk)$ obeys chiral symmetry equation (\ref{chiral symmetry}). Here we use the polar coordinate $\Delta k_x \equiv k_0\cos \theta$ and $\Delta k_y \equiv k_0\sin \theta$. With the integral path encircling $\boldsymbol{K}_{0}$, by Eq.~\ref{winding number eq} the winding number $\nu(\boldsymbol{K}_{0})=n$. ($n=\pm 1$ corresponds to a Dirac point.) This node is equivalent to $|n|$ Dirac nodes with $\nu=\textrm{sign}(n)$. Chiral symmetry quantizes the winding number and forbids the presence of $\tau_z$ destroying the nodal point. Moreover, the zero energy of the nodal point cannot be lifted, since the identity matrix, which is the only one changing the nodal point energy, is forbidden by chiral symmetry. In this regard, the non-zero winding number $n$ indicates the presence of the stable nodal point. We note that without breaking chiral symmetry, the nodal point with winding number $n$ can be deformed to several nodal points and the summation of the winding numbers for the nodal point is still $n$. 

Likewise, the Hamiltonian having another nodal point with $-n$ winding number is written as
\begin{align}
\mathcal{H}_{-n}(\bk)=
k_0^n e^{in\theta} \tau_+ + k_0^n e^{-in\theta} \tau_-,
\end{align}
When we merge these two systems ($\mathcal{H}_n(\bk)$ and $\mathcal{H}_{-n}(\bk)$), the Hamiltonian is given by
\bee
\mH^\pm(\bk)=k_0^n \cos n\theta \sigma_0 \otimes  \tau_x  +k_0^n \sin n\theta \sigma_z \otimes \tau_y,
\ee
which still preserves chiral symmetry with the symmetry operator $S=\sigma_0\otimes \tau_z$. Since the total winding number for the nodal point vanishes, we simply find a symmetry-preserving mass term $m\sigma_y \otimes \tau_y$ destroying the nodal point and the gapped energy dispersion is given by $E=\pm \sqrt{k_0^{2n}+m^2}$. Namely, the two nodal points at the same locations with the opposite winding numbers are unstable.

\begin{figure}[b]
\centerline{\includegraphics[width=0.3\textwidth]{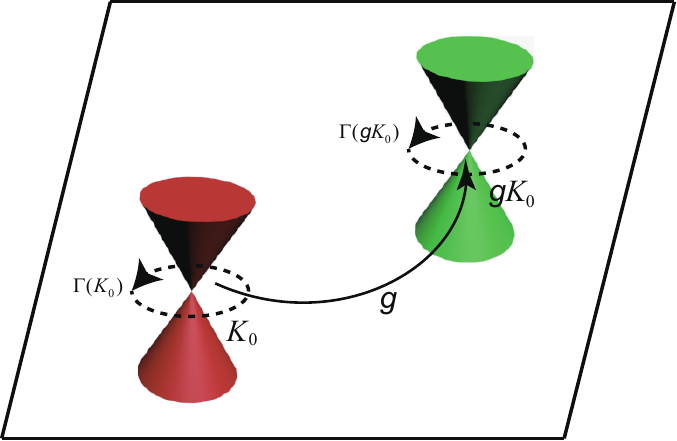}}
\caption{(color online) The two Dirac points at $\bK_0$ and $g\bK_0$ are connected by symmetry $g$. The two dashed circles represent the counterclockwise integral loops $\Gamma (\bK_0)$ and $\Gamma (g\bK_0)$ for the winding number $\nu(\bK_0)$ and $\nu(g\bK_0)$ respectively. }
\label{Diracpoint1}
\end{figure}

A symmetry operator $g$ acts on the k-space in the following form
\bee
g\bk=\bma
g_{11} & g_{12} \\
g_{21} & g_{22}
\ema
\bma
k_x \\
k_y
\ema.
\ee
Suppose there exists a generic nodal point at $\bK_0$, which is protected by chiral symmetry but not necessarily described by the simple model (\ref{simple model}). The winding number of the nodal point is given by
\bee
\nu(\boldsymbol{K}_{0})=\frac{i}{2\pi}\oint_{\Gamma\left(\boldsymbol{K}_{0}\right)} \nabla_{\bm{k}} \big ( \ln\det[h(\bm{k})] \big )\cdot d\bm{k}, \label{winding number}
\ee
where $\Gamma\left(\boldsymbol{K}_{0}\right)$ indicates an infinitesimal loop of the integral encircling $\bK_0$ counterclockwise. The symmetry results in the presence of a nodal point at $g\bK_0$, while this symmetry operator can be time reversal or crystalline. To count the number of the nodal points in the BZ, it is important to study the relation between the two winding numbers $(\nu(\bK_0),\ \nu(g\bK_0))$ of the two nodal points connected by the symmetry $g$; the winding numbers $(\nu(\bK_0),\ \nu(g\bK_0))$ are computed in the integral loops $\Gamma(\bK_0)$ and $\Gamma(g\bK_0)$ encircling $\bK_0$ and $g\bK_0$ points respectively as illustrated in Fig.~\ref{Diracpoint1}.


The algebra of chiral symmetry operator and $g$ symmetry operator determines the connection between the two winding numbers at $\bK_0$ and $g\bK_0$. The derivation details of the connections for crystalline symmetry operators and time reversal symmetry operators are given in appendix \ref{winding connection}. As we first consider crystalline symmetry, the Hamiltonian preserving crystalline symmetry obey (See appendix \ref{Bloch wave} for the derivation)
\begin{align}
\mH(g\bk)=&U_g^\pm(\bk) \mH (\bk) U_g^{\pm\dag}(\bk), \label{crystalline symmetry equations}
\end{align}
where $U_g^\pm$ represents the unitary crystalline symmetry operator in the momentum basis and $\pm$ signs indicate (anti)commutation relation between the two symmetry operators ($U_g^\pm S\mp SU_g^\pm =0$). We note that $\mathcal{H}(\bk)$ and $U(\bk)^\pm$, which are single-valued (appendix \ref{Bloch wave}), obey
\bee
\mathcal{H}(\bk+\boldsymbol{G}_i)=\mathcal{H}(\bk),\ U(\bk+\boldsymbol{G}_i)^\pm=U(\bk)^\pm,
\ee
where $\boldsymbol{G}_i$ is a reciprocal lattice vector.
In the following, we always use $+(-)$ to indicate that the chosen symmetry operator (anti)commutes with chiral symmetry operator ($S$). These two possible forms of $U_g^\pm$ lead to the different connections of the winding numbers between $\bK_0$ and $g\bK_0$
\bee
\nu(g^\pm\bK_0)=\pm \det(g^\pm)\nu(\bK_0). \label{crystalline winding}
\ee
There are two types of crystalline symmetries in wallpaper groups --- (glide) mirror symmetries and rotation symmetries. The mirror operators satisfy $\det g=-1$, whereas the rotation operators obey $\det g=1$. For $g^+$, the two nodal points ($\bK_0$ and $g\bK_0$) connected by the mirror symmetry possess the opposite winding numbers, while the nodal points linked by the rotation symmetry have the identical winding numbers. On the contrary, for $g^-$, the two winding numbers linked by the mirror operator are identical and the ones linked by the rotation operator have the different signs. 


We note that the generic forms of the crystalline symmetry operators in Eq.~\ref{crystalline symmetry equations} are momentum-dependent. That is, nonsymmorphic symmetry operations, which are glide mirror ones only in 2D systems, have the same winding number relations at $\bK_0$ and $g\bK_0$  with mirror symmetry operations. Therefore, in the next subsection when we classify the minimal multiplicities of the nodal points, we treat the WGs \#(3, 4, 5) as one set together since the three WGs have the same effective mirror symmetries in momentum space, regardless of the glide operations. Similarly, WG \#(6, 7, 8, 9) and \#(11, 12) are respectively gathered to two sets for the classification of the generalized no-go theorem in the next subsection.

On the other hand, time reversal symmetry operator can commute ($T^+$) or anticommute ($T^-$) with chiral symmetry operator $S$. For $T^\pm$, the connection between the two winding numbers is given by (see the derivation in appendix \ref{winding connection})
\bee
\nu(-\bK_0)=\mp  \nu(\bK_0). \label{time reversal winding}
\ee
The winding number relations at $\bK_0$ and $g\bK_0$ for different symmetry operators are summarized in Table \ref{winding number table}. This table is our fundamental tool to study the minimal multiplicity of the nodal points for 17 WGs and 5 AZ symmetry classes possessing chiral symmetry.

 \begin{table}[t!]
\begin{tabular}{|c|c|c|}
 \colrule
  $T^\pm$ &  $C_n^\pm$ & $M^\pm$ \\
 \colrule
$\nu(-\bK_0)={\mp\nu}(\bK_0)$ & $\nu(g\bK_0)={\pm\nu}(\bK_0)$ &   $\nu(g\bK_0)={\mp\nu}(\bK_0)$ \\
 \hline
\end{tabular}
\caption{\label{winding number table}%
Symmetry related winding numbers for nodal points. The operator labels $T^\pm,\ C_n^\pm,\ M^{\pm}$ indicate time reversal , $n$-fold rotation, and mirror respectively. The signs $+(-)$ indicate the corresponding symmetry operator (anti)commuting with chiral symmetry operator $S$. }
\end{table}

\subsection{Nodal points for the 17 wallpaper groups and the 5 AZ symmetry classes}

\begin{figure*}[t!]
\centerline{\includegraphics[width=0.85\textwidth]{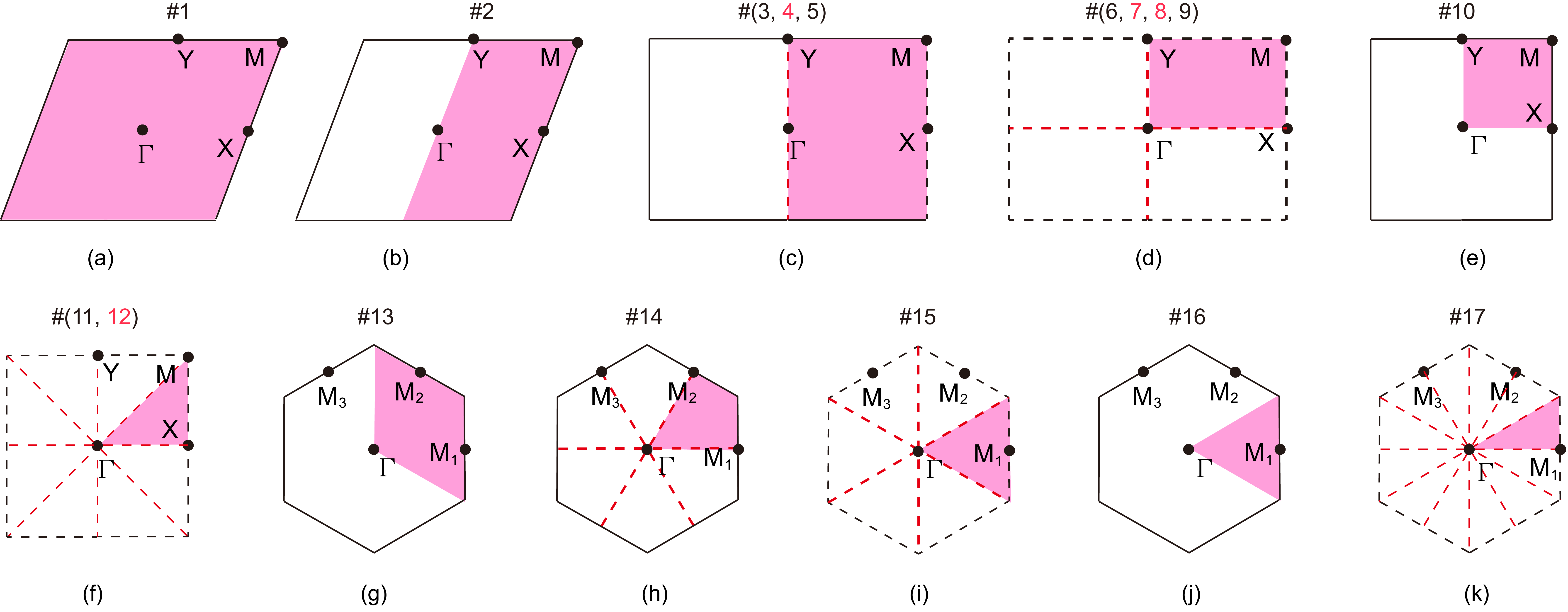}}
\caption{(color online) {\bf The irreducible BZs (purple) for the 17 Wallpaper groups.} Based on the crystalline symmetry operations, the 17 wallpaper groups are classified to 11 equivalent sets for the classification of the no-go theorem. The red numbers denote nonsymmorphic wallpaper groups. The black lines indicate the boundary of the BZ, the solid dots indicate TRI points, and the (red) dashed lines indicate mirror lines (inside the BZ). }
\label{irreducible BZ}
\end{figure*}

Symmetries different from chiral symmetry crucially affect the minimal multiplicity of the stable nodal points in the BZ. For 2D lattice systems with chiral symmetry, symmetries can be classified by 17 WGs and 5 AZ symmetry classes. While the 17 WGs include various crystalline symmetries and translational symmetries in the two spatial directions, the AZ symmetry classes can only have non-spatial (global) symmetries --- chiral symmetry, time-reversal symmetry, and particle-hole symmetry. As we consider the combination of crystalline symmetries and AZ symmetry classes, there are $85$ WG symmetry classes. However, we will show later that the number of the effective WG symmetry classes for the no-go theorem can be reduced. Let us first review the properties of the 17 WGs. Fig.~\ref{wallpaper_structure} illustrates the simplest crystal structures of the 17 WGs and the WGs are classified to four crystal systems (monoclinic, orthorhombic, square and hexagon) as well as five Bravais lattices. In the WGs, $C_{n}$ and $C_{nv}$ (n=1,2,3,4,6) are the only 10 types of the point groups, while mirror $M$ and rotation $C_n$ operations can be the only two generators for these point groups. When conventional cells are chosen, the point groups can reduce the first BZ to the irreducible BZ as shown in Fig.~\ref{irreducible BZ}, since the crystalline symmetry connects any point in the BZ from the irreducible BZ. In this regard, all of the nodal points in the irreducible BZ can be easily extended to the entire BZ by using those two generators of the point group. The irreducible BZ is our building block to study the minimal multiplicity  and configuration of nodal points in the no-go theorem.

	It has been shown in Table \ref{winding number table} that the algebra between the chiral symmetry operator and the other symmetry operators determines the winding numbers of the nodal points extended by the symmetry operators from the irreducible BZ. On the other hand, although the crystalline symmetry operators can be momentum-dependent, regardless of the algebra between the symmetry operators, different forms of the symmetry operators do not affect the winding numbers of the extended nodal points. For example, WG$\#3$ and WG$\#4$ only preserve reflection symmetry and reflection glide symmetry respectively and their symmetry operators are in different forms. Since these two symmetry operations flip momentum only in one direction in the same BZ, these two WGs are classified as the same set to discuss the no-go theorem.
Furthermore, we note that in conventional cell basis, WG$\#5,9$ preserves additional translational symmetry with half-length of the unit cell shift. Since the translational symmetry and chiral symmetry are unrelated, chiral operator $S$ commutes with this shift operator; this additional translation symmetry does not affect the nodal points, we can classify WG$\#3,4,5$ in a set and WG$\#6,7,8,9$ in another set for the no-go theorem.
	Thus, the 17 WGs are classified to 11 equivalent sets for the no-go theorem as shown in Fig.~\ref{irreducible BZ}. Although WG$\#14$ and $\#15$ have the same $C_{3v}$ point group, their mirror lines are along  $\Gamma$-M and $\Gamma$-K differently. Hence, these groups belong to the different sets for the no-go theorem.

	The five AZ symmetry classes (AIII, BDI, DIII, CI, and CII), which preserve chiral symmetry, can possess nodal points characterized by winding numbers. By considering the 11 distinct sets of the WGs, we can directly study the no-go theorem for class AIII, since this symmetry class preserves only chiral symmetry. The remaining four symmetry classes have additional time reversal symmetry. Because chiral symmetry is the combination of time-reversal symmetry and particle-hole symmetry ($S=TC$), we treat particle-hole symmetry as redundancy. The AZ symmetry classes determine the algebra between the chiral operator and the time-reversal operator. While class BDI, CII have $T^+$ time-reversal operator, class DIII, CI have $T^-$ time-reversal operator (see the proof in appendix \ref{AZ chiral}). In two of the following subsections, class BDI, CII have the same minimal configurations for each of the 17 WGs. In total, there are $44(=11\times 4)$ effective WG symmetry classes for the generalized no-go theorem. 

\begin{figure}[b]
\centerline{\includegraphics[width=0.45\textwidth]{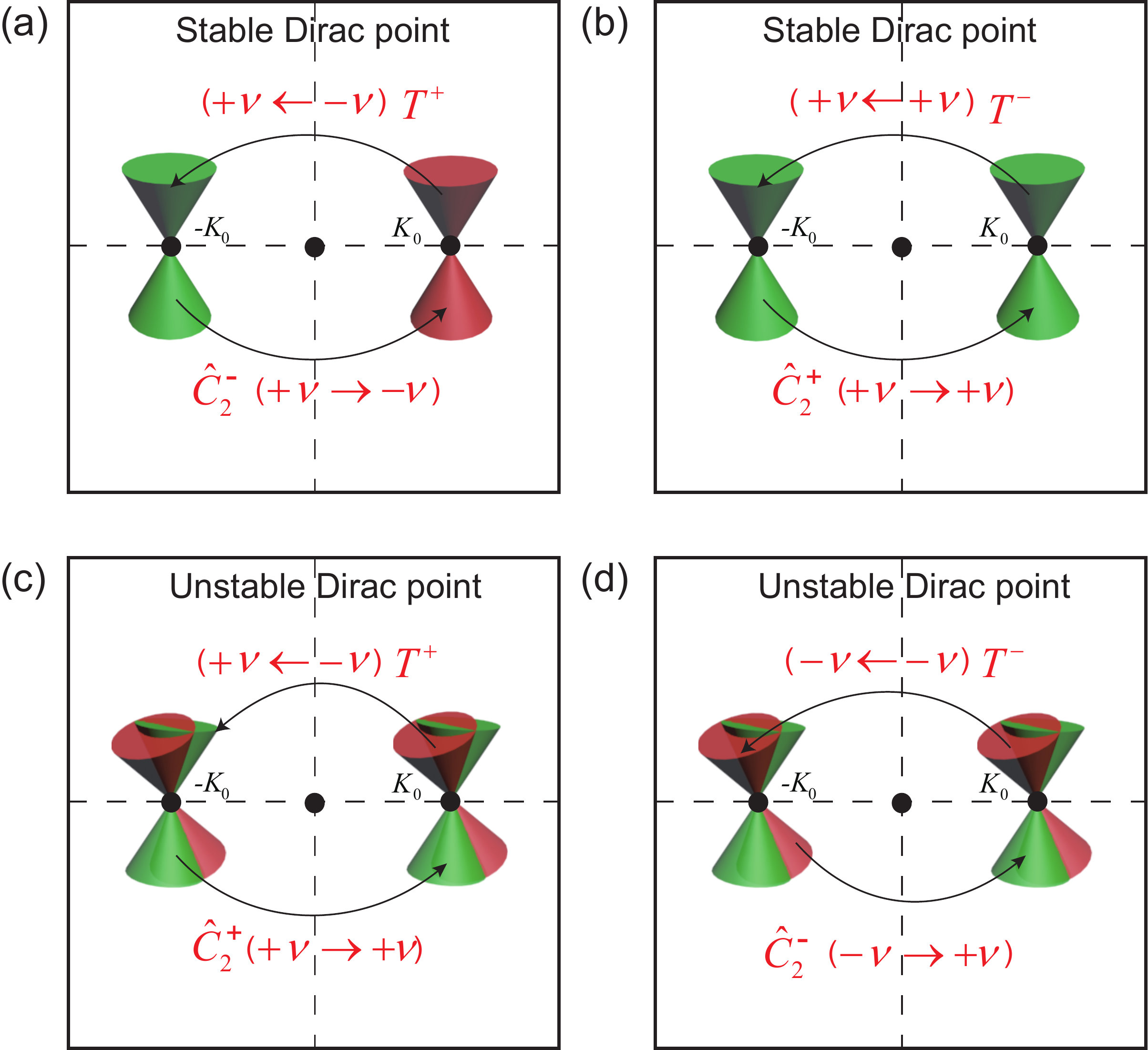}}
\caption{(color online) The survival of nodal points protected by chiral symmetry in the presence of time-reversal symmetry and $C_2$ rotation symmetry. The green and red cones represent nodal points with $+1$ and $-1$ winding numbers respectively. (a,b) $T^\pm$ time-reversal operator and $C^\mp$ rotation operator act on the nodal point at $\bK_0$. The nodal point goes back to the same point with the same winding number.  (c,d) $T^\pm$ time-reversal operator and $C^\pm$ rotation operator act on the nodal point at $\bK_0$. The two nodal points with the opposite winding number at the same point loses the protection from chiral symmetry. }
\label{mapping}
\end{figure}

	Time reversal and $C_2$ symmetry operators both map nodal points at $\pm \bK_0$ to ones at $\mp\bK_0$ respectively. There are two possibilities of the operator algebra arrangements to determine the existence of the protected nodal points. We assume the winding number at $\bK_0$ to be $w$ and time reversal operation $T^\pm$ leads to the winding number at $-\bK_0$ to be $\mp w$. On the one hand, we use $C_2^\mp$ operation, which correspond to the first operation $T^\pm$ respectively, to map the winding number at $-\bK_0$ back to the one at $\bK_0$ as illustrated in Fig.~\ref{mapping}(a,b). After the two symmetry operations, the winding number $w$ at $\bK_0$ is identical to the original one. Hence, this nodal point survives in the symmetry operations. On the other hand, after the $T^\pm$ operation, we exchange the two algebra types of $C_2$ operations so that $C_2^\pm$ correspond to the first operation $T^\pm$ respectively as illustrated in Fig.~\ref{mapping}(c,d). After the two symmetry operations, the mapping winding number $-w$ at $\bK_0$ coexists with the original winding number $w$ at the same location. Since the total winding number at $\bK_0$ vanishes, $T^\pm$ time reversal symmetry and $C_2^\pm$ symmetry always force nodal points to lose the protection from chiral symmetry in the entire BZ. Thus, when a WG in some symmetry class preserves $T^\pm$ and $C_2^\pm$ symmetries, the number of nodal points protected by chiral symmetry is always zero. Similarly, in the presence of $C_3^-$ symmetry, by performing three times of $C_3^-$ rotation operation the nodal point moving back to $\bK_0=(C_3^-)^3 \bK_0$ has $w$, $-w$ winding number together. Since the total winding number vanishes at $\bK_0$, which can be any point in the BZ, a nodal point protected by chiral symmetry is absent in the entire BZ.
	
	The trivialization of the nodal points can be extended to nodal points at symmetry-invariant points. According to Table \ref{winding number table}, any of $T^+,\ C^-_n,$ and $M^+$ operations leads to $\nu(\bK_0)=-\nu(g\bK_0)$. Hence, when a nodal point is located at the corresponding time-reversal, rotation, or mirror invariant points, its winding number must vanish and the nodal point does not have chiral symmetry protection.


The charge of the nodal point is defined by the winding number $C(\bK_0)=\nu(\bK_0)$ and  the absolute charge number $\nu_{\rm{abs}}$ is defined in Eq.~\ref{absolute number}. Given WG symmetry class and the algebra of the symmetry operators, the smallest number of the absolute charge number $\nu_{\rm{abs}}$ indicates the minimal configuration of the nodal points. We note that the minimal configuration of nodal points in BZ might not be unique.
 Since the winding numbers of the nodal points have the weight in this indicator number, the minimal number of the nodal points cannot directly indicate the minimal configuration. The only exception is the all of the nodal points in the BZ are Dirac node with $\nu=\pm 1$. In this case, the number of the Dirac nodes is equal to $\nu_{\rm abs}$.

	Since a nodal point at a time-reversal invariant point inherits additional constraints from the the ten-fold classification of the topological nodal superconductors~\cite{ChiuSchnyder14,RevModPhys.88.035005}, we separate the classification of the no-go theorem to two categories --- \textit{off} and \textit{at} time-reversal invariant (TRI) points. We note that class AIII, which does not have TRI points, is studied only in the \textit{off}-TRI point case due to the absence of time-reversal symmetry.
	
\subsection{Nodal points off time-reversal invariant points} \label{chiral off}


According to the classification of topological semimetals and nodal topological superconductors~\cite{ChiuSchnyder14,RevModPhys.88.035005}, for the five AZ symmetry classes preserving chiral symmetry, nodal points away from TRI points can be characterized by winding number. In the presence of crystalline symmetries, focusing on the irreducible BZs is sufficient to study the minimal configuration of the nodal points in entire BZ. The number of nodal points crucially depends on their locations in the irreducible BZ and the algebra between chiral operator and other symmetry operators. We provide the complete no-go theorem for nodal points away from TRI points in Table \ref{no-go away} by using the following four steps to find the minimal $\nu_{\rm{abs}}$ of the nodal points in the different configurations.


(a) We check if the symmetries force nodal points to possess zero winding number. For example, when the system preserves $T^\pm$ time-reversal symmetry and $C_2^\pm$ symmetry, there are no nodal points protected by chiral symmetry.  Allowing nodal points with non-zero winding number, the symmetries do not restrict the value of the winding number away from any TRI points. To find the minimal configuration of the nodal points, we first place one Dirac point with winding number $\nu=1$ in the irreducible BZ but not at TRI points.

(b) Different placements of the Dirac point lead to different configurations of the Dirac points in the entire BZ. We separately consider the location of the Dirac point at {\it mirror lines, rotation centers, and general points}. Commonly, rotation centers are in mirror lines, although $K$ and $K'$ points for WG$\#14$ are two of the exceptions. If the Dirac point is located at a symmetry-invariant point, we have to further check if the symmetry corresponding to the invariant point trivializes the Dirac point. For example, according to Table \ref{winding number table}, $M^+$ forces the winding number of the Dirac point at the mirror line to vanish since $\bK_0=M^+\bK_0$ and $\nu(\bK_0)=-\nu(M^+\bK_0)$. If the non-zero winding number survives, we can continue to discuss the configuration of the nodal points.


(c) Using the symmetry operation (rotation, mirror, or time-reversal), we extend the nodal point from the irreducible BZ to the remaining area of the BZ. Following table \ref{winding number table}, we determine the winding numbers of the extended nodal points based on the algebra between the chiral symmetry operator and the symmetry operators for the extension.

(d) Since the summation over the winding numbers of all the nodal points in the entire BZ must vanish as the key result of the original Nielsen-Ninomiya theorem (\ref{nogo}) (see appendix \ref{neutralization_proof} for the proof)
\begin{equation}\begin{aligned}
\sum_i\nu(\boldsymbol{K}_{0}^i)&=0, \label{zero winding}
\end{aligned}\end{equation}
we have to check if the summation is zero after the extension of nodal points. If so, this configuration has the minimal configuration of the nodal points. If not, we add an additional nodal point and repeat step (b,c) until the total winding number vanishes (\ref{zero winding}). We note that when we place a new nodal point, its winding number is usually chosen to be either $+1$ or $-1$ to neutralize the entire BZ and to have minimal $\nu_{\rm{abs}}$ but exceptions are possible.   

\renewcommand\arraystretch{1.5}
\begin{table*}
\caption{\label{character2} {\bf The absolute winding number $\nu_{\rm{abs}}$ for nodal points protected by chiral symmetry away from TRI points.} The table exhaustively provides the values of $\nu_{\rm{abs}}$ for all the possible minimal configurations for each WG symmetry class. The two point group generators $(M_x^\pm, C_n^\pm)$ with signs indicates the algebra between the chiral symmetry operator and the generators. While `$0$' indicates the absence of nodal points protected chiral symmetry, the non-zero number $\nu_{\rm{abs}}$ indicates the minimal number of Dirac points with $\pm1$ winding numbers. The non-zero $\nu_{\rm{abs}}$ without location specification denotes that all Dirac points in the minimal configuration are located at general points. Label $\sim$ above some numbers indicates all nodal(Dirac) points connected by symmetries in the minimal configuration and label $*$ indicates some nodal points possess high charges ($|\nu|>1$). Furthermore, while ``MLs" is an abbreviation for mirror lines, label ${\circ}$ indicates that the Dirac points are in the effective mirror lines ($T*M$ invariant lines), and label ${\bullet}$ indicates that Dirac nodes can be located in ordinary and effective mirror lines.
}
\label{no-go away}
\setlength{\tabcolsep}{1mm}{
\begin{tabular}{|c|c|c|c|c|}
  \hline
 WG & Generators & AIII & BDI/CII $T^+$ & DIII/CI $T^-$ \\\hline
$\#$1&  & 2 & $\tilde{2}$ & 4  \\\hline
$\#$2& $C^{+}_2$  &  4 or $4$($\Gamma$, M, X, Y)   & 0 & 4  \\
    & $C^{-}_2$  & $\tilde{2}$ & $\tilde{2}$ & 0  \\\hline
$\#$3,4,5& $M^{+}_{x}$ & $\tilde{2}$ & $\tilde{2}^{\circ} $MLs or $\tilde{4}$  & $\tilde{4}$ \\
     & $M^{-}_{x}$  & 2 MLs including ($\Gamma$, M, X, Y)  or 4   & $\tilde{2}$ MLs or $\tilde{4}$  & 4$^{\bullet}$MLs or 8 \\\hline
$\#$6,7 & $C^{+}_2$ and $M^{+}_{x}$ & $\tilde{4}$ & 0 & $\tilde{4}$  \\
 ~~~8,9  &$C^{+}_2$ and $M^{-}_{x}$  &  4 MLs or $4$($\Gamma$, M, X, Y)  or 8 & 0 & 4 MLs or 8  \\
      & $C^{-}_2$ and $M^{\pm}_{x}$ & $\tilde{2}$ MLs or $\tilde{4}$ & $\tilde{2}$ MLs or $\tilde{4}$ & 0  \\\hline
$\#$10& $C^{+}_4$ &  $4$($\Gamma$, M, X, Y) or 8 & 0 & 8  \\
     & $C^{-}_4$ &  $\tilde{4}$ & 0 & $\tilde{4}$  \\\hline
$\#$11,12 & $C^{+}_4$ and $M^{+}_{x}$ & $\tilde{8}$ & 0 & $\tilde{8}$  \\
     & $C^{+}_4$ and $M^{-}_{x}$ &  $4$($\Gamma$, M, X, Y) or 8 MLs or 16 & 0 & 8 MLs or 16  \\
      & $C^{-}_4$ and $M^{\pm}_{x}$ & $\tilde{4}$ MLs or $\tilde{8}$ & 0  & $\tilde{4}$ MLs or $\tilde{8}$ \\\hline
$\#$13   & $C^{+}_3$ &   $2$(K, K$^{'}$) or $4^{*}$(K, K$^{'}$, $\Gamma$) or $6$(K, K$^{'}$, $\Gamma$, M$_{1/2/3}$) or $6^{*}$($\Gamma$, M$_{1/2/3}$)  or 6 & $\tilde{2}$(K, K$^{'}$) or $\tilde{6}$  &  12 \\
     & $C^{-}_3$ & 0& 0  & 0 \\\hline
$\#$14& $C^{+}_3$ and $M^{+}_{y}$ & $\tilde{2}$(K, K$^{'}$) or $\tilde{6}$ & $\tilde{2}$(K, K$^{'}$) or $\tilde{6}^{\circ}$MLs or $\tilde{12}$  & $\tilde{12}$ \\
     & $C^{+}_3$ and $M^{-}_{y}$ & $4^{*}$(K, K$^{'}$, $\Gamma$) or 6 MLS or $6$(K, K$^{'}$, $\Gamma$, M$_{1/2/3}$) or  $6^{*}$($\Gamma$, M$_{1/2/3}$) or 12 & $\tilde{6}$ MLs or $\tilde{12}$ &  12$^{\bullet}$MLs or 24  \\
      & $C^{-}_3$ and $M^{\pm}_{y}$ & 0 & 0  & 0 \\\hline
$\#$15& $C^{+}_3$ and $M^{+}_{x}$ & $\tilde{6}$ & $\tilde{6}^{\circ}$MLs or $\tilde{12}$ & $\tilde{12}$  \\
      & $C^{+}_3$ and $M^{-}_{x}$ & $2$(K, K$^{'}$), $4^{*}$(K, K$^{'}$, $\Gamma$), 6 MLs, $6$(K, K$^{'}$, $\Gamma$, M$_{1/2/3}$), $6^{*}$($\Gamma$, M$_{1/2/3}$), or 12 & $\tilde{2}$(K, K$^{'}$) or $\tilde{6}$ MLs or $\tilde{12}$ & 12$^{\bullet}$MLs or 24  \\
      & $C^{-}_3$ and $M^{\pm}_{x}$ & 0 & 0  & 0 \\\hline
$\#$16& $C^{+}_6$ & $4^{*}$(K, K$^{'}$, $\Gamma$) or $6$(K, K$^{'}$, $\Gamma$, M$_{1/2/3}$) or $6^{*}$($\Gamma$, M$_{1/2/3}$) or 12& 0  & 12 \\
     & $C^{-}_6$ & $\tilde{2}$(K, K$^{'}$) or $\tilde{6}$& $\tilde{2}$(K, K$^{'}$) or $\tilde{6}$  & 0 \\\hline
$\#$17& $C^{+}_6$ and $M^{+}_{x}$ & $\tilde{12}$ & 0  & $\tilde{12}$ \\
     & $C^{+}_6$ and $M^{-}_{x}$ & $4^{*}$(K, K$^{'}$, $\Gamma$) or $6$(K, K$^{'}$, $\Gamma$, M$_{1/2/3}$) or $6^{*}$($\Gamma$, M$_{1/2/3}$) or 12 MLs or 24 & 0 &  12 MLs or 24  \\
      & $C^{-}_6$ and $M^{+}_{x}$ &  $\tilde{6}$ MLs or $\tilde{12}$ & $\tilde{6}$ MLs or $\tilde{12}$  & 0 \\
       & $C^{-}_6$ and $M^{-}_{x}$ & $\tilde{2}$(K, K$^{'}$) or $\tilde{6}$ MLs or $\tilde{12}$ & $\tilde{2}$(K, K$^{'}$) or $\tilde{6}$ MLs or $\tilde{12}$  & 0 \\\hline
\end{tabular}}
\end{table*}

Table \ref{no-go away} shows that the smallest absolute winding numbers $\nu_{\rm{abs}}$ subtly depend on WG symmetry class, locations of the nodal points, and algebra between the symmetry operators. In particular, choosing different locations of the first Dirac point in the irreducible BZ in step (c) leads to different configurations of nodal points. Table \ref{no-go away} shows all of the minimal nodal point configurations in the BZ for each WG with the five symmetry classes. Namely, the different locations of the nodal points correspond to different numbers $\nu_{\rm{abs}}$ even if the nodal points are away from TRI points.

\begin{figure}
\centerline{\includegraphics[width=0.45\textwidth]{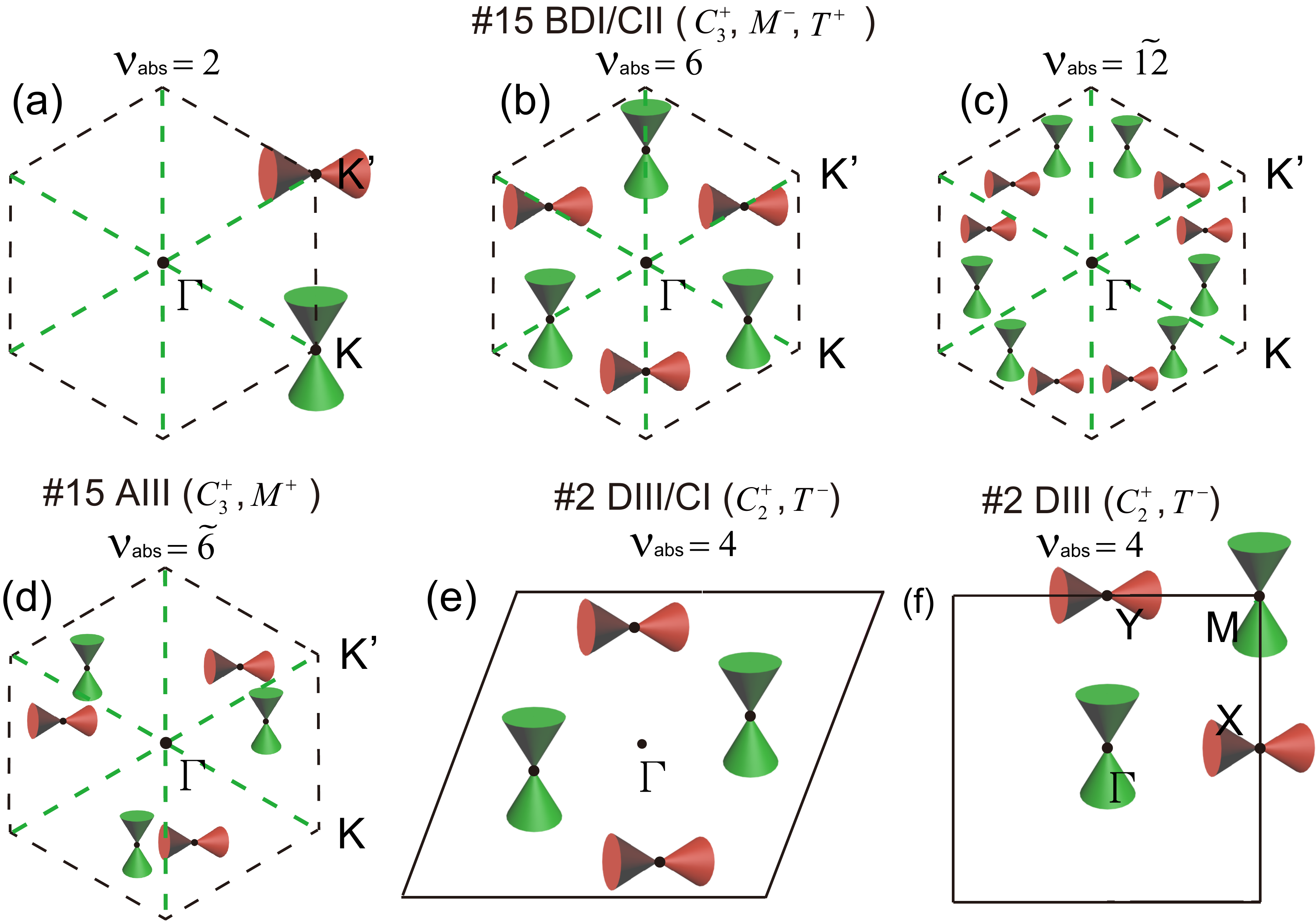}}
\caption{(color online) Examples of the configuration for nodal points away from TRI points. The green and red cones represent nodal points with $+1$ and $-1$ winding numbers respectively, while the dashed lines indicate the mirror lines.  (a-c) show that the nodal points are in the three different types of the locations --- rotation centers, mirror lines, and general points. (a-d) In the configurations all nodal points in the BZ can be connected by symmetries. (e) The symmetry do not link all nodal points since the two red nodal points with $\nu=-1$ connected by the symmetries can move freely without affecting the green nodal points with $\nu=1$. (f) four Dirac points located at ($\Gamma,M,X,Y$) are the minimal configuration for WG$\#2$ with $C_2^+$. {\clr move $\sim$}
\label{eg_off_TRI} }
\end{figure}

	Now we consider a specific WG symmetry class to demonstrate the aforementioned approach for the generalized no-go theorem by using WG$\# 15$ with $C_3^+,\ M^-$ in class BDI/CII ($T^+$) as an example. (a) The symmetries do not kill winding numbers of any nodal points. (b) There are three types of the locations to place the first nodal point --- rotation centers ($\Gamma, K,\ K'$), mirror lines ($\overline{\Gamma K}$, $\overline{\Gamma K'}$), and general points. Let us first focus on the rotation centers by placing a Dirac point at $K$ with $+1$ winding number, since $\Gamma$ is a TRI point, which will be discussed in the next subsection. (c) Time-reversal symmetry $T^+$ maps the Dirac point at $K$ to the Dirac point at $K'$ with $-1$ winding number. In other words, these two Dirac points are connected by time-reversal symmetry. (d) The total winding number at $K$ and $K'$ is zero. Hence, the Dirac points at $K,\ K'$ are the minimal configuration as shown in Fig.~\ref{eg_off_TRI}(a) and $\nu_{\rm{abs}}=2$ is the absolute winding number.

	 Alternatively, when the first Dirac point with $-1$ winding number is placed at one point of the mirror lines ($\overline{\Gamma K}$, $\overline{\Gamma K'}$, not $K,\ K'$), $C_3^+$ rotation symmetry generates 2 additional copies of the Dirac points with the same winding number. Furthermore, from these 3 Dirac points, time-reversal symmetry produces other 3 Dirac points with $-1$ winding numbers as shown in Fig.~\ref{eg_off_TRI}(b). Since the total winding number of the six Dirac points vanishes, the six Dirac points are the minimal configuration for mirror lines. Similarly, for the first Dirac point located at a general point, as illustrated in Fig.~\ref{eg_off_TRI}(c) mirror symmetry doubles the number of the six Dirac points, which are extended by time-reversal symmetry and $C_3$ rotation symmetry. In this case, $\nu_{\rm{abs}}=12$. For the three types of the nodal point locations, each nodal point can connect any other nodal point through symmetries. That is, once we fix the location of the first Dirac point, all of the nodal point locations are determined. Hence, for this WG symmetry class with $T^+, C_3^+, M^-_x$, all nodal points in each configuration are connected by symmetries. 
To mark the connections of the nodal points, we put $\sim$ on the minimal absolute number $2,\ 6,\ 12$ in Table \ref{no-go away}. 

	Let us consider another example for WG$\#15$ in class AIII with $C_3^+,\ M^+$. Reflection symmetry $M^+$ enforces vanishing winding number for any nodal point located at a mirror line ($\overline{\Gamma K}$, $\overline{\Gamma K'}$). In this regard, nodal points protected by chiral symmetry can survive only at general points and then we place the first Dirac point with $+1$ winding number at a general point. Using reflection symmetry and $C_3$ rotation symmetry, there are 3 Dirac points with $+1$ winding number and other three with $-1$ winding number; as shown in Fig.~\ref{eg_off_TRI}(d) this minimal configuration with six nodal points connected by the symmetries is labeled by $\tilde{6}$.
	
	Another example is WG$\# 2$ in class DIII, CI with $T^-$ and $C_2^+$. (a) With these symmetries, nodal points with non-zero winding numbers survive. (b) We place a Dirac point with $+1$ winding number at a general point (say $\bK_0$). (c) These two symmetries ($T^-,C_2^+$) lead to another Dirac point with $+1$ winding number at $-\bK_0$. (d) However, the total winding number is two so we have to place another Dirac point with $-1$ winding number at another general point. Due to the symmetries, there are two additional Dirac points with $-1$ winding numbers. With the four Dirac points as shown in Fig.~\ref{eg_off_TRI}(e), the total winding number is neutralized. In the literature it has been shown that the number of the Dirac nodes must be the multiple of 4 due to $C_2^+$ symmetry~\cite{PhysRevB.102.165150}. The Dirac points with the opposite winding numbers are not directly connected by the symmetries so that we can freely choose their locations even when the Dirac points with $\nu=1$ are spatially fixed.  

	Most of the WG symmetry classes follow the aforementioned recipe to search for the minimal configuration of the protected nodal points. However, for class AIII there are some exceptions, which are nodal points located in the special points ($\Gamma,M,X,Y$) and $(\Gamma,M_i)$ for the two types of the BZs respectively. An additional step of the recipe has to be considered in these special cases.

\begin{figure}
\centerline{\includegraphics[width=0.45\textwidth]{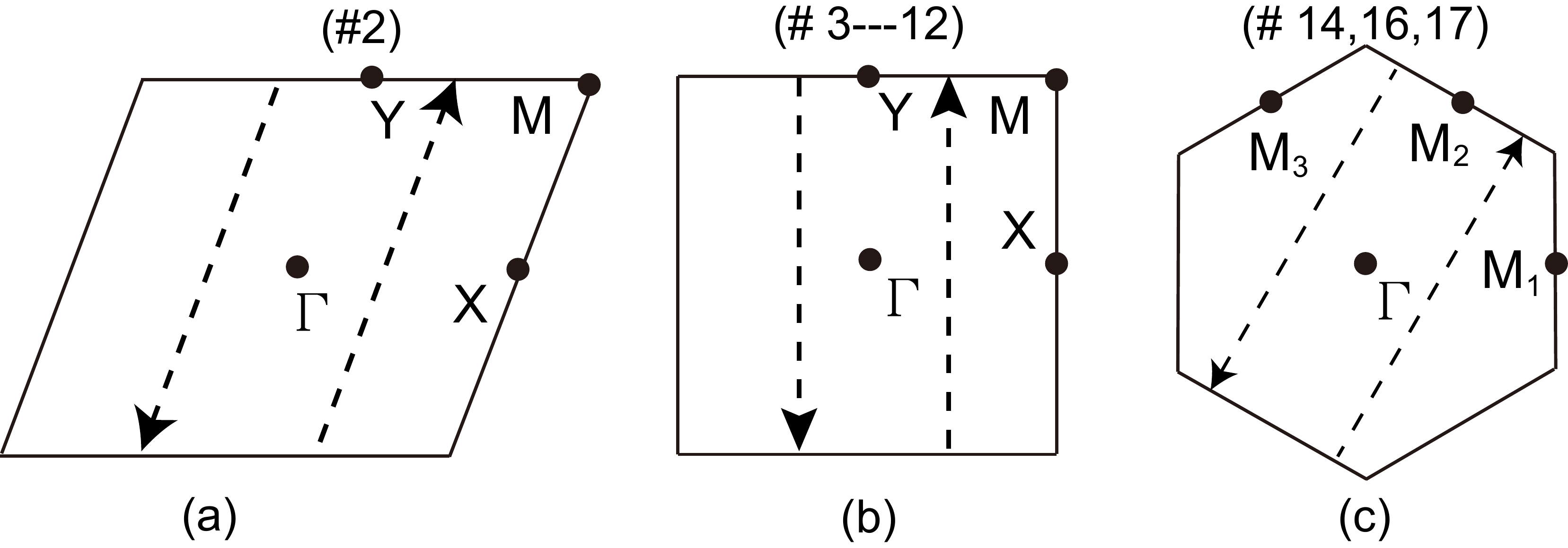}}
\caption{ The dashed lines are the integral paths of the winding numbers through the entire BZs. Due to $C_2^+$ rotation symmetry or $M^-$ reflection symmetry, the two winding numbers of the separate dashed lines, which are quantized, in each BZ are identical. Through the integral path deformation, the summation of the two winding numbers is identical to (a,b) $\nu(\Gamma)+\nu(Y)$ and (c) $\nu(\Gamma)+\nu(M_2)$.  
\label{invariant_BZ} }
\end{figure}

	(e) The winding number with the integral path through the entire BZ (the dash lines in Fig.~\ref{invariant_BZ}) must be quantized. Rotation symmetry ($C_2$) and mirror symmetry ($m$) can connect the two winding numbers with these two integral paths. For $C_2^+$ and $M^-$, these two winding numbers with the two symmetry related paths are identical. Furthermore, by considering nodal points located at the aforementioned special points, these two paths together can deform to two infinitesimal circles surrounding two of the special points, and the winding numbers are unchanged. Hence, the total winding number of the two special points must be even. For the WGs ($\#2,6-12$) possessing $C_2^+$ generator, the winding numbers in two of the special points obey
\bee
\nu(\Gamma)+\nu(Y)=0\ (\textrm{mod}\ 2), \nu(\Gamma)+\nu(X)=0\ (\textrm{mod}\ 2). \label{constraint1}
\ee	
For the WGs ($\#3-9,11,12$) possessing $M^-_x$ generator, the constraint is reduced to
\bee
\nu(\Gamma)+\nu(Y)=0\ (\textrm{mod}\ 2), \label{constraint2}
\ee
On the other hand, the hexagonal BZ has different special points. For WGs ($\#14,17$) possessing $M^-_y$ generator and WGs ($\#16,17$) possessing $C_2^+$ generator, we have
\bee
\nu(\Gamma)+\nu(M_i)=0\ (\textrm{mod}\ 2). \label{constraint3}
\ee
Thus, in this final step we have to check if the configuration of the nodal points satisfies Eqs.~\ref{constraint1}, \ref{constraint2}, or \ref{constraint3} for the corresponding WG symmetry classes. If so, the minimal configuration is found. If not, step (b,c,d) are need to be repeated until the constraint conditions are fulfilled.

	Let us use WG$\#2$ with $C_2^+$ in class AIII as an example. We place two Dirac points with $\pm 1$ winding number at $\Gamma$ and $M$ separately. Although the entire BZ is neutralized, Eq.~\ref{constraint1} is violated. In fact, no matter how the two Dirac points are located at two of the rotation invariant points ($\Gamma,X,Y,M$), Eq.~\ref{constraint1} does not hold. The only way is to have four Dirac points placed at the four invariant points separately and Fig.~\ref{eg_off_TRI}(f) illustrates one of the valid minimal configuration. 


%

\subsection{Nodal points at time-reversal invariant points}


The TRI points are special for nodal points. Since TRI points obey $\bK_0=-\bK_0$, in the presence of $T^+$ time reversal symmetry the winding number of the nodal point at this invariant point vanishes. The reason is that based on table \ref{winding number table} $T^+$ gives the opposite winding number at the same point. Class BDI/CII with $T^+$ never possesses a nodal point with non-zero winding number at any time-reversal invariant points. Hence, we discuss the no-go theorem for only two non-trivial AZ symmetry classes --- DIII and CI. The absolute numbers $\nu_{\rm{abs}}$ of the minimal configurations for the WG symmetry class are shown in Table~\ref{no-go TRS}. This non-trivial property is consistent with the classification of topological semimetals and nodal topological superconductors~\cite{ChiuSchnyder14,RevModPhys.88.035005}, which shows that in these two AZ symmetry classes the nodal points at TRI points can be characterized by non-zero winding number. Namely, class DIII and CI correspond to $\bZ$ and $2\bZ$ invariants respectively.

	Class AIII and CII may possess protected nodal points, since according to the classification table, nodal points in class AIII and CII can be characterized by $\bZ$ and $\bZ_2$ invariants respectively. However, TRI points are  absent in class AIII, which has been exhaustively studied in the previous subsection. In addition, although in class CII the non-zero $Z_2$ invariant can protect the nodal point, here we focus on only nodal points protected by winding numbers. Hence, class AIII and CII are excluded in the no-go theorem for TRI nodal points.

\renewcommand\arraystretch{1.5}
\begin{table*}
\caption{\label{character1} {\bf The absolute winding number $\nu_{\rm{abs}}$ for nodal points protected by chiral symmetry at TRI points.} The table shows the minimal $\nu_{\rm{abs}}$ for the 17 WGs in class DIII and BDI. 
The two point group generators ($C_n^\pm, M_x^\pm$) with signs indicate the algebra between the chiral symmetry operator and the generators; $\nu_{\rm{abs}}=0$ indicates the absence of nodal points protected by chiral symmetry. In the table there are some cases that nodal points are forced to be placed at $K,K'$, which are not TRI points, due to the charge neutralization. In some cases, high-ordered nodal points with $|\nu|>1$ must be present in the minimal configurations; hence, we use * to distinguish the particular cases from the case possessing only Dirac nodes.
}
\label{no-go TRS}
\setlength{\tabcolsep}{1.5mm}{
\begin{tabular}{|c|c|c|c|}
  \hline
 WG & Generators &  DIII $T^-$ & CI $T^-$ \\\hline
$\#$1&  &  $4$($\Gamma$, M, X, Y) & $4^{*}$ \\\hline
$\#$2& $C^{+}_2$  & $4$($\Gamma$, M, X, Y) & $4^{*}$  \\
    & $C^{-}_2$  &  0 & 0 \\\hline
$\#$3,4,5& $M^{+}_{x}$  &  0 & 0\\
     & $M^{-}_{x}$  &  $4$($\Gamma$, M, X, Y) & $4^{*}$ \\\hline
$\#$6,7,8,9 & $C^{+}_2$ and $M^{+}_{x}$ &  0 & 0 \\
      &$C^{+}_2$ and $M^{-}_{x}$  & $4$($\Gamma$, M, X, Y) & $4^{*}$ \\
      & $C^{-}_2$ and $M^{\pm}_{x}$ &  0  & 0 \\\hline
$\#$10& $C^{+}_4$ & $4$($\Gamma$, M, X, Y) & $4^{*}$($\Gamma$, M) or  $8^{*}$  \\
     & $C^{-}_4$ &  0  & $4^{*}$(X, Y) \\\hline
$\#$11,12 & $C^{+}_4$ and $M^{+}_{x}$  &  0 & 0 \\
     & $C^{+}_4$ and $M^{-}_{x}$ &  $4$($\Gamma$, M, X, Y) & $4^{*}$($\Gamma$, M) or $8^{*}$ \\
      & $C^{-}_4$ and $M^{\pm}_{x}$ &  0 & 0 \\\hline
$\#$13   & $C^{+}_3$ &  $4^*$(K, K$^{'}$, $\Gamma$),  $6$(K, K$^{'}$, $\Gamma$, M$_{1/2/3}$), or $6^{*}$($\Gamma$, M$_{1/2/3}$) & $4^{*}$(K, K$^{'}$, $\Gamma$), $12^{*}$\{(K, K$^{'}$, $\Gamma$, M$_{1/2/3}$), or ($\Gamma$, M$_{1/2/3}$)\}  \\
     & $C^{-}_3$ &  0 & 0\\\hline
$\#$14& $C^{+}_3$ and $M^{+}_{y}$ &  0 & 0 \\
     & $C^{+}_3$ and $M^{-}_{y}$ &   $4^*$(K, K$^{'}$, $\Gamma$),  $6$(K, K$^{'}$, $\Gamma$, M$_{1/2/3}$), or $6^{*}$($\Gamma$, M$_{1/2/3}$) & $4^{*}$(K, K$^{'}$, $\Gamma$), $12^{*}$\{(K, K$^{'}$, $\Gamma$, M$_{1/2/3}$), or ($\Gamma$, M$_{1/2/3}$)\}\\
      & $C^{-}_3$ and $M^{\pm}_{y}$ &  0 & 0 \\\hline
$\#$15& $C^{+}_3$ and $M^{+}_{x}$ &  0 & 0 \\
      & $C^{+}_3$ and $M^{-}_{x}$ &  $4^*$(K, K$^{'}$, $\Gamma$),  $6$(K, K$^{'}$, $\Gamma$, M$_{1/2/3}$), or $6^{*}$($\Gamma$, M$_{1/2/3}$) & $4^{*}$(K, K$^{'}$, $\Gamma$), $12^{*}$\{(K, K$^{'}$, $\Gamma$, M$_{1/2/3}$), or ($\Gamma$, M$_{1/2/3}$)\}\\
      & $C^{-}_3$ and $M^{\pm}_{x}$ &  0 & 0\\\hline
$\#$16& $C^{+}_6$ &  $4^*$(K, K$^{'}$, $\Gamma$),  $6$(K, K$^{'}$, $\Gamma$, M$_{1/2/3}$), or $6^{*}$($\Gamma$, M$_{1/2/3}$) & $4^{*}$(K, K$^{'}$, $\Gamma$), $12^{*}$\{(K, K$^{'}$, $\Gamma$, M$_{1/2/3}$), or ($\Gamma$, M$_{1/2/3}$)\} \\
     & $C^{-}_6$ &  0 & 0\\\hline
$\#$17& $C^{+}_6$ and $M^{+}_{x}$ &  0 & 0\\
     & $C^{+}_6$ and $M^{-}_{x}$ &  $4^*$(K, K$^{'}$, $\Gamma$), $6$(K, K$^{'}$, $\Gamma$, M$_{1/2/3}$), or $6^{*}$($\Gamma$, M$_{1/2/3}$) &  $4^{*}$(K, K$^{'}$, $\Gamma$), $12^{*}$\{(K, K$^{'}$, $\Gamma$, M$_{1/2/3}$), or ($\Gamma$, M$_{1/2/3}$)\} \\
      & $C^{-}_6$ and $M^{\pm}_{x}$ &  0 & 0\\\hline
\end{tabular}}
\end{table*}

	
	The recipe to find the smallest absolute winding number $\nu_{\rm{abs}}$ for each WG symmetry class is identical to the previous subsection for off-time-reversal-invariant-point. Additional cares are needed for class DIII with $T^2=-1$ and for class CI with $T^2=+1$.
First, due to $T^2=-1$, the Kramers' degeneracy leads to double degeneracy for each band at all of the TRI points. In addition, chiral symmetry pairs two energy bands with $\pm E$. Hence, for the odd number of the energy band pairs in the system there must exist an energy band pair at $E=0$ at all of the four time-reversal invariant points in the BZ. Since we assume that nodal lines are absent, the nodal points always appear at the four time-reversal invariant points. On the other hand, since in class CI $2Z$ invariant given by the classification table~\cite{ChiuSchnyder14,RevModPhys.88.035005} indicates that the winding number of the nodal point located at a TRI point must be even. Hence, in step (b) the winding number of the first nodal point should be chosen to be $+2$, instead of $+1$. The Kramers' degeneracy in class DIII and even winding numbers in class CI automatically fulfill the constraints in Eq.~\ref{constraint1}, \ref{constraint2}, \ref{constraint3} in step (e). The remaining steps to find the minimal configuration of the nodal points are identical to the previous subsection.


\begin{figure}
\centerline{\includegraphics[width=0.45\textwidth]{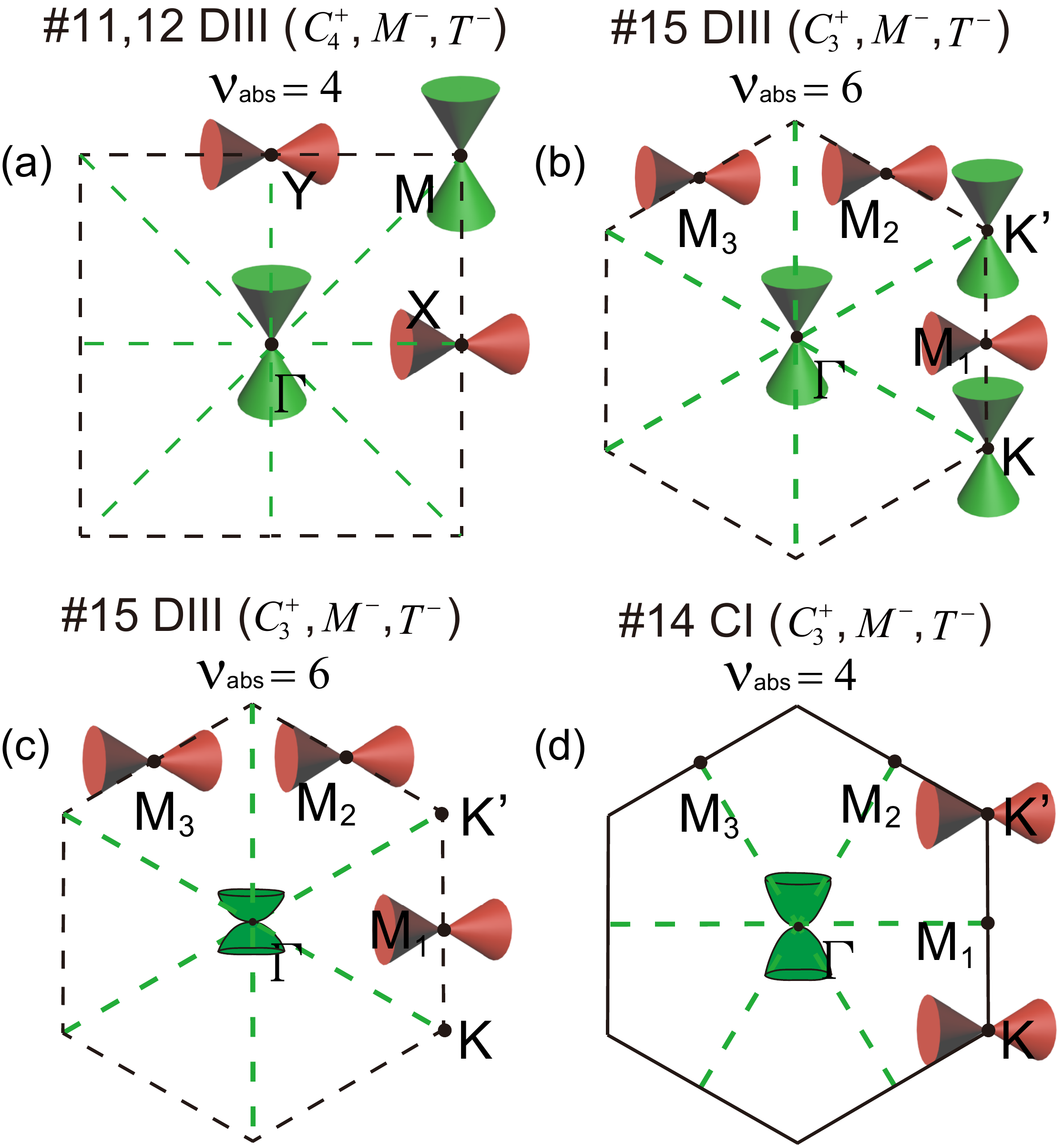}}
\caption{(color online) {\bf Examples of the configuration for nodal points at the TRI points.} The green and red cones with linear dispersions represent nodal points with $+1$ and $-1$ winding numbers respectively, the green cones with quadratic and cubic dispersions represent nodal points with $+2$ and $+3$ winding numbers in (d) and (c) respectively. The dashed lines indicate the mirror lines. {\clr no $\sim$}
\label{eg TRI} }
\end{figure}

 	We demonstrate three examples to show how to find the minimal configurations of the nodal points. We first consider WG$\#11,12$ with $C_4^+$ and $M^-$ in class DIII ($(T^-)^2=-1$). These two WGs share the same point groups. In the BZ, $\Gamma, X, Y, M$ are TRI points and three of them (except for $Y$) are in the irreducible zone as shown in Fig.~\ref{irreducible BZ}(g). 
	As discussed previously, due to the Kramers' degeneracy, in class DIII once a nodal point at $E=0$ appears at one of the TRI points, all of the TRI points must be nodal points. Since $C_4^+$ rotation symmetry or $M^-$ mirror symmetry forces the winding numbers at $X, Y$ to be identical. The only minimal configuration is formed by Dirac points with $+1$ winding numbers at $\Gamma, M$ and ones with $-1$ winding numbers at $X, Y$ as shown in Fig.~\ref{eg TRI}(a).

 	Another example is WG$\#15$ in class DIII with $T^-,$ $C^+_3,\ M^-$. The four points ($\Gamma,M_1,M_2,M_3$) are TRI points in the hexagonal zone as shown in Fig.~\ref{irreducible BZ}(j). The Kramers' degeneracy leads to nodal points located at these four points together. Due to $C_3^+$ rotation symmetry, $M_1,M_2,M_3$ possess the same winding number (say $-1$). However, $\Gamma$ with $+1$ winding number cannot neutralize the entire zone. Therefore, we place two Dirac point with $+1$ winding numbers at $K,K'$ and the total winding number vanishes as shown in Fig.~\ref{eg TRI}(b). However, the minimal configuration with $\nu_{\rm{abs}}$ is not unique. Alternatively, we choose a nodal point with $+3$ winding number at $\Gamma$ and three Dirac points with $-1$ winding number at three $M_i$ points as shown in Fig.~\ref{eg TRI}(c); this choice is the other minimal configuration.

	We consider WG$\#14$ in class CI with $T^-, C^+_3, M^-$ as the last example. We place the first nodal point at $\Gamma$. Since in class CI any nodal point at a TRI point must possess even winding number, the simplest choice for the nodal point at $\Gamma$ is $+2$ winding number. With $C_3^+$ rotation symmetry, a nodal point in the most area of the BZ has three copies with the same winding number, except for $K', K, \Gamma$. Therefore, a Dirac point with $-1$ winding number is placed at $K$. Due to $T^-$ time-reversal symmetry or $M^-$ reflection symmetry, another Dirac point appear at $K'$ with $-1$ winding number. The total winding number vanishes and as shown in Fig.~\ref{eg TRI}(d) this configuration with $\nu_{\rm abs}=4$ is minimal.

\subsection{Examples}
	
	The symmetry constraints control the configurations of the nodal points protected by chiral symmetry. Table~\ref{no-go away},\ref{no-go TRS} provide all of the possible minimal configurations for each WG symmetry class. The minimal configurations can be building blocks to understand realistic systems possessing the nodal points. Here we study several examples in condensed matter systems to show that the examples of the nodal points cannot escape from the predicted configurations in our generalized no-go theorem. 

\begin{center}
{\bf Off time-reversal invariant points}
\end{center}

	Nodal points protected by chiral symmetry away from TRI points can be realized in 2D Dirac materials and time reversal symmetric superconductors. 2D Dirac materials, such as graphene, may preserve accidental chiral (sublattice) symmetry, whereas time reversal symmetric superconductors naturally preserve chiral symmetry stemming from in intrinsic time reversal symmetry and particle-hole symmetry.  


\subsubsection{D-wave superconductors: WG$\# 11$ in class CI}

We first consider a two-dimensional nodal Cu-based superconductor with $d$-wave pairing~\cite{RevModPhys.67.515,RevModPhys.72.969} with spin $SU(2)$ symmetry. The Bogoliubov-de Gennes Hamiltonian~\cite{sachdev_2011} in the $2\times 2$ Nambu basis can be written in the second quantization form of
\begin{equation}\begin{aligned}
H&=\sum_{\bm{k}}\psi^{\dag}(\bm{k})\mathcal{H}(\bm{k})\psi(\bm{k})\\
&=\sum_{\bm{k}}\hat{\psi}^{\dag}(\bm{k})[\varepsilon(\bm{k})\tau_z+\Delta(\bm{k})\tau_x]\hat{\psi}(\bm{k}),
\end{aligned}\end{equation}
where $\hat{\psi}(\bm{k})=(\hat{c}_{\bm{k},\uparrow},\hat{c}_{-\bm{k},\downarrow}^\dag)$, $\varepsilon(\bm{k})=-2t_1(\cos k_x+\cos k_y)-4t_2\cos k_x\cos k_y-\mu$, and $\Delta(\bm{k})=\Delta_0(\cos k_x-\cos k_y)$. We choose the parameters  as $t_1=0.25$, $t_2=-0.05$, $\mu=-0.18$ and $\Delta_0=0.1$. 
First, the system preserves particle-hole symmetry (\ref{particle-hole symmetry}) with $C=\tau_y\mathcal{K}$. 
Secondly, time-reversal symmetry (\ref{time reversal symmetry}) is also preserved with $T=\mathcal{K}$. By combining the two symmetry operators above, chiral symmetry (\ref{chiral symmetry}) is also preserved with $S=\tau_y$. We have $C^2=-1$ and $T^2=1$ corresponding to class CI, which is consistent with the symmetry class of time-reversal symmetric superconductor with spin $SU(2)$ symmetry in the literature~\cite{Schnyder2008};  furthermore, the time-reversal operator $T^-$ anti-commutes with the chiral symmetry operator $S$. On the other hand, the system belongs to WG$\#11$ possessing $C_{4v}$ point group symmetry with generators $C_{4}=\tau_z$ and $M_{x}=\tau_0$, which obey $C_{4}H_{BdG}(k_x,k_y)C_4^{-1}=H_{BdG}(-k_y,k_x)$ and $M_{x}H_{BdG}(k_x,k_y)M_{x}^{-1}=H_{BdG}(-k_x,k_y)$, and the (anti)commutation relations of the generator operators are given by $C^{-}_{4}$ and $M^{+}_{x}$.

\begin{figure}
\centerline{\includegraphics[width=0.5\textwidth]{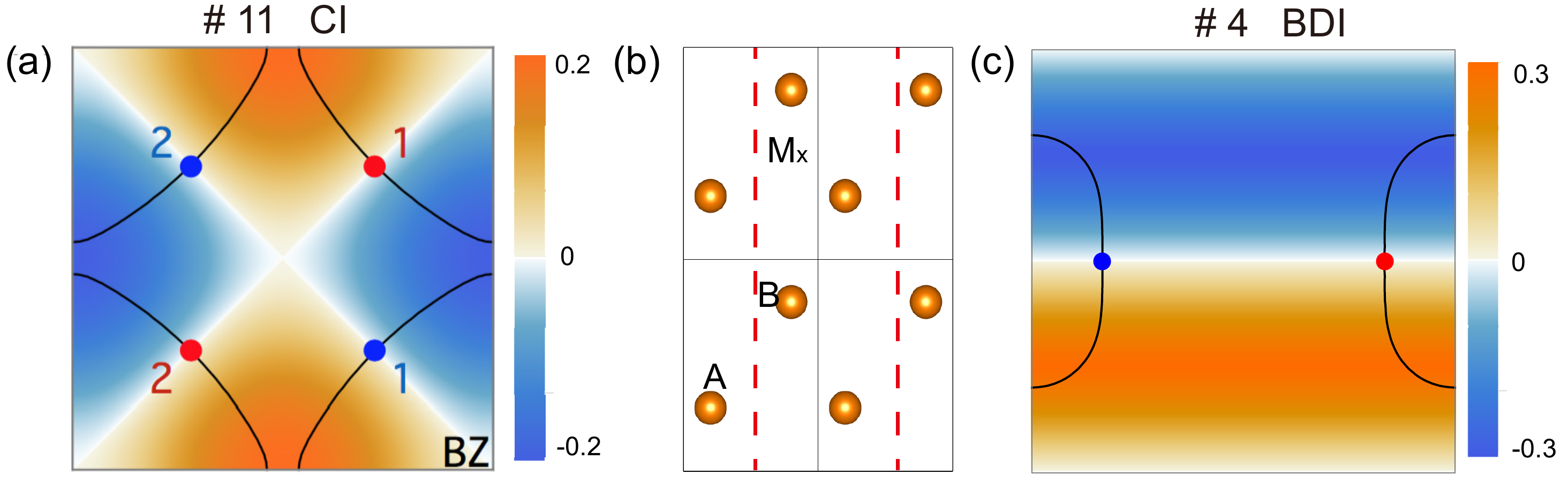}}
\caption{(color online) Time-reversal symmetric superconductors possess protected nodal points. The colorbar from blue to yellow indicates the value of the order parameter $\Delta(\bm{k})$, and the black lines represent the Fermi lines in the absence of the superconductivity. Their crossings form the nodal points with $\nu = +1$(red) and $\nu=-1$(blue). (a) The d-wave SC belonging to WG$\#11$ in class CI has 4 nodal points in the BZ. (b) The crystal structure for the toy model of the glide-reflection symmetric superconductor belongs to WG$\# 4$ in class BDI. (c) There are 2 nodal points with $\nu=\pm 1$ in the BDI superconductor.
\label{d-wave} }
\end{figure}

This $d-$wave SC has four nodes located at the mirror lines as shown in Fig.~\ref{d-wave}(a). The Hamiltonian of the low energy expansion near the four nodal points $\bm{k_{0\pm}}=($0.44$\pi$,$\pm$ 0.44$\pi$), $\bm{k_{0\pm}}'=(-$ 0.44$\pi$,$\mp$ 0.44$\pi$) are in the form of 
\begin{equation}\begin{aligned}
&\mathcal{H}^{\rm{r}}_{\pm}(\bm{k})\simeq \alpha (\delta k_{x\pm} \mp \delta k_{y\pm})\sigma_x+\beta (\delta k_{x\pm} \pm \delta k_{y\pm})\sigma_y,\\
&\mathcal{H}^{\rm{l}}_{\pm}(\bm{k})\simeq - \alpha (\delta k_{x\pm}'\pm \delta k_{y\pm}')\sigma_x- \beta (\delta k_{x\pm}'\mp \delta k_{y\pm}')\sigma_y,
\end{aligned}\end{equation}
where $\delta \bm{k}_\pm=\bm{k}-\bm{k_{0\pm}}$, $\delta \bm{k}'_\pm=\bm{k}-\bm{k_{0\pm}}'$, $\alpha=0.98$, and $\beta=0.45$.
Since chiral symmetry is still preserved with $S=\sigma_z$ and the linear dispersions lead to the four Dirac points with $\pm 1$ winding numbers, with the low-energy Hamiltonians $H_{\pm}^{\rm{r/l}}(\bk) = \delta k_i A_{ij} \delta k_j$,  $\rm{sign}(\det A)$ corresponds to $\pm 1$ winding numbers of the Dirac points. Therefore, $\nu$=+1 for $\bk_{0+}, \bk_{0+}'$ and $\nu$=-1 for $\bk_{0-}, \bk_{0-}'$ as shown in Fig.~\ref{d-wave}(a).



 Back to our no-go theorem in Table \ref{character2}, for WG $\#11$ with ${C}_4^-$ and $M_{x}^+$ in class CI, we have $\nu_{\rm{abs}}=4$ for nodal points located at mirror lines. This is consistent with the presence of the 4 nodal points with $\pm 1$ winding numbers in the $d-$wave SC model. By using ${C}_4^-$ and $M_{x}^+$ operations, we can generate the same distribution of the four Dirac points.

%

\subsubsection{Glide reflection symmetric superconductors: WG$\#4$ in class BDI}

	The classification of our no-go theorem depends on point groups, regardless of symmorphic and non-symmorphic symmetries. Glide reflection symmetry is the only type of the non-symmorphic symmetries in the 17 wallpaper groups. We study a toy model of a 2D glide plane symmetric superconductor in class BDI. Fig.\ref{d-wave} (b) shows 2D orthorhombic lattice of non-symmorphic WG$\#4$ with A and B sites in the primitive cell. The glide reflection symmetry, which is the only crystalline symmetry in WG$\#4$, has its operator $\hat{g}_{\delta}$= $\{M_{x}|\delta= (0, \frac{b}{2})\}$ exchanging site A and B with a reflection $M_{x}$ in the $x$ direction and half-lattice-constant movement along the $y$ direction. Then, we provide a toy model of the spinless fermion in the lattice
\begin{equation}\begin{aligned}
&H_{\rm{BdG}}(\bm{k})= \\
&(\epsilon(\bm{k})-\mu)\tau_{z}\sigma_{0}+\beta(\bm{k})\tau_{z}\sigma_{x}+\gamma(\bm{k})\tau_{z}\sigma_{y}-\Delta(\bm{k}) \tau_{y} \sigma_{0},
\end{aligned}\end{equation}
in the basis of $\psi(\bm{k})=(C_{\bm{k},A}, C_{\bm{k},B},C^{\dagger}_{-\bm{k},A}, C^{\dagger}_{-\bm{k},B} )^{T}$, where $\epsilon(\bm{k})=m_0+2t_2\cos(k_x)+2t_3\cos(k_y)$, $\beta(\bm{k})=2t_{1}\cos(\frac{k_{y}}{2})\cos(k_x+\frac{k_{y}}{2})$, $\gamma(\bm{k})=2t_{1}\cos(\frac{k_{y} }{2})\sin(k_x+\frac{k_{y}}{2})$, and superconductor gap $\Delta(\bm{k})=\Delta_{0}\sin(k_{y})$. We choose the parameters as $t_1=1.2$, $t_2=0.2$, $t_3=0.3$, $m_0=2$, $\mu=0$ and $\Delta_{0}=0.3$ so that two nodal points appear at  ($\pm 2\pi/3$,0). 

The spinless superconductor system naturally preserves particle-hole symmetry (\ref{particle-hole symmetry}) with its operator $C=\tau_{x} \sigma_{0}\mathcal{K}$ and time-reversal symmetry (\ref{time reversal symmetry}) with its operator $T=\tau_{0} \sigma_{0}\mathcal{K}$. Hence, this belongs to class BDI and chiral symmetry (\ref{chiral symmetry}) is preserved with $S=\tau_{x} \sigma_{0}$. By using chiral symmetry operator, the winding numbers for nodal points at ($\pm 2\pi/3$,0) are given by $\pm 1$; time-reversal symmetry $T^+$ connects these two nodal points with the opposite winding numbers. Due to the glide reflection symmetry, the BdG Hamiltonian obeys $M_g H(k_x, k_y)M_g^{-1}=H(-k_x, k_y)$, where the symmetry operator $M_g=e^{-i\frac{k_y}{2}}(\cos(\frac{k_y}{2})\tau_0\sigma_x+\sin(\frac{k_y}{2})\tau_0\sigma_y)$. Since this operator $M_g^+$ commutes with $S$, $M_g^+$ also connects the two nodal points with the opposite winding numbers. Away from TRI points, the two nodal points with $\pm 1$ winding numbers are consistent with Table~\ref{no-go away} for WG$\#4$ in class BDI.

%
%



\subsubsection{Graphene: WG$\#17$ in class BDI}\label{graphene section}

	It is known that there are two Dirac points in the graphene model, which can be simply described by the $p_z$ orbital containing nearest neighbor hoppings~\cite{graphene}. The spinless Hamiltonian in the basis of $(C^{A}_{k},C^{B}_{\bk})$ can be written as
\begin{eqnarray}
H_g(\bk)= \left(\begin{array}{cc}
 0 & h_{12}(\bk)   \\
   h^{*}_{12}(\bk) &   0    \\
\end{array}\right)
\end{eqnarray}
where $h_{12}(\bk)=t\left[e^{i(\frac{1}{2}k_y-\frac{\sqrt{3}}{6}k_x)}+e^{i\frac{\sqrt{3}}{3}k_x}+e^{i(-\frac{1}{2}k_y-\frac{\sqrt{3}}{6}k_x)}\right]
$ and $t$ is hopping parameter. Since the system preserves sublattice chiral symmetry $S=\sigma_z$, time reversal symmetry $T=\mathcal{K}$, effective particle-hole symmetry $C=\sigma_z\mathcal{K}$, we have class BDI with $T^+$. We note that the graphene is a non-interacting electron system in this case so the effective particle-hole symmetry, which stems from the combination of the sublattice symmetry and time reversal symmetry, is unrelated to superconductivity. On the other hand, the graphene belongs to WG$\#17$ with two generators  $M_{x}^-$=$\sigma_x$ and $C_6^-$=$\sigma_x$. The crystalline symmetries lead to
\begin{align}
M_{x}H_g(-k_x,k_y)M^{-1}_{x}&=H_g(k_x,k_y), \label{reflection} \\
C_6H_g(\frac{k_x}{2}+\frac{\sqrt{3}k_y}{2},\frac{k_y}{2}-\frac{\sqrt{3}k_x}{2})C^{-1}_6&=H_g(k_x,k_y). \label{C6}
\end{align}
The nodal points are present at $K:(2\pi/\sqrt{3},2\pi/3)$ and $K':(2\pi/\sqrt{3},-2\pi/3)$ with $+1$ and $-1$ winding numbers respectively.
Table \ref{no-go away} indicates $\nu_{\rm{abs}}=2$ at $K, K'$ for WG$\# 17$ with $M_x^-$ and $C_6^-$. It is not surprising that our no-go theorem covers the two Dirac points in the graphene.

\begin{figure}
\centerline{\includegraphics[width=0.45\textwidth]{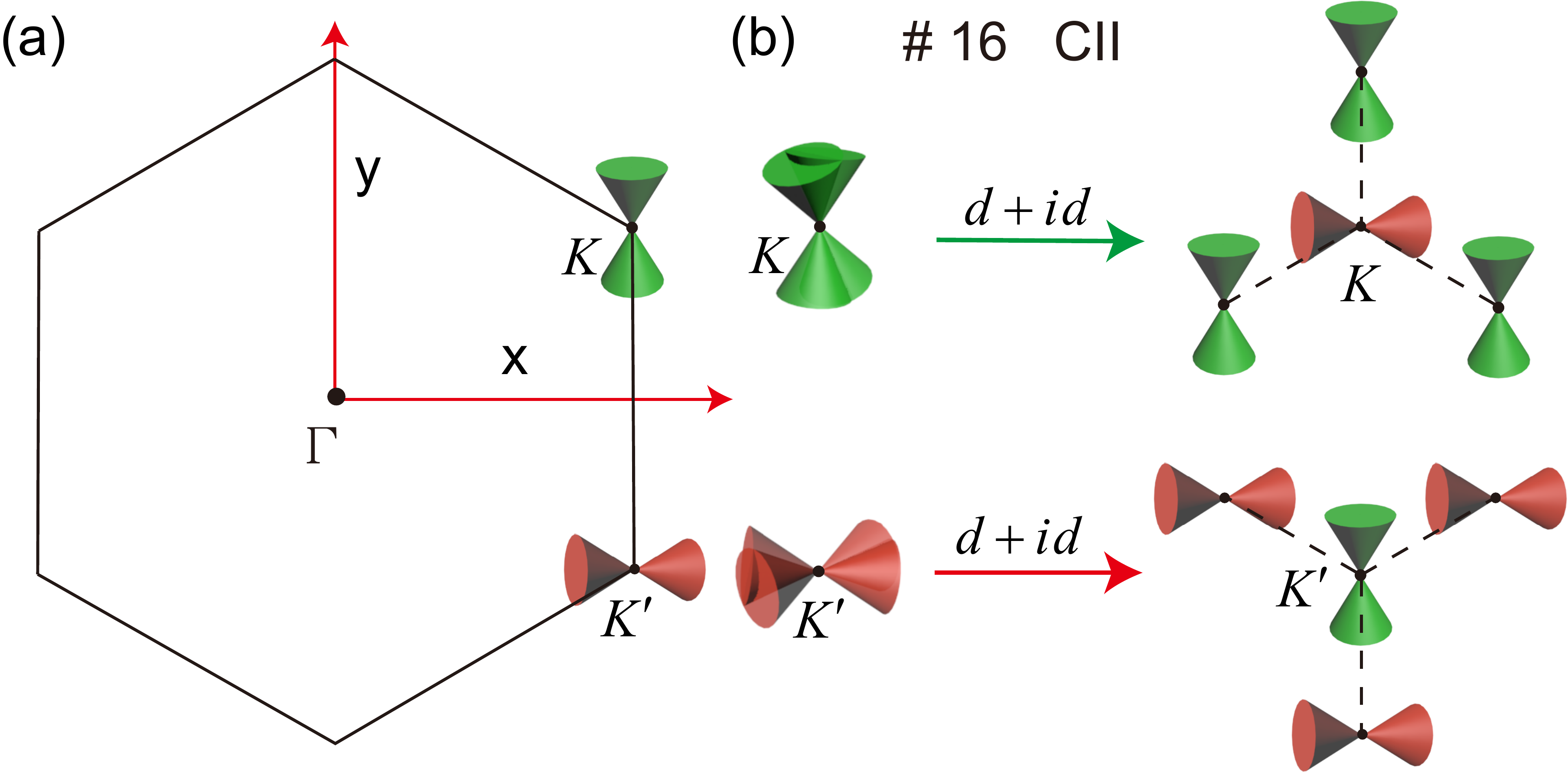}}
\caption{(color online) The nodal points in the graphene without or with $d+id$-wave SC. (a) The graphene Hamiltonian $H_g$ hosts two Dirac nodes at $K, K^{'}$. (b) By introducing $d+id$-wave SC pairing, each of the nodal points at $K, K^{'}$ become four separate Dirac nodes. by using $K$ point as an example, three of them move along $\overline{K\Gamma}$ with $\nu=1$ in the three directions and the remaining Dirac point stays at $K$ with $\nu=-1$.
\label{graphene} }
\end{figure}

\subsubsection{$d+id$ superconductors: WG$\#17$ in class CII}


	We add $d+id$-wave superconductor pairing in the graphene. Its BdG Hamiltonian in the basis of $\Psi(k)=(C^{A}_{k\uparrow}, C^{B}_{k\uparrow},C^{A,\dagger}_{-k\downarrow},C^{B,\dagger}_{-k\downarrow})^{T}$ is written as~\cite{Zhangqiang}
\begin{equation}H_{BdG}(\bk)  =  \left(\begin{array}{cccc}
0 & h_{12}(\bk) & 0 & \Delta(\bk) \\
h^{*}_{12}(\bk) & 0 & \Delta(-\bk) &0 \\
0 & \Delta^{*}(-\bk) & 0 & -h^{*}_{12}(-\bk) \\
\Delta^{*}(\bk) & 0 & -h_{12}(-\bk) & 0 \\
\end{array}\right),
\end{equation}	
where the $d+id$-wave pairing by nearest-neighbor antiferromagnetic exchange coupling in a honeycomb lattice~\cite{Hu2012} is
\bee
\Delta(\bk)\equiv \frac{\Delta_{0}}{3}e^{i\frac{k_x}{\sqrt{3}}}\left[2\cos(-\frac{k_y}{2}+\frac{2\pi}{3})e^{-i\frac{\sqrt{3}k_x}{2}}+1\right].
\ee
Since this spinless SC model stems from the spinful system with spin $SU(2)$ symmetry, we have particle-hole symmetry with $C=\tau_y\sigma_0 \mathcal{K}$ and $C^2=-1$~\cite{Schnyder2008}. The SC system inherits chiral symmetry from the graphene with $S=\tau_0\sigma_z$; hence, effective time reversal symmetry is preserved with $T^+=\tau_y \sigma_z \mathcal{K}$, although the physical time-reversal symmetry is broken in the $d+id$-wave superconductor. Furthermore, the graphene structure leads to the same WG$\#17$ with two extended generators  $M_x^-=\tau_y \sigma_y$ and
\bee
C_6^-  =  \left(\begin{array}{cc}
 1 & 0  \\
 0 & e^{i 2\pi/3} \\
\end{array}\right)\otimes \sigma_x.
\ee
Therefore, the Hamiltonian $H_{BdG}(\mathbf{\bk})$ preserves reflection symmetry (\ref{reflection}) and 6-fold rotation symmetry (\ref{C6}). The systems belongs to WG$\#17$ with $M_x^-$ and $C_6^-$ in class CII in Table \ref{no-go away}.

	Back to nodal points, when the superconducting gap vanishes, in the BdG Hamiltonian each of $K$ and $K'$ possesses two Dirac points with the same winding numbers --- one from particle part and the other from the hole part. Let us focus on $K$ possessing the two Dirac points with $+1$ winding number. As the superconductor gap gradually increases from zero,  a Dirac nodal point with $-1$ winding number stays at $K$ and three Dirac nodal points with $+1$ winding number move away from $K$ along three  $\overline{\Gamma K}$ lines separately as illustrated in Fig.~\ref{graphene}(b). The reason is that  the combination symmetry of the time-reversal $T^+$ and the reflection $M_x^-$ symmetries locks the three Dirac points at $\overline{\Gamma K}$. The total winding number of the four Dirac points near $K$ is still 2, whereas the four Dirac points near $K'$ with the opposite winding numbers have the identical distribution. The SC system possesses three Dirac points with $-1$ winding numbers in the three {\it effective} mirror lines ($\overline{\Gamma K}$) and three Dirac points with $-1$ winding numbers the three mirror lines ($\overline{\Gamma K'}$), while two other Dirac points with $+1$ and $+1$ are located at $K$ and $K'$. On the other hand, for WG$\#17$ with $M_x^-$ and $C_6^-$ in class CII Table \ref{no-go away} shows $\nu_{\rm{abs}}=2$ in $K, K'$ and $\nu_{\rm{abs}}=6$ in the mirror lines. Thus, the table of the no-go theorem shows the building blocks of the nodal points for this $d+id$ wave superconductor.

\begin{center}
{\bf At time-reversal symmetry points}
\end{center}

	The key difference between nodal points located at and away from TRI points is that time reversal symmetry imposes an additional constraint at TRI points. For class DIII, nodal points exhibit Kramers' degeneracy at TRI points, while for class BDI, the winding numbers of the nodal points are always even. We provide several examples to show these special properties.

\subsubsection{ $p+ip$ superconductors: WG$\#6$ in class DIII}

\begin{figure}
\centerline{\includegraphics[width=0.5\textwidth]{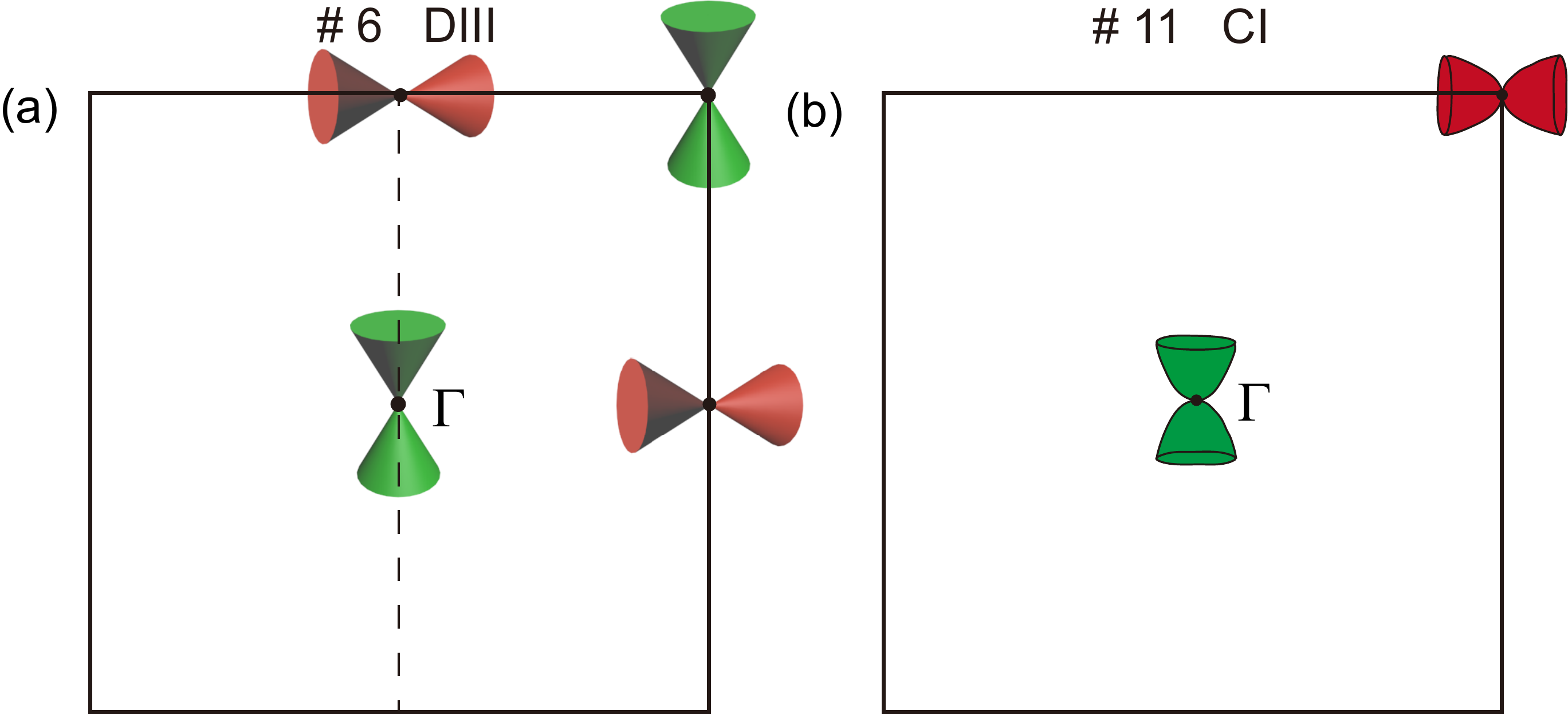}}
\caption{(color online) The nodal points at TRS points (a) class DIII with $C_2^{+}$ and $M_{x}^{-}$, and (b) class CI .
\label{fig5} }
\end{figure}

To illustrate the Dirac points at TRI points, we study an effective Hamiltonian of a $p+ip$ wave superconductor with $C_{2v}$ point group symmetry in the rectangular lattice
\begin{eqnarray}
H(\bk)=a\sin k_{x} \sigma_{x}+b\sin k_{y} \sigma_{y}. \label{C2H}
\end{eqnarray}
Particularly, we remove the diagonal term so that the Hamiltonian preserves chiral symmetry (\ref{chiral symmetry}) with $S=\sigma_z$.
Four Dirac points are located at TRI momenta (0,0), (0,$\pi$), ($\pi$,0), ($\pi$,$\pi$) because Kramers' degeneracy and chiral symmetry force two-fold degenerate states with zero energy at all of the TRI points. The system preserves time reversal symmetry (\ref{time reversal symmetry}) with $T^-=i\sigma_y\mathcal{K}$ and particle-hole symmetry (\ref{particle-hole symmetry}) with $C=\sigma_x\mathcal{K}$. Since $T^2=-1$ and $C^2=+1$, the system belongs to class DIII. On the other hand, the system possesses $C_{2v}$ point group symmetry with generators $C_{2}=\sigma_z$ and $M_{x}=\sigma_y$, which obey  $C_{2}H(\bk)C_2^{-1}=H(-\bk)$ and $M_{x}H(k_x,k_y)M_{x}^{-1}=H(-k_x,k_y)$, and the algebra of the crystalline symmetry operators is given by $C^{+}_{2}$ and $M^{-}_{x}$. According to Table \ref{winding number table}, the algebra of the symmetry operators ($T^-,M^{-}_{x},C^{+}_{2}$) does not trivialize winding numbers at TRI points, where nodal points are located.

Table~\ref{character1} shows that in class DIII WG \#6 with $C^{+}_{2}$ and $M^{-}_{x}$ has $\nu_{\rm{abs}}=4$ indicating 4 nodal points with $\pm 1$ winding numbers at TRI points.
As expected, the distribution of the nodal points in the model is in agreement with the generalized no-go theorem.
Furthermore, we find that the winding number of the system is in the form of $v=\frac{i}{2 \pi} \oint h^{-1} d h$, with
\bee
h(\bk)=a\sin k_{x}-i b\sin k_{y} .
\ee
By choosing the integral path encircling one of the nodal points, we obtain $\nu$=+1 for the nodes at (0,0) and ($\pi$,$\pi$), whereas $\nu$=-1 for the nodes at (0,$\pi$) and ($\pi$,0).

\subsubsection{Quadratic nodal superconductors: WG$\#11$ in class CI}

	The symmetries in class CI force nodal points at TRI points to have even winding numbers. Since $T^2=1$, Kramers' degeneracy is absent at TRI points; hence, unlike class DIII, it is not necessary that all of the TRI points in the BZ possess nodal points at zero energy when the number of the energy band pairs is odd. To demonstrate these properties, we consider a simple 2D tight-binding Hamiltonian
\begin{eqnarray}
H(\bm{k})=(\cos k_{y}-\cos k_{x}) \sigma_{x}+\sin k_{x}\sin k_{y} \sigma_{y}.
\end{eqnarray}
The system preserves chiral symmetry (\ref{chiral symmetry}) with $S=\sigma_z$, time-reversal symmetry (\ref{time reversal symmetry}) with $T=\sigma_x\mathcal{K}$, and particle-hole symmetry (\ref{particle-hole symmetry}) with $C=\sigma_y\mathcal{K}$. Furthermore, the Hamiltonian obeys $C_{4}H(k_x,k_y)C_4^{-1}=H(-k_y, k_x)$ and $M_{x}H(k)M_{x}^{-1}=H(-k_x, k_y)$, where symmetry operators $C^{+}_{4}=\sigma_z$ and $M^{-}_{x}=\sigma_x$. Since $T^2=1$ and $C^2=-1$, the system belongs to SG$\#11$ in class CI.

Two nodal points with quadratic energy dispersions are located at TRI momenta $\Gamma$:(0,0) and M:($\pi$,$\pi$). The Hamiltonian of the low energy expansion near the nodal points $\Gamma$ and $M$ is in the form of
\begin{equation}\begin{aligned}
&H_{\Gamma/M}(\delta\bm{k})\simeq \pm \frac{1}{2}(\delta k^2_x-\delta k^2_y)\sigma_x+\delta k_x\delta k_y\sigma_y,\\
\end{aligned}\end{equation}
where $\delta\bm{k}=\bm{k}-\bm{k}_{\Gamma/M}$ indicates momentum expansion at the nodal points. Using $S=\sigma_z$, we have the values of the winding number $\nu_\Gamma=2$ and $\nu_M=-2$. The distribution of the nodal points with $\pm 2$ winding numbers is consistent with table~\ref{no-go TRS} for WG$\#11$ in class CI with $C^{+}_{4}$ and $M^{-}_{x}$.

\begin{figure}
\centerline{\includegraphics[width=0.5\textwidth]{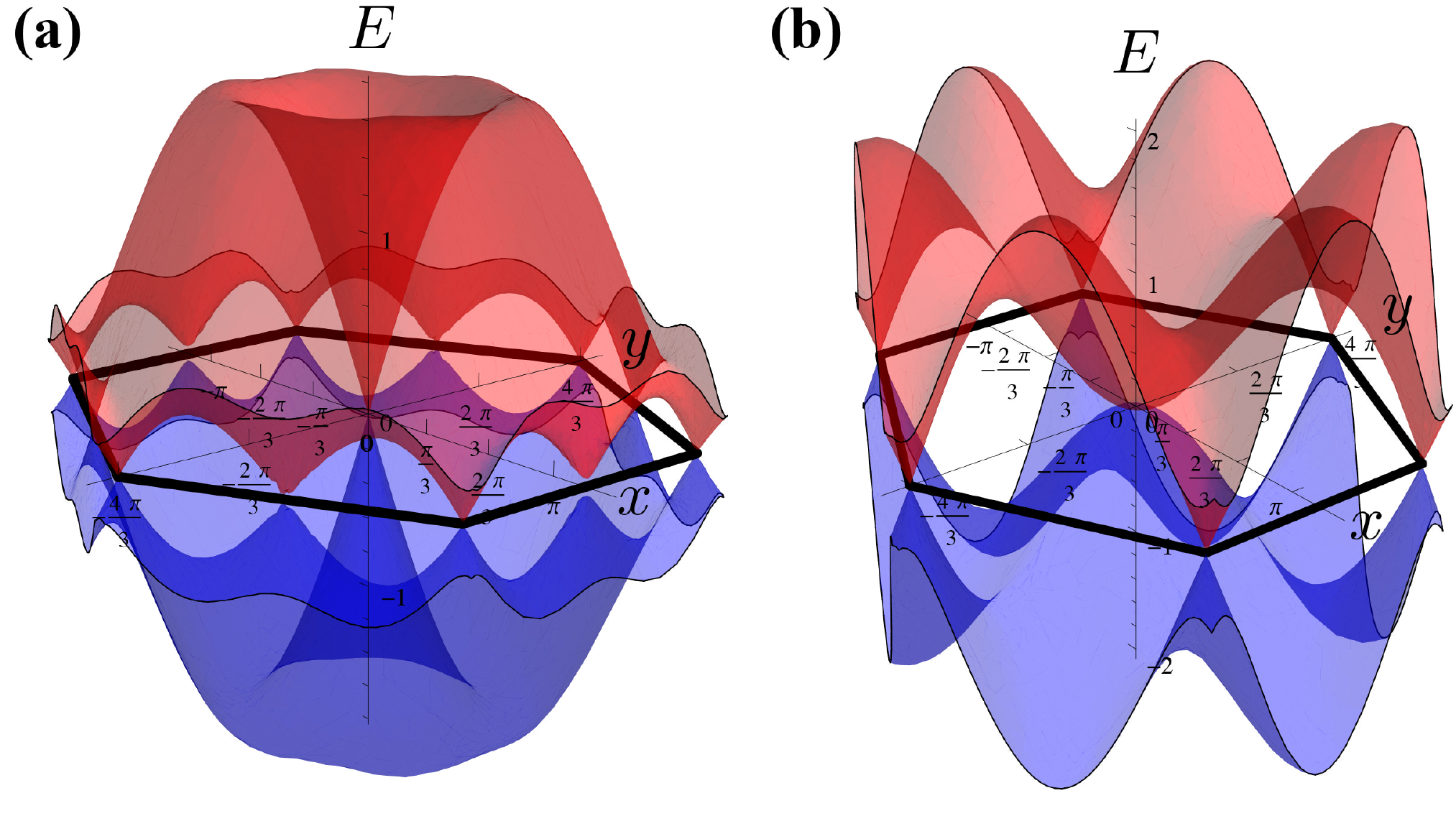}}
\caption{(color online) The energy dispersion of the hexagonal BZ for the nodal points located at TRI points. (a) Three Dirac points with $\nu=1$ are separately located at $\Gamma,K,K'$, whereas other three with $\nu=-1$ are separated located at $M_1,M_2,M_3$. (b) One nodal point with quadratic dispersion at $\Gamma$ carries $-2$ winding number, while the remaining two Dirac points with $\nu=1$ is separately located at $K,K'$.
\label{special nodes} }
\end{figure}


\subsubsection{Six Dirac nodes: WG$\#17$ in class DIII}

	We provide a 2-band tight binding model possessing six Dirac nodes located at high symmetry points ($\Gamma,K,K',M_1,M_2,M_3$) as shown in Fig.~\ref{eg TRI}(b). Importantly, the Dirac semimetal is first predicted in the manuscript and has never been observed. Starting with each site having spin-$1/2$ degree of freedom in triangular lattice, we consider only the nearest neighbor hoppings with the Pauli matrices depending on the directions. The Hamiltonian in the real space is written in the form of the second quantization
\begin{align}
\hat{H}_6= &\frac{1}{2}\sum_{\br} \Big \{ i D^\dagger_{\br+a\hat{y}} \sigma_y  D_{\br} +i D^\dagger_{\br+C_6(a\hat{y})} C_6(\sigma_y )  D_{\br} \nonumber \\
&+i D^\dagger_{\br+C_6^2(a\hat{y})} C_6^2(\sigma_y)  D_{\br} +h.c. \Big \}, \label{H6}
\end{align}
where the rotation operator obeys $C_6(\hat{x})=(\hat{x}+\sqrt{3} \hat{y})/2$ and $C_6(\hat{y})=(\hat{y}-\sqrt{3} \hat{x})/2$. Furthermore, the rotation operation $C_6$ acting on $\sigma_x$ and $\sigma_y$ exhibits the same basis changing as $\hat{x}$ and $\hat{y}$. We rewrite the Hamiltonian in the momentum space
\bee
H_6(\bk)=
\bma
0 & h_6(\bk) \\
h^*_6(\bk) & 0
\ema, \label{H6}
\ee
where $h_{6}(\bk)=-i\sin k_y -i \sin (-(k_y+\sqrt{3}k_x)/2)e^{-2\pi i/3}-i \sin ((-k_y+\sqrt{3}k_x)/2)e^{2\pi i/3}$. Since chiral symmetry with $S=\sigma_z$, time reversal symmetry with $T=\sigma_y\mathcal{K}$, and particle-hole symmetry with $C=\sigma_x\mathcal{K}$ are preserved, $T^2=-1$ and $C^2=1$ lead to class DIII.
Moreover, the system preserves reflection symmetry $M_x H_6(-k_x, k_y) M_x^{-1}=H_6(k_x,k_y)$ with  $M_x=\sigma_y$ and $C_6 H_6(k_x/2+\sqrt{3}k_y/2,k_y/2-\sqrt{3}k_x/2) C_6^{-1}=H (k_x,k_y)$ with
\bee
C_6=
\bma
e^{-\pi i/6} & 0 \\
0 & e^{\pi i /6}
\ema
\ee
Based on the algebra between the crystalline symmetry operators and the chiral symmetry operator, the model belongs to WG$\#17$ with $M_x^-$ and $C_6^+$. The energy dispersion in Fig.~\ref{special nodes}(a) shows that three Dirac nodes with $\nu=1$ are located at $\Gamma,K,K'$ and the other three with $\nu=-1$ are located at $M_1,M_2,M_3$. We note that this tight-binding model can be used in any WG symmetry class possessing the same minimal configuration of the nodal points in Tables \ref{no-go away}, \ref{no-go TRS}. For example, the model can be attached to another breaking the time reversal symmetry and the $C_2$ rotation symmetry but preserving the remaining symmetries. The WG symmetry class, which is class AIII with $C_3^+$ and $M_x^-$, possesses the six nodal points at ($\Gamma,K,K',M_{1/2/3}$) as predicted in Table \ref{no-go away}.

\subsubsection{Quadratic node at $\Gamma$: WG$\#17$ in class CI}

	We build a tight binding model hosting a quadratic node at $\Gamma$ and two Dirac nodes at $K,K'$ separately. Consider each site has spin-1/2 degree of freedom in triangular lattice. The $2\times2$ Hamiltonian in the real space is written as
\begin{align}
\hat{H}_4= &\frac{1}{2}\sum_{\br} \Big \{ D^\dagger_{\br+a\hat{y}} \sigma_y  D_{\br} + D^\dagger_{\br+C_3(a\hat{y})} C_3(\sigma_y)  D_{\br} \nonumber \\
&+D^\dagger_{\br+C_3^2(a\hat{y})} C_3^2(\sigma_y)  D_{\br} +h.c. \Big \}, \label{H4}
\end{align}
where $C_3=C_6^2$. In this regard, the Hamiltonian in the momentum space is given by
\bee
H_4(\bk)=
\bma
0 & h_4(\bk) \\
h^*_4(\bk) & 0
\ema, \label{H4}
\ee
where $h_4(\bk)=-i\big (\cos k_y + \cos  ((k_y+\sqrt{3}k_x)/2)e^{-2\pi i/3}+ \cos ((-k_y+\sqrt{3}k_x)/2)e^{2\pi i/3}  \big )$.  The symmetry class corresponds to CI, since one can check that chiral symmetry with $S=\sigma_z$, time reversal symmetry with $T=\sigma_x\mathcal{K}$, and particle-hole symmetry with $C=\sigma_y\mathcal{K}$ are preserved. Moreover, the system preserves reflection symmetry $M_x H_4(-k_x, k_y) M_x^{-1}=H_4(k_x,k_y)$ with $M_x^-=\sigma_y$ and $C_6 H_4(k_x/2+\sqrt{3}k_y/2,k_y/2-\sqrt{3}k_x/2) C_6^{-1}=H_4 (k_x,k_y)$ with
\bee
C_6=
\bma
e^{\pi i/3} & 0 \\
0 & e^{-\pi i /3}
\ema.
\ee
As shown in Fig.~\ref{special nodes}(b), one quadratic node with $\nu=-2$ is located at $\Gamma$, while two Dirac nodes with $\nu=1$ are separately located at $K, K'$.

	This tight-binding model can be used for the same minimal configuration in different WGs($\#13-17$) in class CI as listed in Table \ref{no-go TRS} and in class AIII as listed in Table \ref{no-go away}. That is, since $H_4(\bk)$ belongs to WG$\#17$ in class CI, we can simply break selected symmetries without destroying the Dirac nodes so that the WG symmetry class can be changed to any of the aforementioned symmetry classes with the same minimal configuration. However, the only problem is that $H_4(\bk)$ cannot be directly transformed from class CI to class DIII, although the same minimal configuration with $\nu_{\rm{abds}}=4$ in class DIII is listed in Table \ref{no-go away}. Instead, we provide the 4-band tight-binding model $H_{4\times 4}(\bk)$, which belongs to WG$\#17$ in class DIII, having the same minimal configuration in Appendix \ref{special four}.
\\
	
	The tight-binding models $H_6(\bk), H_4(\bk), H_{4\times 4}(\bk)$ in the example and Appendix \ref{special four} are new types of topological nodal systems, which have not been predicted and observed till now.  Being different from graphene, the new nodal systems can host a Dirac node at $\Gamma, M_{1,2,3}$ or a quadratic node at $\Gamma$. Furthermore, for hexagonal BZ (WGs($\#13-17$)) these three models are the only building blocks for any minimal configuration with nodal points located at TRI points. For example, a nodal point carries $\nu=3$ at $\Gamma$ and three nodal points separately at $M_{1,2,3}$ carry $\nu=-1$ as the minimal configuration belonging to WG$\#17$ in class DIII. This configuration can be realized by putting $H_6(\bk)$ and $H_{4\times 4}(\bk)$ together.

%
%
%

%
%
%
%

\section{Dirac nodes protected by time-space inversion symmetry} \label{time-space inversion section}



	For 2D lattices, chiral symmetry is not the only symmetry protecting nodal points. Space-time inversion symmetry, which is the combination of time-reversal symmetry and inversion symmetry, is another option and its symmetry operator obeying $(C_{2}T)^2=\pm 1$ represents the two distinct types of the space-time inversion symmetry. Particularly, the space-time inversion symmetry with $(C_2T)^2=1$ can protect nodal points. While chiral symmetry intrinsically appears in time-reversal-symmetric superconductors, the space-time inversion symmetry with $(C_2T)^2=1$ can naturally arise in most of the inversion-symmetric materials preserving time-reversal symmetry; the symmetry-preserving Hamiltonian in the momentum space obeys
\bee
V_\bk H^*(\bk)V_{\bk}^*=H(\bk),
\ee
where the symmetry operator $C_{2}T=V_\bk\mathcal{K}$ and $V_\bk V_\bk^*=1$. In this regard, to search for new Dirac materials it is important to study the generalized no-go theorem of the space-time inversion symmetry for different WGs. However, this no-go theorem is distinct from the aforementioned theorem of the chiral symmetry, because there are two main differences between chiral symmetry and space-time inversion symmetry. First, the $Z_2$ Berry phase quantized by space-time inversion symmetry characterizes the stability of the Dirac nodes, while the $\bZ$ winding number quantized by chiral symmetry characterizes the stability of the node points.
Second, the Dirac nodes protected by space-time inversion symmetry can move freely at any energy level, whereas the node points protected by chiral symmetry are locked at zero energy.


	To preserve space-time inversion symmetry, we consider inversion symmetry $C_{2}$ and time-reversal symmetry $T$ both are preserved. That is, to simplify the problem, we consider type-II magnetic wallpaper groups ($G+TG$). Since crystalline symmetry operator $C_{2}$ and time-reversal symmetry operator $T$ always commute, we have $(C_{2}T)^2=C_{2}^2T^2=+1$. For spinless, $C_2^2=1$ corresponds to class AI ($T^2=1$), while for spin-$1/2$, $C_2^2=-1$ corresponds to class AII ($T^2=-1$). Furthermore, only 10 of the 17 WGs preserve inversion symmetry and their corresponding point groups are $C_n$ and $C_{nv}$ with $n=2, 4, 6$. Hence, the generalized no-go theorem of space-time inversion symmetry in class AI and AII is much simpler than the one of chiral symmetry.



	To study the generalized no-go theorem of space-time inversion symmetry, in the following we review that the symmetry quantizes the $\bZ_2$ Berry phase, and build the relation between the $\bZ_2$ Berry phases for two nodal points connected by crystalline symmetries or time reversal symmetry. Then, we discuss the minimal configurations of the Dirac nodes for different WGs by borrowing the idea from the discussion of chiral symmetry. Lastly, we implement the no-go theorem on the study of the two-dimensional Dirac semimetal materials.

	




\subsection{$\bZ_2$ Berry phases}
	
	For the space-time inversion symmetry, the Berry phase characterizing the nodal points is the key to its generalized no-go theorem. The Berry phase for the nodal point $K_0$ can be in form of
\begin{eqnarray}
\gamma\left(\boldsymbol{K}_{0}\right)=  \sum_{n \in {\rm occupied}}  \oint_{\Gamma\left(\bK_{0}\right)}\boldsymbol{A}_n(\boldsymbol{k}) \cdot \mathrm{d} \boldsymbol{k}, \label{Berry}
\end{eqnarray}	
where $\boldsymbol{A}_n(\boldsymbol{k})= i \left\langle \bu_{n}(\boldsymbol{k})\left|\partial_{\boldsymbol{k}}\right | \bu_{n}(\boldsymbol{k})\right\rangle$ is the Berry connection, $\ket{\bu_{n}(\boldsymbol{k})} $ is the periodic function of the Bloch wavefunction, the summation is for all the occupied states below the energy level of $\bK_{0}$, and $\Gamma\left(\bK_{0}\right)$ is the closed integral path encircling $K_0$. In the literature, the space-time inversion symmetry quantizes this Berry phase with the only two possible values $(0,\pi)$ as an $\bZ_2$ invariant~\cite{2018arXiv181004094C} (see the derivation in appendix \ref{PT symmetry}). Comparing with the definition of the winding number in Eq.~\ref{winding number}, we find the wavefunction of the Berry phase can go through gauge transformation so that $\gamma\left(\boldsymbol{K}_{0}\right)=0, 2\pi$ are equivalent. On the contrary, for chiral symmetry, any two different winding numbers correspond inequivalent topological systems. Furthermore, since $\gamma\left(\boldsymbol{K}_{0}\right)=\pm\pi$ are identical, choosing the orientation of the integral path does not affect the value of the Berry phase.

	As the Berry phase $\gamma\left(\boldsymbol{K}_{0}\right)=\pi$, the integral path cannot smoothly shrink to a single point and then vanish, due to the singularity of the nodal point. This leads to the protection of the nodal point against gap opening without breaking space-time inversion symmetry. This protected nodal point with the lowest order of energy dispersion is a Dirac node. Although there exists a high-ordered nodal point with $\gamma=\pi$, without breaking the symmetry, the node can be smoothly deformed to a Dirac node~\cite{PhysRevLett.121.126402}. This is the reason that space-time inversion symmetry is the essential symmetry for Dirac materials. Here we focus on Dirac nodes only.

	


Next, in order to include all the nodal points in the BZ, we consider the relation between the two $\bZ_2$ Berry phases $(\gamma\left(\boldsymbol{K}_{0}\right), \gamma\left(g\boldsymbol{K}_{0}\right))$ of two nodal points connected by crystalline symmetry or time-reversal symmetry $g$ and this symmetry operation can be symmorphic or nonsymmorphic. The berry phase at $gK_0$ point is given by
\begin{eqnarray}
\gamma(g\boldsymbol{K}_{0})= \sum_{n\in \rm{occupied}} \oint_{\Gamma(g\boldsymbol{K}_{0})}  \boldsymbol{A}_n(\boldsymbol{k}) \cdot \mathrm{d} \boldsymbol{k}
\end{eqnarray}
Since the value of the Berry phase is either $0$ or $\pi$, it is expected that the two Berry phases connected by the symmetry $g$ are identical
\be
\gamma\left(g\boldsymbol{K}_{0}\right)=\gamma\left(\boldsymbol{K}_{0}\right)
\ee
(see the proof in appendix \ref{PT symmetry}). While the relations of the winding numbers connected by a symmetry in Table \ref{winding number table} are complicated, we find that this simple relation for the Berry phases can be easily applied for the generalized no-go theorem.

%



\subsection{Minimal configurations of nodal points}

\begin{figure}
\centerline{\includegraphics[width=0.5\textwidth]{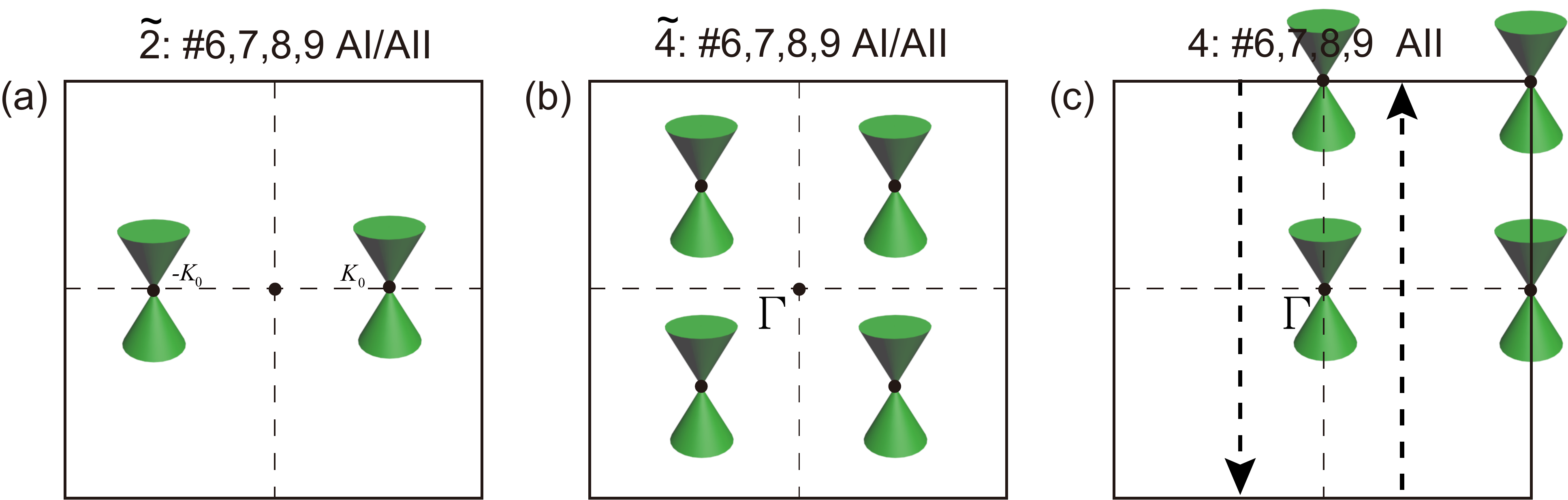}}
\caption{(color online)~ The minimum number of nodal points (a) at the mirror lines, (b) at general points and (c) at TRI points in the WG$\#6, 7, 8, 9$. The dashed lines are the integral paths of the quantized Berry phase through the entire BZs. Due to $C_2$ rotation symmetry or $m$ reflection symmetry, the quantized Berry phases of the separate dashed
lines are identical. 
\label{PT_figure} }
\end{figure}

	The $\bZ_2$ property of the Berry phase significantly simplifies the process to find the minimal configurations of Dirac nodes as the generalized no-go theorem, when we compare this with the complexity of chiral symmetry with the $\bZ$ invariant. In sec.~\ref{chiral off}, the five steps for nodal points protected by chiral symmetry can be simplified for space-time inversion symmetry. Let us first provide the recipe for Dirac nodes not located at TRI points.

(a) We start to place a Dirac node with $\gamma=\pi$ at any location of the irreducible BZ, except for TRI points, as shown in Fig.~\ref{irreducible BZ}. Similarly, different placement leads to different minimal configuration; hence, we have to consider all placement possibilities. Moreover, the Dirac node always survives at any location even if crystalline symmetries or time-reversal symmetry is introduced. The reason is that the symmetry operation does not bring the opposite charge at the same momentum point. This ubiquity of the Dirac node is different from the chiral-symmetric systems, since the symmetries can trivialize nodal points protected by chiral symmetry at some locations with zero winding number in total.

(b) This step is similar to the one for chiral symmetry. That is, using all of the symmetry operations, we duplicate the node point inside of the irreducible BZ to the remaining area of the BZ. We note that space-time inversion symmetry is used to quantize the Berry phase, only one of time-reversal symmetry and inversion symmetry can duplicate the Dirac node. We always use inversion for duplication unless mention otherwise.

(c) Finally, we have to check if the summation of the Berry phases from all the Dirac nodes in the BZ obeys
\begin{equation}\begin{aligned}
\sum_i\gamma(\boldsymbol{K}_{0}^i)&=0\ ({\rm{mod}}\ 2\pi), \label{zero Z2}
\end{aligned}\end{equation}
which is the Nielsen-Ninomiya theorem for the $Z_2$ charges. The derivation is similar with the one for the $\bZ$ charges in appendix \ref{neutralization_proof}. Since inversion symmetry is always preserved, the duplication in step (b) leads to even number of the Dirac nodes. Hence, the summation condition is naturally satisfied. This is the end of searching for the minimal configuration of the Dirac nodes. In the minimal configurations, all the Dirac nodes are connected by symmetries; to mark this symmetry connection, we put $\sim$ on the corresponding minimal numbers in Table \ref{space time inversion table}. 

On the other hand, for the Dirac node placed at one of the TRI points, the procedure is identical to the point off TRI points till step (c). It has be proved that for class AI any Dirac node cannot be located at any TRI points~\cite{ChiuSchnyder14}. Hence, only class AII is considered for TRI points. Inversion symmetry cannot directly lead to even number of the Dirac nodes, since inversion symmetry operation cannot duplicate any nodal point located at a TRI point.
Therefore, we have to add another Dirac node and repeat the procedure for the charge neutralization. At the same time, we need to make sure the constraint conditions similar to Eq.~\ref{constraint1},\ref{constraint2},\ref{constraint3} for specific Dirac nodes are satisfied. That is, the Berry phases at some TRI points obey
\bee
\gamma(\Gamma)+\nu(Y)=0\ ({\rm mod}\ 2\pi), \gamma(\Gamma)+\nu(X)=0\ ({\rm mod}\ 2\pi), \label{Z2constraint1}
\ee	
for WG$\#2,6-12$ and
\bee
\nu(\Gamma)+\nu(M_i)=0\ ({\rm mod}\ 2\pi), \label{Z2constraint2}
\ee
for WG$\#16,17$. The reason is that the two quantized Berry phases calculated in the two dashed lines as  shown Fig.~\ref{invariant_BZ} must be quantized separately and identical due to $C_2$ rotation symmetry and the two integral paths together can be deformed to the paths encircling two TRI points; hence, the summation of these two Berry phases, characterizing two TRI points, must vanish.
Once the summation condition (\ref{zero Z2}) holds, searching for the minimal configurations ends.

	After considering all possible placement of Dirac nodes, we can build Table \ref{PT} to show the minimal numbers and configurations of the Dirac nodes for the ten WGs with time-reversal symmetry. Furthermore, the minimal configurations in Table \ref{PT} are the basic building blocks for any configuration of Dirac nodes in any Dirac semimetals. That is, the combinations of the minimal configurations exhaustively cover all possible configurations of Dirac nodes, due to the absence of the high-ordered nodes.

\renewcommand\arraystretch{2.0}
\begin{table*}
\caption{\label{PT} {\bf The  minimal configurations of Dirac nodes protected by PT-symmetry off- points for the 10 WGs preserving $C_2$ symmetry.} The table exhaustively provides all the possible minimal configurations with the minimal numbers. The minimal numbers for off-TRI points without location specification indicate Dirac points located at general points, while the minimal numbers for at-TRI points without location specification indicate Dirac points located at all the TRI points. Lastly, $\sim$ indicates all nodal points connected by symmetries. } 
\begin{center} \label{space time inversion table}
\setlength{\tabcolsep}{3.5mm}{
\begin{tabular}{|c|c|c|c|c|}
  \hline
WG & Generators & Off-TRI (AI/AII) & At-TRI (AII) & Materials \\\hline
\#2 & $C_{2}$ & $\tilde{2}$ & 4 & \\\hline

\#6,7,8,9 & $C_{2}$ and $m_{x}$ & $\tilde{2}$ MLs or $\tilde{4}$ & 4 & Pmmn boron\cite{Zhou2014}, 6,6,12-graphyne\cite{Malko2012} \\\hline

\#10 & $C_{4}$ & $\tilde{4}$  & 4 & \\\hline

\#11,12 & $C_{4}$ and $m_{x}$ & $\tilde{4}$ MLs or $\tilde{8}$ & 4 & Square-graphyne\cite{Zhang2015} \\\hline

\#16 & $C_{6}$ & $\tilde{2}(K,K')$ or $\tilde{6}$ & 4($M_{1,2,3}$, $\Gamma$) &   \\\hline

\#17 & $C_{6}$ and $m_{x}$ & $\tilde{2}(K,K')$ or $\tilde{6}$ MLs or $\tilde{12}$ & 4($M_{1,2,3}$, $\Gamma$) &$\beta$-graphyne\cite{Malko2012}, TiB$_2$\cite{Zhang2014}, NiPSe$_2$\cite{Gu2019}  \\\hline
\end{tabular}}
\end{center}
\end{table*}
	
	Now let us use WG$\#6,7,8,9$ to demonstrate the procedure obtaining the three minimal configurations of the Dirac nodes with Hamiltonians. (a) We place a Dirac point with $\gamma=\pi$ in one of the mirror lines at momentum $\boldsymbol{K}_{0}$. (b) Since $C_2$ inversion symmetry is the only symmetry for duplication, the duplicated Dirac node is located at $-\boldsymbol{K}_{0}$. (c) There are two Dirac nodes with $\gamma=\pi$ as shown in Fig.~\ref{PT_figure}(a), the total Berry phase in the BZ is neutralized. Hence, the presence of the two Dirac nodes is the minimal configuration. The example Hamiltonian is given by $H_2(\bk)=(\cos k_x + 1 - \cos k_y)\sigma_x + \sin k_y \sigma_y$. The two Dirac nodes with $\nu=\pi$ are located at $(\pm\pi/2,0 )$ in the $M_y$ mirror line. It is easy to check the Hamiltonian preserving space-time inversion symmetry, inversion symmetry, and mirror symmetry with $C_2T=\mathcal{K},\ C_2=\sigma_x, M_y=\sigma_x$ respectively. On the other hand, if the first Dirac node is placed at a general point, $C_2$ inversion symmetry and $M_x$ mirror generate 3 other Dirac nodes as shown in Fig.~\ref{PT_figure}(b). Hence, there are 4 Dirac nodes as the minimal configuration. The Hamiltonian $H_4(\bk)=\cos k_x\sigma_x + \cos k_y \sigma_y$ is a simple example. Four Dirac nodes with $\nu=\pi$ are located at $\pm(\pi/2,\pi/2)$ and $\pm(\pi/2,-\pi/2)$ respectively. The three symmetries are preserved with symmetry operators $C_2T=\mathcal{K}$, $C_2=\bI$, $M_x=\bI$. Furthermore, in these two configurations all Dirac nodes are connected by symmetries. Thus, the minimal number of the Dirac nodes off TRI points is either $\tilde{2}$ or $\tilde{4}$ as listed in Table \ref{PT}.
	
	When we place a Dirac node at TRI point, say $\Gamma$, any symmetry is unable to duplicate the Dirac point, which violates the neutralization (\ref{zero Z2}). Furthermore, the distribution of the Dirac nodes must obey the additional condition (\ref{Z2constraint1}) so that two other Dirac nodes have to be present at $X$ and $Y$ respectively. Due the charge neutralization (\ref{zero Z2}), another Dirac must appear at $M$. Thus, there are four Dirac nodes separately located at the four TRI points as shown in Fig.~\ref{PT_figure}(c). The example Hamiltonian is identical to Eq.~\ref{C2H} ($H_{\rm{TRI}}(\bk)=a \sin k_x \sigma_x + b\sin k_y \sigma_y $). It is known that $H_{\rm{TRI}}(\bk)$ preserves chiral symmetry, inversion symmetry, and mirror symmetry with symmetry operators $S=\sigma_z,\ C_2=\sigma_z,\ M_x=\sigma_y$. Also, space-time inversion symmetry is preserved with $C_2T=\sigma_x \mathcal{K}$. Four Dirac nodes with $\nu=1$ are located at the four TRI points ($\Gamma, X, Y, M$).

\subsection{Examples}

\begin{figure}
\centerline{\includegraphics[width=0.45\textwidth]{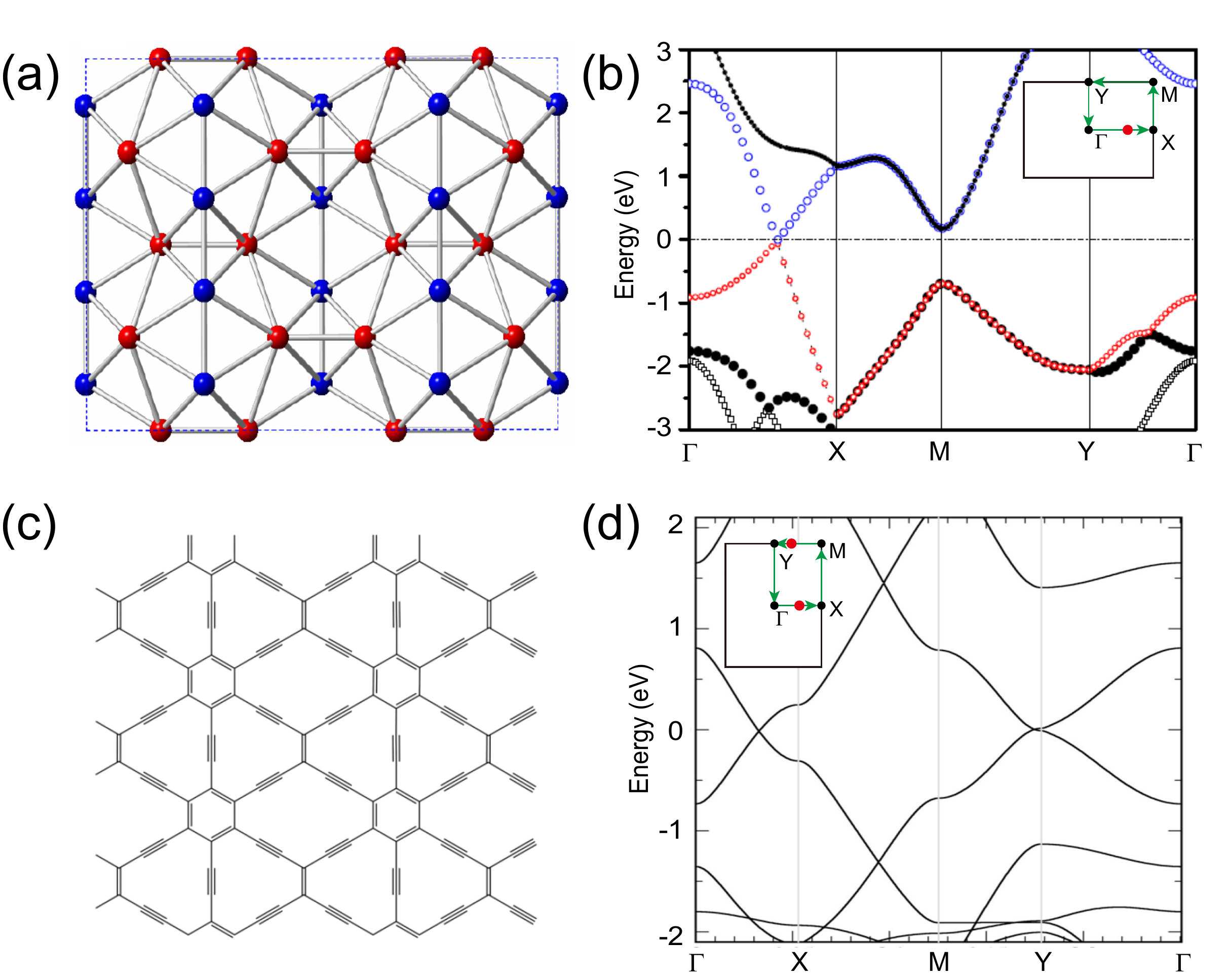}}
\caption{(color online)~ The crystal and band structures of (a, b) Pmmn boron\cite{Zhou2014}, (c, d) 6,6,12-graphyne \cite{Malko2012}. The red dots denote the locations of Dirac points.}  \label{material_1}
\end{figure}

\begin{figure}
\centerline{\includegraphics[width=0.45\textwidth]{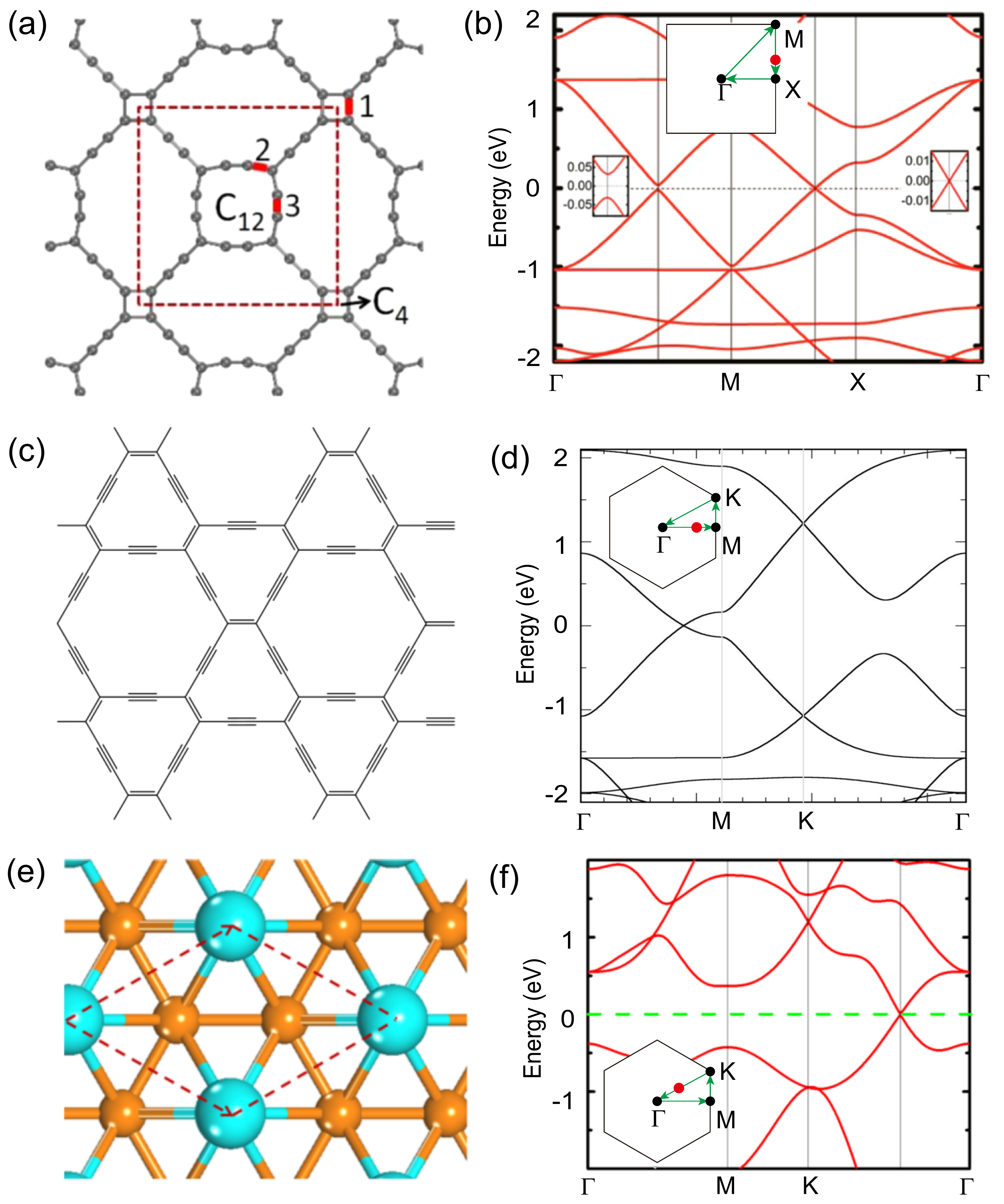}}
\caption{(color online)~ The crystal and band structures of (a, b) Square-graphyne\cite{Zhang2015}, (c, d) $\beta$-graphyne\cite{Malko2012}, (e, f) TiB$_2$\cite{Zhang2014}.\label{material_2} The red dots denote the locations of Dirac points }
\end{figure}

Over hundreds of 2D and quasi-2D materials, graphene\cite{Novoselov2005, 1Zhang2005, Wallace1947}, silicene and germanene\cite{Cahangirov2009}, and several graphynes\cite{Malko2012,Huang2013} have been predicted to be Dirac materials. Particularly, Dirac cones in graphene have been truly confirmed experimentally. Since graphene preserves time-space inversion symmetry and chiral (sublattice) symmetry, its Dirac nodes are under double protection. However, chiral symmetry is fragile for most of the free-fermion materials, while space-time inversion symmetry can be naturally preserved. By implementing our generalized no-go theorem (Table \ref{PT}), we review Dirac semimetals in the literature~\cite{Wang2015}, and show that the minimal configurations of our theorem includes the configurations of the Dirac nodes in the material examples.



(a) WG\#6 -- Pmmn boron and 6,6,12-graphyne:
The Pmmn boron with WG$\#6$ is a 2D layer structure as shown in Fig.\ref{material_1}(a). Fig.\ref{material_1}(b) shows two Dirac points connected by $C_{2}$ rotational symmetry in the mirror line $\overline{\rm \Gamma X}$ in agreement with Table~\ref{PT}. Similarly, 6,6,12-graphyne also belongs to WG$\#6$, see Fig.\ref{material_1}(c). Two Dirac nodes connected by $C_{2}$ rotational symmetry are separately located in the mirror line $\overline{\rm \Gamma X}$, while two others are present in the mirror line $\overline{\rm YM}$. The material possesses the combination of the two minimal configurations of the Dirac nodes in Table~\ref{PT}.


(b) WG\#11 -- Square-graphyne:
The crystal structure of S-graphyne in Fig.\ref{material_2}(a) indicates the system belongs to WG$\#11$. Fig.\ref{material_2}(b) shows a Dirac point located at $\overline{\rm XM}$ as the mirror line. $C_4$ rotation symmetry produces three other Dirac nodes in the mirror lines. The total number of the Dirac nodes is four in agreement with Table~\ref{PT}.


(c) WG\#17 -- $\beta$-graphyne and TiB$_2$: Fig.~\ref{material_2}(c) shows that the crystal structures of $\beta$-graphyne belongs to WG$\#17$. For $\beta$-graphyne, the band structure in Fig.~\ref{material_2}(d) shows that a Dirac point located in the mirror line $\overline{\rm \Gamma M}$, and there are six Dirac points in the BZ due to $C_{6}$ rotation symmetry. On the other hand, TiB$_2$ is a 3D layer structure described by layer group P6/mmm. The 2D projection in Fig.~\ref{material_2}(e) shows the subgroup of the layer group is WG$\#17$.
In the band structure of Fig.~\ref{material_2}(f), six Dirac nodes are located in the six different mirror lines ($\overline{\rm \Gamma K}$ and $\overline{\rm \Gamma {K'}}$) are connected by $C_{6}$ rotational symmetry. Indeed, the number of the Dirac nodes is consistent with Table~\ref{PT}.



\begin{figure}
\centerline{\includegraphics[width=0.5\textwidth]{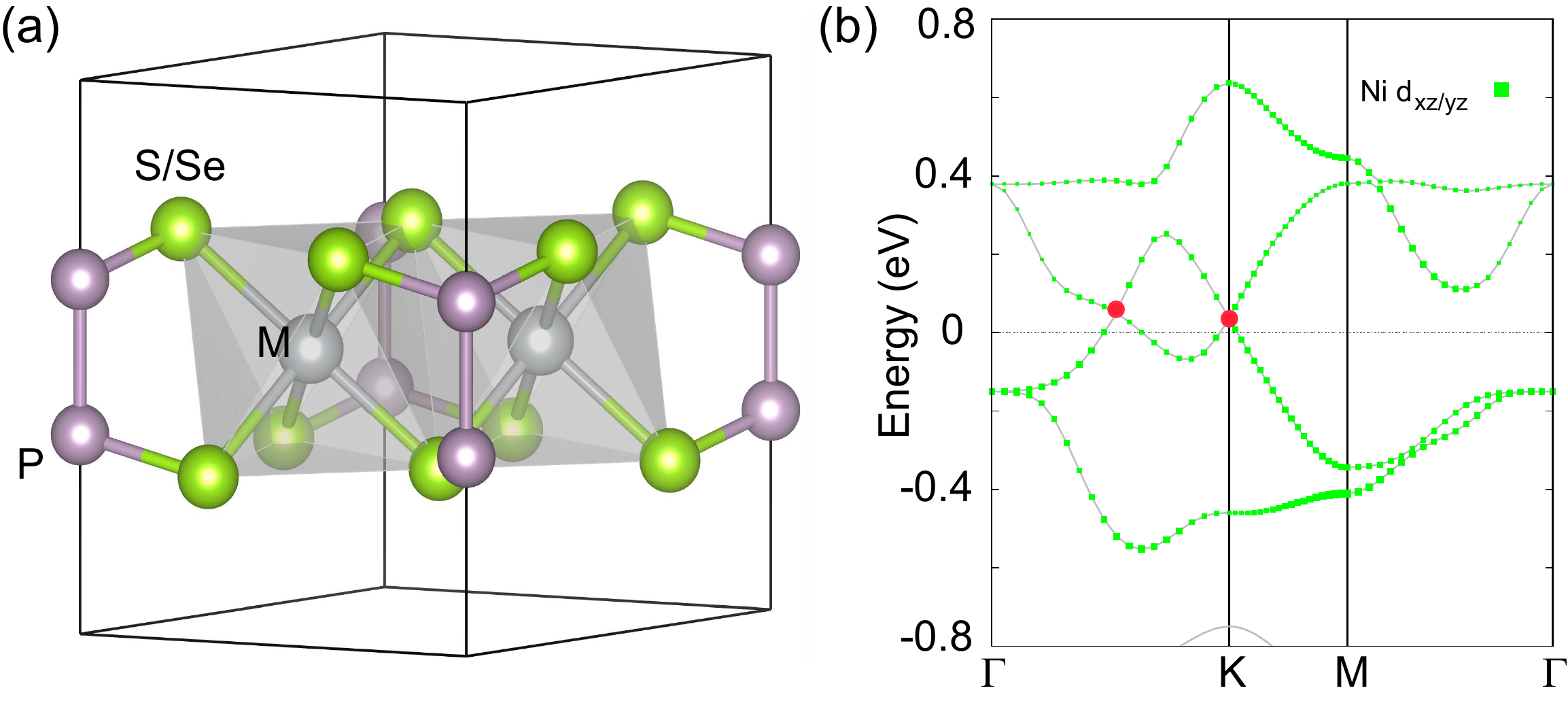}}
\caption{(color online)~ The crystal and band structures of the monolayer NiPSe$_3$\cite{Gu2019}. The red dots denote Dirac points located at non-zero energy.} \label{NiPSe}
\end{figure}

(d) WG\#17 -- NiPSe$_2$: The monolayer NiPSe$_3$ belongs to the layer group P$\bar{3}1$m including point group D$_{3d}$ and its crystal structure is shown in Fig.~\ref{NiPSe}(a). Although the layer does not include $C_2$ rotation symmetry, its inversion symmetry can be treated as an effective $C_2$ symmetry after the monolayer is projected to the 2D plane. Hence, the 2D projection belongs to WG\#17, which is a subgroup of P$\bar{3}1$m with additional $C_2$ rotation symmetry. Since the monolayer in the paramagnetic state preserves time reversal symmetry and spin-orbital coupling is absent, the system is equivalent to two identical subsystems preserving time reversal symmetry with its symmetry operator $T=\mathcal{K}$. Hence, space-time inversion symmetry ($C_2T$) is preserved to protect Dirac nodes.

Near the Fermi level, the valence and conduction bands are predominantly attributed to the Ni-3$d_{xz/yz}$ orbitals. With Ni-3$d_{xz/yz}$ orbitals and two bases of the sublattice the effective 4-band tight binding model~\cite{Gu2019,Li2019} can capture the Dirac points near Fermi level. The band structures of the effective model (Fig.~\ref{NiPSe}(b)) show that two Dirac nodes are separately located at $K$ and $K'$ as well as six Dirac nodes are separately located at three $\overline{\Gamma K}$ lines and three $\overline{\Gamma K'}$ lines. By choosing the integral path of nodal points, we obtain $\mathbb{Z}_{2}$ Berry phase $\nu=\pi$ for every Dirac point. The total number of Dirac points for this system is the sum of the two minimum number [2($K, K^{'}$) and 6 MLs] in Table \ref{PT}.

%
%
%
%
%
%
%

\section{2D Non-Hermitian lattices} \label{non-Hermitian section}


Beyond the Hermitian systems, non-Hermitian physics~\cite{benderMakingSenseNonHermitian2007b,moiseyevNonHermitianQuantumMechanics2011,rotterReviewProgressPhysics2015a,fengNonHermitianPhotonicsBased2017b,el-ganainyNonHermitianPhysicsPT2018d,ozdemirParityTimeSymmetry2019a,miriExceptionalPointsOptics2019a,guptaParitytimeSymmetryNonHermitian2018,martinezalvarezTopologicalStatesNonHermitian2018b,ghatakNewTopologicalInvariants2019,PhysRevLett.80.5243,PhysRevLett.89.270401,PhysRevLett.76.4472,PhysRevLett.77.570,PhysRevLett.102.065703,PhysRevLett.105.013903,leeAnomalousEdgeState2016a,leykamEdgeModesDegeneracies2017e,PhysRevLett.120.146601,2017arXiv170805841K,PhysRevB.99.201107,PhysRevLett.120.146402,PhysRevLett.121.026403,yaoEdgeStatesTopological2018b,yaoNonHermitianChernBands2018b,PhysRevLett.123.170401,PhysRevLett.123.246801,kunstBiorthogonalBulkBoundaryCorrespondence2018a,yokomizoNonBlochBandTheory2019a,2019arXiv191205499Y,zhangCorrespondenceWindingNumbers2019,okumaTopologicalOriginNonHermitian2019,PhysRevLett.118.045701,	PhysRevB.97.075128, PhysRevA.98.042114, PhysRevB.99.081102, PhysRevLett.123.066405,2019arXiv191202788Y,	PhysRevLett.122.076801,PhysRevLett.122.195501,PhysRevLett.123.073601,PhysRevLett.123.016805,PhysRevX.8.031079,PhysRevB.99.235112,PhysRevB.99.125103,PhysRevLett.123.206404,PhysRevLett.123.090603,PhysRevLett.122.237601,PhysRevLett.123.190403,shenTopologicalBandTheory2018a,kawabataClassificationExceptionalPoints2019,leykamEdgeModesDegeneracies2017c,xuWeylExceptionalRings2017,PhysRevX.9.041015}, has emerged in a variety of quantum platforms, such as open quantum systems~\cite{rotterReviewProgressPhysics2015a}, wave systems with gain and loss~\cite{,fengNonHermitianPhotonicsBased2017b,el-ganainyNonHermitianPhysicsPT2018d,ozdemirParityTimeSymmetry2019a,miriExceptionalPointsOptics2019a,guptaParitytimeSymmetryNonHermitian2018} and interacting electron systems~\cite{PhysRevLett.121.026403}. The origin of non-Hermitian terms can stem from the self-energy correction of the Green's function~\cite{rotterReviewProgressPhysics2015a,PhysRevLett.121.026403,2017arXiv170805841K,PhysRevB.99.201107} as an example. We focus on a 2D non-Hermitian lattice model in periodic boundary condition. 
The eigenenergy of the system can be complex $E_n(\bm{k})=\varepsilon_n(\bm{k})+i\Gamma_n(\bm{k})$, where the real part $\varepsilon_n(\bm{k})$ represents the renormalized band dispersion and the imaginary part $\Gamma_n(\bm{k})$ represents the inverse of quasi-particle lifetime, which is affected by environment dissipation~\cite{rotterReviewProgressPhysics2015a} or many body interactions~ \cite{PhysRevLett.121.026403,2017arXiv170805841K,PhysRevB.99.201107}.


In a 2D non-Hermitian Hamiltonian $\mathcal{H}_{nH}(\bm{k})$, Fermi points (FPs) and exceptional points (EPs) can be topologically protected and immune from any perturbations~\cite{PhysRevLett.121.026403,PhysRevLett.123.066405,2019arXiv191202788Y}. The topological invariants, which identify the robustness of these two distinct points, share the mathematical similarity of the winding number~\cite{2019arXiv191202788Y} characterizing Dirac nodes in Hermitian chiral symmetric systems. In this regard, the non-Hermitian no-go theorems for FPs and EPs can smoothly inherit the Hermitian no-go theorem. Due to the constraints of the WGs, the minimal number of the protected points might be greater than 2. We first study the no-go theorem for the EPs and then extend the discussion to the FPs.


In the following, we first review the topological invariants of EPs and then show that the relation of the EPs connected by crystalline symmetries is identical to the chiral-symmetric lattices. Being different from the Hermitian chiral-symmetric systems, rotation symmetry in the non-Hermitian systems imposes an additional constraint on EPs located at rotation centers. Finally, we generalize the corresponding no-go theorems to the 17 WGs. Likewise, the no-go theorem of FPs follows the same recipe in the next subsection.

\subsection{Non-Hermitian exceptional points}

In 2D BZ, non-Hermitian degeneracy points can be topological stable points and are categorized into exceptional (defective) points and non-defective points. That is, at an EP the corresponding eigenstates coalesce~\cite{kankiExactDescriptionCoalescing2017} (become identical), while at a non-defective degeneracy point the eigenstates remain distinct.
It has been shown that without any symmetry a degeneracy point can be characterized by an integer discriminant number ($\nu_{\rm{disc}}$) and this non-zero number leads to the robustness of the degeneracy point~\cite{2019arXiv191202788Y}. However, the presence of the EPs with high topological charges 
($|\nu_{\rm{disc}}|>1$) or non-defective degeneracy points requires fine-tuned parameters. In the presence of arbitrarily small perturbations, these points can be easily deformed to EPs with $\nu_{\rm{disc}}= \pm 1$. Since the only EPs with $\nu_{\rm{disc}}= \pm 1$ are stable, we focus on the physics of these EPs.
In the literature~\cite{2019arXiv191202788Y}, the EPs obey the Fermion doubling theorem
\bee
\sum_j \nu_{\rm{disc}}(\bK_{\rm{EP}}^j)=0 \label{EP neutralization},
\ee
where $\bK_{\rm{EP}}^j$ is the momentum of the EP in the BZ.

Before extending the no-go theorem of the EPs to the 17 wallpaper groups, let us review the definition of the discriminant number $\nu_{\rm{disc}}(\bK_{\rm{EP}})$.  For a generic $m$-band non-Hermitian Hamiltonian $\mathcal{H}_{\rm{nH}}(\bm{k})$, the characteristic equation determines all of the eigenenergies
\begin{equation}
f(E,\bm{k})=\det[E-\mathcal{H}_{\rm{nH}}(\bm{k})]=\prod_{\alpha=1}^m[E-E_{\alpha}(\bm{k})]=0,
\label{E7}
\end{equation}	
where $E_{\alpha}(\bm{k})$ is the band dispersion of the $\alpha$-th band and is the roots of the polynomial $f(E,\bm{k})$. To identify EPs, we use the discriminant of the polynomial $f(E,\bm{k})$
\begin{equation}
\operatorname{Disc}_{E}[\mathcal{H}_{\rm{nH}}](\bk)= \prod_{\alpha<\beta}\left[E_\alpha(\bm{k})-E_\beta(\bm{k})\right]^{2}. \label{discriminant E}
\end{equation}
The EP momentum $\bK_{\rm{EP}}$ corresponds to the vanishing of the discriminant $\operatorname{Disc}_{E}[\mathcal{H}_{\rm{nH}}](\bm{k}_{\rm{EP}})=0$. We note that although this equation can also identify non-defective degeneracy points, the points are unstable and can be easily deformed to EPs~\cite{2019arXiv191202788Y}. The discriminant $\operatorname{Disc}_{E}[\mathcal{H}_{\rm{nH}}](\bm{k})$ for EPs and $\det[h(\bk)]$ for Hermitian Dirac nodes share similar mathematical roles. 
That is, when the discriminant vanishes, its vanishing real and imaginary parts form 0D points in the 2D BZ.
Furthermore, the important property of the discriminant is a single-valued function of $\bk$. The reason is that the discriminant, which is the determinant of the Sylvester matrix of the polynomials $f(E,\bm{k})$ and $\partial_Ef(E,\bm{k})$~\cite{Discriminant,Resultant1,Resultant2}, inherits the single-valued property from the Hamiltonian $\mathcal{H}_{\rm{nH}}(\bm{k})$. In this regard, by comparing with the winding number (\ref{winding number}) for the Dirac node, one can define the discriminant number
\begin{equation}
\nu_{\rm{disc}}(\bK_{\rm{EP}})=\frac{i}{2\pi}\oint_{\Gamma(\bK_{\rm{EP}})} d\bm{k} \cdot \nabla_{\bm{k}} \ln \operatorname{Disc}_{E}[\mathcal{H}_{\rm{nH}}](\bm{k}),
\label{discriminant number}
\end{equation}
where the integration path $\Gamma(\bK_{\rm{EP}})$ is a loop encircling $\bK_{\rm{EP}}$ counterclockwise. Due to the single-valued discriminant function, the discriminant number is always an integer. Since the topological numbers from the two distinct origins (chiral-symmetric systems and EPs) share the same mathematical structure, we can derive the doubling theorem for EPs by following the same recipe in chiral-symmetric system (Appendix \ref{neutralization_proof}). That is, the sum over the discriminant numbers of all the EPs vanishes as previously shown in Eq.~\ref{EP neutralization}.

\subsubsection{Conditions from crystalline symmetries}



To extend the the EP no-go theorem to the 17 wallpaper groups, we study the relation of the discriminant numbers for EPs connected by symmetries. It is known that the two symmetry generators $C_n$ and $M$ form the 17 wallpaper groups, with crystalline symmetry operator matrix $U_g(\bk)$. The symmetry-preserved Hamiltonian obeys
\begin{equation}\begin{aligned}
\mathcal{H}_{\rm nH}(g\bm{k})=U_g(\bm{k})\mathcal{H}_{\rm nH}^{\eta}(\bm{k})U_g^{-1}(\bm{k}),
\label{non-Hermitian symmetry}\end{aligned}\end{equation}
where $U_g$ is a unitary matrix and particularly $\eta=1$ for the conventional crystalline symmetry. The crystalline symmetry generators can be extended to the combination symmetry of crystalline operation and complex-conjugation/transpose~\cite{PhysRevX.9.041015}. In this regard, $U_g(\bk)\mathcal{K}^t,\ U_g(\bk)\mathcal{K}^*,$ and $U_g(\bk)\mathcal{K}^\dag$ indicate the combination crystalline operators of the transpose/complex-conjugation/conjugate-transpose respectively. The corresponding Hamiltonians obey the symmetry equation (\ref{non-Hermitian symmetry}) with $\eta=t,*,\dag$.
Hence, conventional ($U_g$), transpose ($U_g\mathcal{K}^t$), conjugation ($U_g\mathcal{K}^*$), and conjugate-transpose ($U_g\mathcal{K}^{\dagger}$) crystalline generators correspond to the four types of the symmetries.

	For $\eta=1,t$, using the symmetry equation, we have $E_n(g\bm{k})=E_{n}(\bm{k})$. Similarly, for $\eta=*,\dag$, $E_n(g\bm{k})=E_{n}^*(\bm{k})$. In the following discussion, we label $g$ to indicate $U_g(\bm{k})$ and $U_g(\bm{k})\mathcal{K}^{t}$ representations and $\bar{g}$ to indicate  $U_g(\bm{k})\mathcal{K}^{*/\dag}$ representations.  We are interested in WGs formed by one or two generators and the generators can either $g$ or $\bar{g}$. The generators are either rotation or mirror. To be specific, the WG with one generator is in form of $g$ or $\bar{g}$, while the WG with two generators is in form of $(C_n,M), (C_n,\bar{M}), (\bar{C}_n,M),(\bar{C}_n,\bar{M})$. We study the no-go theorem for the WGs in those forms and build Table \ref{no-go FPs} in the following. Particularly, the EP no-go theorem for the 17 conventional WGs correspond to the cases with the generator $g$.

%
%


We use examples to illustrate the distinct combination symmetries above. Consider a Hermitian Hamiltonian $\mathcal{H}_{\rm H}(\bm{k})=\mathcal{H}_{\rm H}^\dag(\bm{k})$ which preserves inversion symmetry, i.e. $\mH_{\rm{H}}(-\bk)=U_I \mH_{\rm{H}} (\bk) U_I^{-1}$. When a non-Hermitian term $i\lambda(\bm{k})\Gamma$ (which is typically a perturbation) is added, where $\lambda(\bm{k})=\lambda(-\bm{k})$ is a real even function and $\Gamma$ is a Hermitian matrix, the Hamiltonian becomes $\mathcal{H}_{\rm nH}(\bm{k})=\mathcal{H}_{\rm H}(\bm{k})+i\lambda(\bm{k})\Gamma$. One can easily check that when $[\Gamma,U_I]_\pm=0$, the corresponding inversion symmetry representation becomes $U_I$/$U_I\mathcal{K}^\dag$, respectively. On the other hand, $U_I(\bk)\mathcal{K}^*$ is equivalent to time reversal symmetry operator. We start with a Hermitian Hamiltonian preserving time reversal symmetry ($\mathcal{H}_{\rm H}(-\bm{k})=U_I(\bm{k})\mathcal{H}_{\rm H}^{*}(\bm{k})U_I^{-1}(\bm{k})$). Similarly, when a non-Hermitian perturbation is introduced, the Hamiltonian reads $\mathcal{H}_{\rm nH}(\bm{k})=\mathcal{H}_{\rm H}(\bm{k})+i\lambda(\bm{k})\Gamma$. As $[\Gamma,U_I]_\pm=0$, transpose inversion symmetry($U_I(\bk)\mathcal{K}^t$)/time-reversal symmetry($U_I(\bk)\mathcal{K}^*$) is preserved respectively.

Before building the generalized no-go theorem for $g$ and $\bar{g}$, where $g=C_n,M$, in Table \ref{no-go FPs}, we provide a guide to read the non-Hermitian table. We use the absolute number $\nu_{\rm{abs}}$ to indicate the minimal configuration of the EPs, where $\nu_{\rm{abs}}=\sum_{\bk_{\rm{EP}}^i\in \rm{BZ}}|\nu_{\rm{disc}}(\bk_{\rm{EP}}^i)|$. For all of the WGs, $\nu_{\rm{abs}}$ can indicate the number of the EPs with $\nu_{\rm{disc}}=\pm 1$ in the minimal configuration, except for EPs located at rotation centers. It will be shown later that a charge for an EP located at a $C_n$ rotation center must be a multiple of $n$. For a WG with one/two symmetry generator $C_n$ or/and $\bar{M}$, symmetries can not connect all nodal points in the minimal configuration. This is an important property leading to anomalous surface BZ violating the no-go theorem. The details are discussed in sec.~\ref{anomalous surface}. For the remaining WGs, all nodal points are connected by symmetries. After this overview of the table, we provide a step-by-step instruction for the non-Hermitian no-go theorem table in the following.


(a) We first consider the conventional crystalline symmetries and the transpose symmetries with the generators all indicated by $g$. Since no global symmetries are required to protect EPs in 2D non-Hermitian systems, unlike Hermitian Hamiltonian preserving chiral symmetry, there is no algebra between crystalline symmetry operators and other global symmetry operators. By using the symmetry equation (\ref{non-Hermitian symmetry}),  the Hamiltonians at these two points possess the identical characteristic polynomial
\begin{align}
\det \big[ E-\mathcal{H}_{\rm{nH}}(\bm{k})\big ] &= \det \big [E -  U_g (\bk) \mH_{\rm{nH}}^{1/t}  (\bk) U_g^{-1}(\bk) \big ] \nonumber \\
&= \det \big [E - \mH_{\rm{nH}}  (g\bk)  \big ].
\end{align}
Hence, the discriminants at these two points are identical
\bee
\operatorname{Disc}_{E}[\mathcal{H}_{\rm{nH}}](\bk) = \operatorname{Disc}_{E}[\mathcal{H}_{\rm{nH}}](g\bk). \label{discriminant relation}
\ee
As symmetry $g$ connects two EPs ($g\boldsymbol{K}_{\rm{EP}}$ and $\boldsymbol{K}_{\rm{EP}}$), the relation of the discriminant numbers at these two points is given by
\begin{align}
\nu_{\rm{disc}}(g\boldsymbol{K}_{\rm{EP}})
&=\frac{i}{2\pi}\oint_{\Gamma\left(\boldsymbol{K}_{\rm{EP}}\right)} \det(g) d \Big ( \ln \big ( \operatorname{Disc}_{E}[\mathcal{H}_{\rm{nH}}](g\bk) \big) \Big ) \nonumber \\
&=\frac{i}{2\pi}\oint_{\Gamma\left(\boldsymbol{K}_{\rm{EP}}\right)} \det(g) d \Big ( \ln \big ( \operatorname{Disc}_{E}[\mathcal{H}_{\rm{nH}}](\bk) \big ) \Big ) \nonumber \\
&=\det(g) \nu_{\rm{disc}}(\bK_{\rm{EP}}) \label{EP relation}
\end{align}	
The explicit relations between the two EPs for the two different types of the crystalline symmetries are given by
\begin{small}
\begin{equation}
\nu_{\rm{disc}}(\hat{C}_n\bm{K}_{\rm{EP}})=\nu_{\rm{disc}}(\bm{K}_{\rm{EP}}), \nu_{\rm{disc}}(\hat{M}\bm{K}_{\rm{EP}})=-\nu_{\rm{disc}}(\bm{K}_{\rm{EP}}), \label{discriminant number relation}
\end{equation}
\end{small}
which are identical to the relation of the generators $C_n^+$ and $M^+$ in Table \ref{winding number table} for the Hermitian chiral-symmetric systems without any time-reversal symmetry. Thus, as shown in Table \ref{no-go FPs}, the list of the EP no-go theorem for the 17 WGs formed by $U_g$ and $U_g \mathcal{K}^t$ can inherit the cases of $C_n^+$ and $M^+$ in class AIII in Table \ref{no-go away}, except for EPs located at rotation centers.

To understanding an EP located at a rotation center, we consider $n$-fold rotation symmetry $C_n$. Being different from the Hermitian chiral symmetric systems, the charge of the non-Hermitian EP at a rotation center, say $\bm{K}_{\rm{EP}}^r$, is limited by an additional restriction. 
In order to show the restriction, in the expression (\ref{discriminant number}) of the discriminant number, the integral path $\Gamma$ enclosing $\boldsymbol{K}_{\rm{FP}}^r$ is divided into $n$ integral paths related by the $C_n$ rotation symmetry
\bee
\Gamma\left(\boldsymbol{K}_{\rm{FP}}^r\right) = \cup_{l=0}^{n-1} \hat{C}_n^l\Gamma_{1/n}\left(\boldsymbol{K}_{\rm{FP}}^r\right),
\ee
where $\Gamma_{1/n}\left(\boldsymbol{K}_{\rm{FP}}^r\right)$ denotes the one n-th of the closed integral path $\Gamma\left(\boldsymbol{K}_{\rm{FP}}^r\right)$ with the two end points $\bk_o$ and $\hat{C}_n \bk_o$. After decomposing the integral of the winding number to $n$ separate integral paths, we can reorganize the discriminant number equation
\begin{align}
\nu_{\rm{disc}}(\boldsymbol{K}_{\rm{EP}}^r)
&=\frac{i}{2\pi}\int_{\Gamma_{1/n}\left(\boldsymbol{K}_{\rm{EP}}^r\right)} \sum_{l=1}^{n-1 } d \Big ( \ln \big ( \operatorname{Disc}_{E}[\mathcal{H}_{\rm{nH}}](\hat{C}_n^l\bm{k}) \big ) \Big ) \nonumber  \\
&=n \times  \frac{i}{2\pi}\oint_{\Gamma_{1/n}\left(\boldsymbol{K}_{\rm{EP}}^r\right)}d \Big ( \ln \big ( \operatorname{Disc}_{E}[\mathcal{H}_{\rm{nH}}](\bm{k}) \big ) \Big ) \nonumber \\
&=n j. \label{rotation constraint EP}
\end{align}
where $j$ is an integer. The second line stems from that the symmetry relation (\ref{non-Hermitian symmetry}) between the two points ($\bk,\hat{C}_n\bk$) leads to
$\operatorname{Disc}_{E}[\mathcal{H}_{\rm{nH}}](\bk) = \operatorname{Disc}_{E}[\mathcal{H}_{\rm{nH}}](\hat{C}_n\bk)$. Since the integrands at $\bk_o$ and $\hat{C}_n \bk_o$ are identical, the integral path of $\Gamma_{1/n}\left(\boldsymbol{K}_{\rm{EP}}^r\right)$ is effectively closed so that the integral is quantized. Hence, as an EP is located at a $n-$fold rotation symmetry center, its discriminant number is a multiple of $n$. This constraint for the value of the discriminant number does not allow the EPs to inherit some of the minimal configurations from the chiral symmetric systems. For example, for WG$\#10$ with $C_4^+$, one of the chiral symmetric minimal configurations is that four Dirac points with $\pm 1$ charges quantized can be separately located at $\Gamma, X, Y, M$, which are rotation centers. However, the $C_4$ rotation symmetry from WG$\#10$ forbids any non-Hermitian EP with $\pm 1$ charge to appear at any rotation center. Hence, this configuration is excluded in Table \ref{no-go FPs}. On the other hand, we note that the non-Hermitian cases in Table \ref{Summary table} only show the no-go theorem for the 17 conventional WGs ($C_n,M_x$).

For the transpose symmetries with $U_{g}U_{g}^*=-1$, we have to take a special care of degeneracy points. Back to the ten-fold classification of the Hermitian systems, the different values of the symmetry operator squares ($T^2,C^2=\pm1$) correspond to distinct AZ symmetry classes. In particular, class AII ($T^2=-1$) leads to the Kramers degeneracy at TRI points. On the contrary, the Kramers degeneracy does not hold in the time-reversal symmetric non-Hermitian systems. Instead, only when transpose symmetry with $U_{g}U_{g}^*=-1$ leads to $\mathcal{H}_{\rm nH}(g\bm{k})=U_{g}\mathcal{H}_{\rm nH}^{t}(\bm{k})U_{g}^{-1}$, where $g=C_n, M_i$, these $g$-symmetry-invariant points must exhibit at least two-fold {\it non-defective} degeneracy~\cite{PhysRevX.9.041015}. In other words, the invariant points cannot be a two-fold defective degeneracy EP. Although without any symmetry constraint a non-defective degeneracy point can be easily deformed to an EP~\cite{2019arXiv191202788Y}, under this transpose symmetry with $U_{g}U_{g}^*=-1$ the degeneracy point in the $g$-symmetry-invariant location is non-defective. 




(b) Consider the conjugation crystalline symmetries and the conjugate-transpose ones; their generators are all indicated by $\bar{g}$. By using the symmetry equation (\ref{non-Hermitian symmetry}), we can directly connect the determinant at $\bk$ and $\bar{g}\bk$
\begin{align}
\det \big[ E-\mathcal{H}_{\rm{nH}}(\bar{g}\bm{k})\big ] &= \det \big [E -  U_{\bar{g}} (\bk) \mH_{\rm{nH}}^{*/\dag}  (\bk) U_{\bar{g}}^{-1}(\bk) \big ] \nonumber \\
&= \det \big [E - \mH_{\rm{nH}}^*  (\bk)  \big ].
\end{align}
The conjugation adds an extra minus sign to the relation equation (\ref{EP relation}) between the two symmetry-connected discriminant numbers	
\begin{align}
	\nu_{\rm{PF}}(\bar{g}\boldsymbol{K}_{\rm{EP}}) =- \det(\bar{g}) \nu(\bK_{\rm{EP}}) \label{bar relation}
\end{align}
Hence, in Table \ref{no-go FPs} the EP no-go theorem for the conjugate 17 WGs  ($\bar{C}_n, \bar{M}$) corresponds to the Hermitian no-go theorem for $C_n^-$ and $M^-$ in class AIII as listed in Table \ref{no-go away}. 	
		
(c) For a WG containing two generators, there are four groups based on the combinations --- $(C_n,M), (C_n,\bar{M}), (\bar{C}_n,M),(\bar{C}_n,\bar{M})$. The last two can be easily extended by the recipe above. Due to the relations between the symmetry-connected EPs (\ref{EP relation},\ref{bar relation}), the non-Hermitian no-go theorem for $C_n, \bar{M}$/$\bar{C}_n, M$ inherits the Hermitian no-go theorem for $C_n^+, M^-$/$C_n^-, M^+$ in class AIII. However, we note that the constraint (\ref{rotation constraint EP}) for EP at rotation centers also excludes some minimal configurations from the inheritance of $C_n^+$ in Table \ref{no-go away}. For example, WG$\#14$ with $C_3$ and $M_y$, WG$\#16$ with $\bar{C}_6$, and WG$\#17$ with $\bar{C}_6$ and $\bar{M}_x$ can have an EP at $K$ with $+3$ charge and another EP at $K'$ with $-3$ charge and the the minimal configuration of the two EPs connected by symmetries labeled by $\tilde{6}(K,K')$. 

\renewcommand\arraystretch{1.5}
\begin{table*}
	\caption{{\bf The minimal absolute topological number $\nu_{\rm{abs}}$ for non-Hermitian ``Exceptional points" and ``Fermi Points".} In the column of rotation centers, $\bigtimes$ indicates the absence of the topological points with non-zero charges in the rotation centers, while the labels of the rotation centers with $n\bZ$ indicates that topological points can be located at the listed centers and the topological numbers must be multiples of integer $n$. Label $\sim$ above some absolute numbers $\nu_{\rm{abs}}$ indicates all topological points connected by symmetries in the minimal configuration. ``MLs" is an abbreviation for mirror lines and $G$ indicates topological points at general points. The non-zero $\nu_{\rm{abs}}$ without location specification denotes that all topological points in the minimal configuration are located at general points.   }
	\label{no-go FPs}
	\begin{center}
		\setlength{\tabcolsep}{4mm}{
			\begin{tabular}{|c|c|c|c|}
				\hline
				WG & Generators & Rotation centers & $\nu_{\rm{abs}}$ \\\hline
				$\#$1&  & No centers  & $2$  \\\hline
				$\#$2& $C_2$  & $(\Gamma,X,Y,M):2\bZ$ & 4 ($G$, $\Gamma,\ X,\ Y,$ or $M)$     \\
				& $\bar{C}_2$  & $\bigtimes$ & $\tilde{2}$  \\\hline
				$\#$3,4,5& $M_{x}$ & $\bigtimes$  & $\tilde{2}$ \\
				& $\bar{M}_{x}$ & No centers & 2 (MLs, $\Gamma,\ X,\ Y,$ or $M)$ or $4$  \\\hline
				$\#$6,7,8,9 & $C_2$ and $M_{x}$ & $\bigtimes$ & $\tilde{4}$ \\	
				& $C_2$ and $\bar{M}_{x}$ & $(\Gamma,X,Y,M):2\bZ$ & 4 (MLs, $\Gamma,\ X,\ Y,$ or $M)$   or $8$ \\
				& $\bar{C}_2$ and $M_{x}/\bar{M}_{x}$ & $\bigtimes$ & $\tilde{2}$ MLs or $\tilde{4}$ \\
				\hline	
				$\#$10& $C_4$ & $(\Gamma,M):4\bZ$~\&~$(X,Y):2\bZ$   & 8 ($G,\ \Gamma,\ X,\ Y,$ or $M)$   \\
				& $\bar{C}_4$ & $\bigtimes$ &  $\tilde{4}$ \\
				\hline
				$\#$11,12 & $C_4$ and $M_{x}$ & $\bigtimes$ & $\tilde{8}$  \\
				& $C_4$ and $\bar{M}_{x}$ & $(\Gamma,M):4\bZ$~\&~$(X,Y):2\bZ$  & 8 (MLs, $\Gamma,\ X,\ Y,$ or $M)$ or 16   \\ 	
				& $\bar{C}_4$ and $M_{x}/\bar{M}_{x}$ & $\bigtimes$  & $\tilde{4}$ MLs or $\tilde{8}$  \\
				\hline
				$\#$13   & $C_3$ & $(\Gamma, K, K'):3\bZ$ & 6 $(G,\ \Gamma,\ K,\ K',$ or $M_{1/2/3})$ \\
				& $\bar{C}_3$ & $\bigtimes$ &  $0$  \\
				\hline
				$\#$14& $C_3$ and $M_{y}$ & ($K,K'):3\bZ$  &  $\tilde{6}$ or $\tilde{6}(K,K')$  \\
				& $C_3$ and $\bar{M}_{y}$ & $(\Gamma, K, K'):3\bZ$ &  6 $($MLs$,\ \Gamma,$ or $M_{1/2/3})$ or 12 $(G,\ K,$ or $K')$ \\
				& $\bar{C}_3$ and $M_{y}/\bar{M}_{y}$ & $\bigtimes$ &  $0$ \\
				\hline
				$\#$15& $C_3$ and $M_{x}$ &$\bigtimes$  &  $\tilde{6}$ \\
				& $C_3$ and $\bar{M}_{x}$ & $(\Gamma, K, K'):3\bZ$ &  6 $($MLs$,\ \Gamma,\ K,\ K',$ or $M_{1/2/3})$ or $12$ \\
				& $\bar{C}_3$ and $M_{x}/\bar{M}_{x}$ & $\bigtimes$ &  $0$ \\
				\hline
				$\#$16& $C_6$ & $(\Gamma):6\bZ~\&~(K, K'):3\bZ~\&~(M_{1/2/3}):2\bZ$ & 12 $(G,\ \Gamma,\ K,\ K',$ or $M_{1/2/3})$ \\
				& $\bar{C}_6$ & $(K, K'):3\bZ$ & $\tilde{6}$ or $\tilde{6}(K,K')$  \\
				\hline
				$\#$17& $C_6$ and $M_{x}$ & $\bigtimes$ & $\tilde{12}$ \\
				& $C_6$ and $\bar{M}_{x}$ & $(\Gamma):6\bZ~\&~(K, K'):3\bZ~\&~(M_{1/2/3}):2\bZ$ & 12 $($MLs$,\ \Gamma,\ K,\ K',$ or $M_{1/2/3})$ or  $24$  \\
				& $\bar{C}_6$ and $M_{x}$ & $\bigtimes$ & $\tilde{6}$ MLs or $\tilde{12}$  \\
				& $\bar{C}_6$ and $\bar{M}_{x}$ & $(K, K'):3\bZ$ & $\tilde{6}$ MLs or $\tilde{6}(K,K')$   or $\tilde{12}$  \\
				\hline
		\end{tabular}}
	\end{center}
\end{table*}

\subsubsection{Examples}

To host EPs, the minimal dimension of the non-Hermitian Hamiltonian must be $2\times2$, since at least two states coalesce. First, we calculate the explicit form of the discriminant for a generic $2\times 2$ non-Hermitian Hamiltonian
\bee
\mathcal{H}_{2\times 2}(\bk) = h_0(\bk) \sigma_0 + h_x(\bk) \sigma_x + h_y(\bk) \sigma_y + h_z(\bk) \sigma_z.
\ee	
The eigenenergies are given by $E_\pm = h_0 \pm \sqrt{h_x^2 + h_y^2 + h_z^2}$. Using Eq.~\ref{discriminant E}, we have a simple form of the discriminant
\begin{align}
&\operatorname{Disc}_{E}[\mathcal{H}_{2\times 2}](\bk) = (E_+ - E_-)^2 \nonumber \\
=& 4(h_x^2 + h_y^2 + h_z^2) = 4(h_0^2 - \det \mathcal{H}_{2\times 2}). \label{2by2 discriminant}
\end{align}
This form is employed to compute the discriminant for the following examples.


(a) We start with a $2\times 2$ off-diagonal Hamiltonian
\bee
\mathcal{H}(\bk) =
\bma
0 &  1 \\
\cos k_x + \cos k_y + i (\cos k_x - \cos k_y) & 0
\ema .
\ee
The Hamiltonian preserves the following symmetries
\begin{align}
\bar{C}_4\mathcal{H}^*(k_x,k_y)\bar{C}_4^{-1} &= \mathcal{H}(k_y,-k_x), \\
M_x\mathcal{H}(k_x,k_y)M_x^{-1} &= \mathcal{H}(-k_x,k_y),
\end{align}
where $\bar{C}_4=\sigma_0$ and $M_x =\sigma_0$. Table~\ref{no-go FPs} shows the minimal configuration of WG$\#11$ is either 4 EPs at the mirror lines or 8 EPs at general points.

To find EPs in this model, we obtain the discriminant $\operatorname{Disc}_{E}[\mathcal{H}](\bk)=4 [\cos k_x + \cos k_y + i (\cos k_x - \cos k_y)]$ by using Eq.~\ref{2by2 discriminant}. As the discriminant vanishes, the Hamiltonian becomes a Jordan canonical form and the two eigenstates coalesce as an EP.
In this regard, the four EPs are located at $(\pm\pi/2,\pm\pi/2)$ and $(\pm\pi/2,\mp\pi/2)$ in the mirror lines.  
To compute the discriminant number of the EPs, we particularly consider the EP at $(\pi/2,\pi/2)$. Near the EP, the leading order of the discriminant is given by
\bee
\operatorname{Disc}_{E}[\mathcal{H}](\bk) = -(1+i)dk_x - (1-i)dk_y + O(d\bk^2),
\ee
where $dk_{x/y}=k_{x/y}-\pi/2$. By the definition (\ref{discriminant number}) of the discriminant number, we have $\nu_{\rm disc}=+1$ for the EP at $(\pi/2,\pi/2)$. The $\bar{C}_4$ rotation symmetry extends to this EP to the remaining three by implementing the discriminant number relation (\ref{bar relation}). Hence, the EPs at $(\pm \pi/2,\pm \pi/2)$/$(\pm \pi/2,\mp \pi/2)$ possess $+1/-1$ charge.


(b) Consider a non-Hermitian Hamiltonian preserving $C_4$ rotation symmetry and $\bar{M}_x$ reflection symmetry
\bee
\mathcal{H}(\bk) =
\bma
0 &  1 \\
(h(\bk) - m)(h(\bk) + m) & 0
\ema ,
\ee
where $h(\bk)=\cos k_x - \cos k_y + i\sin k_x \sin k_y $. We can check that the Hamiltonian obeys
\begin{align}
C_4\mathcal{H}(k_x,k_y)C_4^{-1} &= \mathcal{H}(k_y,-k_x), \\
\bar{M}_x\mathcal{H}^*(k_x,k_y)\bar{M}_x^{-1} &= \mathcal{H}(-k_x,k_y),
\end{align}
where $C_4=\sigma_0$ and $\bar{M}_x=\sigma_0$. According to Table \ref{no-go FPs}, for WG$\#11$ there are three types of the EP minimal configurations --- two EPs at $\Gamma,\ M$ with $\nu_{\rm{disc}}=\pm 4$, eight EPs at mirror lines with $\nu_{\rm{disc}}=\pm 1$, and sixteen EPs at general points with $\nu_{\rm{disc}}=\pm 1$.

	To study EPs in this model, we use Eq.~\ref{2by2 discriminant} and then obtain the discriminant $\operatorname{Disc}_{E}[\mathcal{H}](\bk)=4(h(\bk) - m)(h(\bk) + m)$. First, starting with $m=0$, we find two EPs are located at $\Gamma$ and $M$. The discriminant near these two points can be written as
\begin{align}
\operatorname{Disc}_{E}[\mathcal{H}](\bk)=& (k_x -  ik_y)^4 + O(\bk^5) \\
=& (d k_{x\pi} +i d k_{y\pi})^4 + O(d\bk_\pi^5),
\end{align}	
where $d k_{x\pi} = k_x - \pi, d k_{y\pi} = k_y - \pi$. Therefore, by using the definition (\ref{discriminant number}) of the discriminant number, $\nu_{\rm disc}=+4$ for the EP at $\Gamma$ and $\nu_{\rm disc}=-4$ for the EP at $M$. This minimal configuration possesses the EPs located at the $C_4$ rotation centers. The $C_4$ rotation symmetry (\ref{rotation constraint EP}) forces the discriminant numbers to be a multiple of $4$.

	When we increase $m$ from $0$, the EP at $\Gamma$ is split into four EPs at $(0, \pm \cos^{-1} (1-m))$ and $(\pm \cos^{-1} (1-m), 0)$ in the mirror lines. Since the original EP at $\Gamma$ possesses $+4$ discriminant number and the four EPs connected by the $C_4$ rotation symmetry share the same charge, each EP has $+1$ discriminant number. Similarly, the EP at $M$ is split into four EPs with $-1$ discriminant number at $(\pm \cos^{-1} (-1+m), \pi)$ and $(\pi,\pm \cos^{-1} (-1+m))$.
	
\begin{figure*}[t]
	\centerline{\includegraphics[width=0.9\textwidth]{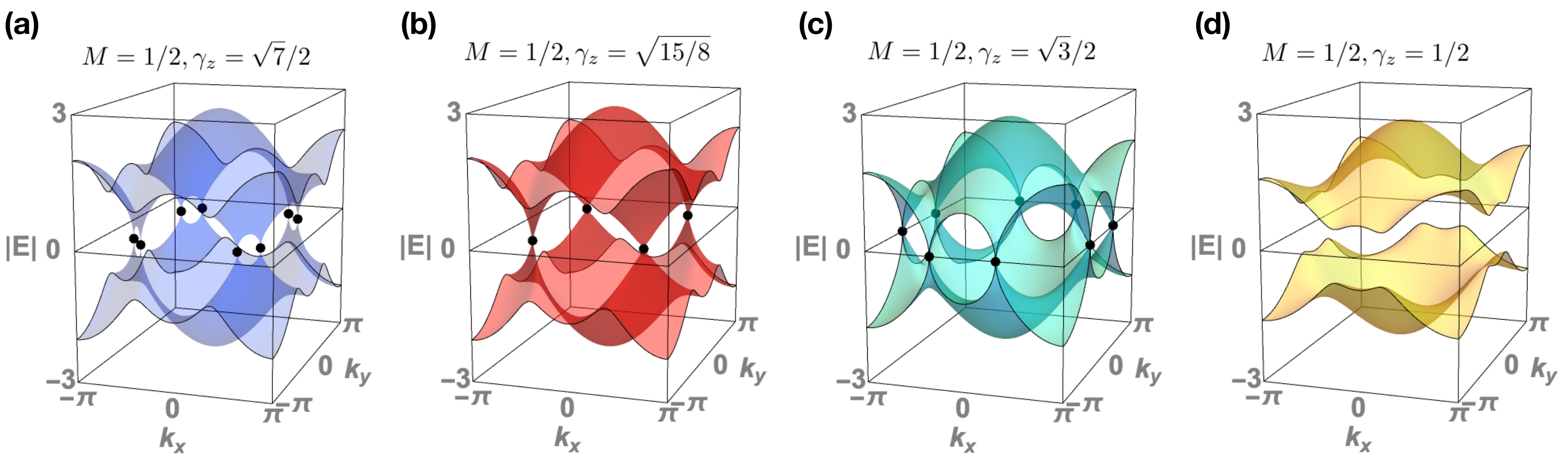}}
	\caption{ The spectrum of the dissipative Chern insulator shows the locations of the EPs. (a) Eight EPs with $\nu_{\rm abs}=\pm 1$ appear as $\sqrt{3}/2< \gamma_z <\sqrt{15/8}$. (b,c) The eight EPs merge to four EPs with $\nu_{\rm abs}=0$ as $\gamma=\sqrt{15/8}$ and $\gamma=\sqrt{3}/2$ respectively. (d) In the region of $|\gamma_z| <\sqrt{3}/2 $, the non-Hermitian phase with a line gap is topological. }
	\label{Dissipative}
\end{figure*}

(c) Being a fundamental 2D example with non-trivial band topology, the Chern insulator has been widely studied in Hermitian topological band theory. One of the simplest Hermitian Hamiltonians describing the Chern insulator~\cite{Qi2008sf} is given by
\begin{eqnarray}\begin{aligned}
\mathcal{H}_{\rm C}(\bm{k})&=\sin k_x\sigma_x+\sin k_y\sigma_y\\
&+(\cos k_x+\cos k_y+M)\sigma_z.
\end{aligned}\end{eqnarray}
In the region $0<|M|<2$, the nonzero Chern number leads to the presence of the chiral edge states and the quantized conductance.
Then we add on-site dissipations uniformly to the Chern insulator so that the Hamiltonian becomes non-Hermitian
\begin{eqnarray}\begin{aligned}
\mathcal{H}_{\rm nC}(\bm{k})=\mathcal{H}_{\rm C}(\bm{k})-i\gamma_z\sigma_z-i \gamma_0\sigma_0, \label{non-Hermitian Chern insulator}
\end{aligned}\end{eqnarray}
where $\gamma_0\pm\gamma_z$ are the on-site dissipations for the different orbitals. Due to the dissipative nature, $\gamma_0\pm\gamma_z$ must be greater than zero, namely $\gamma_0\pm\gamma_z\geq 0$. The reason is that due to the dissipative energies $-i(\gamma_0\pm\gamma_z)$, the corresponding time evolution operator on each site exhibits exponential decay trend ($e^{-(\gamma_0\pm\gamma_z)t}$) as time evolves.



In the presence of the on-site dissipation, this non-Hermitian Chern insulator preserves the following two symmetries $C_4\mathcal{H}_{\rm nC}(k_x,k_y)C_4^{-1}=\mathcal{H}_{\rm nC}(k_y,-k_x)$ and $M_x\mathcal{H}^t_{\rm nC}(k_x,k_y)M_x^{-1}=\mathcal{H}_{\rm nC}(-k_x,k_y)$, where $C_4=e^{i\pi\sigma_z/4}$ and $M_x=\sigma_z$, which belongs to WG$\#11$. That is, the minimal number of EPs with charge $\pm 1$ is given by 8 according to Table \ref{no-go FPs}.

	By Eq.~\ref{2by2 discriminant}, the discriminant can be written in the explicit form
\begin{align}
\operatorname{Disc}_{E}[\mathcal{H}](\bk)= & \\ \nonumber
-4[\sin^2 k_x+\sin^2 k_y&+(\cos k_x+\cos k_y+M-i\gamma_z)^2].
\end{align}	
To demonstrate the evolution of the EPs, we particularly choose $\gamma_0=0$ and $M=1/2$ and start with $\gamma_z=\sqrt{7}/2$. The vanishing discriminant determines eight EPs in the 2D BZ --- ($\pm 2\pi/3,\pm \pi/2$), ($\pm 2\pi/3,\mp \pi/2$), ($\pm \pi/2,\pm 2\pi/3$), ($\pm \pi/2,\mp 2\pi/3$) as shown in Fig.~\ref{Dissipative}(a). 
Knowing one of the EP discriminant numbers, we can obtain the other discriminant numbers by using $C_4$ rotation symmetry and $M_x$ transpose reflection symmetry, because the symmetry relations (\ref{discriminant number relation}) connect the discriminant numbers of the eight EPs. The function of the discriminant is expanded to the linear order near $\bk_1=(2\pi/3,\pi/2)$
\begin{align}
\operatorname{Disc}_{E}[\mathcal{H}](\bk) =&   - 2 \sqrt{3} \left(1-i \sqrt{7}\right) (k_x-\frac{2 \pi }{3}) +4 i \sqrt{7} \left(k_y-\frac{\pi }{2}\right) \nonumber \\
&+ O\big( (\bk - \bk_1)^2 \big )
\end{align}
Using the definition (\ref{discriminant number}) of the discriminant number, we have $\nu_{\rm{disc}}(\bk_1)=1$. In this regard, the $C_4$ rotation symmetry duplicates the three EPs at ($- 2\pi/3,- \pi/2$), ($\pm \pi/2,\pm 2\pi/3$) with $\nu_{\rm{disc}}(\bk_1)=1$, while $M_x$ transpose reflection symmetry leads to the four remaining EPs with $\nu_{\rm{disc}}(\bk_1)=-1$. The presence of the eight EPs is the minimal configuration consistent with Table \ref{no-go FPs}.

	On the one hand, as we increase $\gamma_z$ to $\sqrt{15/8}$, the two EPs from ($2\pi/3,\pi/2$) and $(\pi/2,2\pi/3)$ are merged into one EP at $(\cos^{-1}( -1/4) , \cos^{-1}( -1/4))$ as shown in Fig.~\ref{Dissipative}(b). The charge at the merging point is neutralized. The $C_4$ rotation symmetry duplicates the merging to the six remaining EPs; there are four EPs with zero discriminant numbers located at $(\pm \cos^{-1}(-1/4) ,\pm \cos^{-1}(-1/4))$ and $(\pm \cos^{-1}(-1/4) , \mp \cos^{-1}(-1/4))$. When $\gamma_z>\sqrt{15/8}$, the four neutralized EPs vanish and the non-Hermitian system has a line gap.


	On the other hand, as $\gamma_z$ is decreased to $\sqrt{3}/2$, the two EPs at  ($\pm 2\pi/3,\pi/2$) merged into one EP at ($\pi,\pi/3$) and the merge neutralizes the charge. Similarly, by merging the six remaining EP are reduced to three neutralized EPs separately located at ($\pi,-\pi/3$) and ($\pm \pi/3, \pi$) as shown in Fig.~\ref{Dissipative}(c). As $\gamma_z<\sqrt{3}/2$, the four EPs vanish and the bulk spectrum has a line gap as shown in panel (d). In the open boundary condition the robust edge states connect the two separate bulk region in the complex plane; hence, this gapped phase is topologically non-trivial~\cite{PhysRevLett.120.146402}.
	
	In short, $|\gamma_z| < \sqrt{3}/2$ is the topological region, while $\gamma_z > \sqrt{15/8}$ is the trivial region. As $\sqrt{3}/2<\gamma_z<\sqrt{15/8}$, the 8 EPs with $\nu_{\rm{disc}}=\pm 1$ represent the topological phase transition region. Therefore, the EP minimal configurations listed in Table \ref{no-go FPs} can indicate the topological phase transition for the non-Hermitian line-gap phases.





\subsection{Non-Hermitian FPs}


We briefly introduce the topological protection for FPs in 2D non-Hermitian models, which was recently studied in the literature~\cite{2019arXiv191202788Y,PhysRevLett.123.066405}. Next, we directly extend our new no-go theorem of the 17 WGs to the FPs.


The FPs are the direct generalization of Fermi surface from Hermitian to non-Hermitian systems. We first recall the Fermi surface defined for a Hermitian Hamiltonian $\mathcal{H}_{\rm{H}}(\bm{k})$. In the 2D BZ the Fermi surface, which is determined by $\det[\mathcal{H}_{\rm{H}}(\bm{k})-\mu]=0$, is 1D closed lines. Given a complex chemical potential $\mu$ and a non-Hermitian Hamiltonian $\mathcal{H}_{\rm{nH}}(\bm{k})$, the non-Hermitian Fermi surface is located at $\bk$ satisfying $E_n(\bm{k})-\mu=0$. That is, the two constraints of the equation in the real and imaginary parts form two separate closed lines in the BZ and the non-Hermitian Fermi surface is the crossing points of the two lines; 
hence, we call Fermi {\it points} (FPs) for non-Hermitian Fermi surface in 2D non-Hermitian systems. This is different from 1D Fermi lines in 2D Hermitian systems, since there exists a well-defined invariant characterizing any non-Hermitian FP in the BZ.  To identify all of the FPs $\bK_{\rm{FP}}$ in the BZ, we use determinant
\begin{equation}
\det[\mathcal{H}_{\rm{nH}}(\bK_{\rm{FP}})-\mu]= \prod_n (E_n (\bK_{\rm{FP}})- \mu) = 0.
\label{FPeq}
\end{equation}
%
%
%
These FPs $\bK_{\rm{FP}}$ can be characterized by winding number as an invariant  
\begin{equation}
\nu_{\rm{FP}}\left(\bK_{\rm{FP}} \right)=\frac{i}{2 \pi} \oint_{\Gamma\left(\bK_{\rm{FP}} \right)}  \nabla_{\bk} (\ln \det[\mathcal{H}_{\rm{nH}}(\bm{k})-\mu ] )\cdot d\bk , \label{Fermi winding number}
\end{equation}
where $\Gamma(\bK_{\rm{FP}})$ is a loop counterclockwise encircling the FP at $\bK_{\rm{FP}}$.
Since in a proper basis the Hamiltonian $\mathcal{H}_{nH}(\bm{k})$ is single-valued in the entire BZ and the integrand is a singularity at the FP, the winding number $\nu_{\rm{FP}}$ of the FP is always quantized as an integer. For any non-zero winding number, the integral loop is not contractable so that the FP $\bK_{\rm{FP}}$ inside the loop is a stable singularity point, which cannot be removed. Furthermore, only FPs with non-zero winding numbers are singularities in the entire BZ.

Here we point out that the (non-Hermitian) FPs and the (Hermitian) Dirac points share the identical mathematical structure. In order to see this, we map the non-Hermitian Hamiltonian $\mathcal{H}_{\rm{nH}}(\bm{k})$ to a Hermitian Hamiltonian with chiral symmetry~\cite{PhysRevX.8.031079}
\begin{equation}
\mathcal{H}_{\rm{H}}(\bm{k})=\left(\begin{array}{cc}{0} & {\mathcal{H}_{\rm{nH}}(\bm{k})-\mu} \\{\mathcal{H}_{\rm{nH}}^\dag(\bm{k})-\mu^*} & {0}\end{array}\right).
\end{equation}
The zero energy nodes of the Hermitian Hamiltonian $\mathcal{H}_{\rm{H}}(\bm{k})$ appear when the determinant of the off-diagonal matrix vanishes $\det[\mathcal{H}_{\rm{nH}}(\bm{k})-\mu]=0$. The equation is exactly identical to Eq.~\ref{FPeq} determining the non-Hermitian FPs. Furthermore, the definitions of the topological invariants (\ref{winding number}) for the Hermitian nodes and the non-Hermitian FPs are the same. Hence, these non-Hermitian FPs in $\mathcal{H}_{nH}(\bm{k})$ are mathematically equivalent to the nodal points in $\mathcal{H}_H(\bm{k})$ with the same winding numbers. The no-go theorem for chiral symmetry can be directly extended to the non-Hermitian FP no-go theorem~\cite{2019arXiv191202788Y} described by
\bee
\sum_j \nu_{\rm{FP}}(\bK_{\rm{FP}}^j)=0 \label{FP neutralization}
\ee
This charge neutralization manifests that in the BZ an FP with non-zero winding number must be accompanied by another FP with the opposite winding number.


\subsubsection{Conditions from crystalline symmetries}

	To study the relations of FPs connected by symmetries, we consider four types of the combination crystalline symmetries --- conventional crystalline, transpose crystalline, complex-conjugation, and conjugate-transpose symmetries. We use $g$ to indicate the generators of the first two symmetries ($U_g, U_g\mathcal{K}^t$), while $\bar{g}$ is used to indicate the generators of the last two symmetries ($U_g^*, U_g\mathcal{K}^\dagger$). Transferring from the EP no-go theorem for the 17 WGs to the FP cases, we simply replace the discriminant in the integrand of the discriminant number (\ref{discriminant number}) by determinant for the FP winding number (\ref{Fermi winding number})
\bee
\operatorname{Disc}_{E}[\mathcal{H}_{\rm{nH}}](\bk)\rightarrow \det[\mathcal{H}_{\rm{nH}}(\bm{k})-\mu].
\ee	
It is straightforward that the FP no-go theorem follows the same rules of the EPs as listed in Table \ref{no-go FPs} and we use the absolute number $\nu_{\rm{abs}}=\sum_{\bk_{\rm{FP}}^i\in \rm{BZ}}|\nu_{\rm{FP}}(\bk_{\rm{FP}}^i)|$ to express the FP no-go theorem. The reason of the identical rules is that the relations between two FPs ($\bK_{\rm{FP}}$ and $g\bK_{\rm{FP}}$) connected by the symmetries are identical to the EP cases (\ref{EP relation},\ref{bar relation})
\begin{align}
\nu_{\rm{FP}}(g\boldsymbol{K}_{\rm{FP}})=&\det(g) \nu(\bK_{\rm{FP}}) \label{FP g relation} \\
\nu_{\rm{FP}}(\bar{g}\boldsymbol{K}_{\rm{FP}}) =&- \det(\bar{g}) \nu(\bK_{\rm{PF}}) \label{FP gbar relation}
\end{align}
and for $C_n$ the charges of the FPs at rotation centers ($\boldsymbol{K}_{\rm{FP}}^r$) are limited by the same constraint (\ref{rotation constraint EP})
\begin{align}
\nu_{\rm{FP}}(\boldsymbol{K}_{\rm{FP}}^r)=n j. \label{FP rotation constraint}
\end{align}
The only difference between the FPs and the EPs is that the chemical potential has to be included in the symmetry equations (\ref{non-Hermitian symmetry}) to build the FP connections. For $g$,
\begin{align}
\det[\mathcal{H}_{nH}(g\bm{k})-\mu]&=\det[U_g\mathcal{H}_{nH}^{1/t}(\bm{k})U^{-1}_g-\mu] \nonumber \\
&=\det[\mathcal{H}_{nH}(\bm{k})-\mu].
\end{align}
The chemical potential $\mu$ can be any complex number, then the FP relation (\ref{FP g relation}) still holds. On the other hand, for $\bar{g}$, there is an additional complex conjugation so that
\begin{align}
	\det[\mathcal{H}_{nH}(\bar{g}\bm{k})-\mu]&=\det[U_{\bar{g}}\mathcal{H}^{*/\dag}_{nH}(\bm{k})U^{-1}_{\bar{g}}-\mu] \nonumber \\
	&=\det[\mathcal{H}_{nH}(\bm{k})-\mu^*]^*,
\end{align}
To satisfy the FP relation (\ref{FP gbar relation}), the chemical potential must be real; hence, the Fermi energy of the FP has to be a real number.

\subsubsection{Examples}

We use three examples to demonstrate that our no-go theorem in table \ref{no-go FPs} covers the minimal configurations of the non-Hermitian FPs.

\begin{figure*}[t]
	\centerline{\includegraphics[width=1\textwidth]{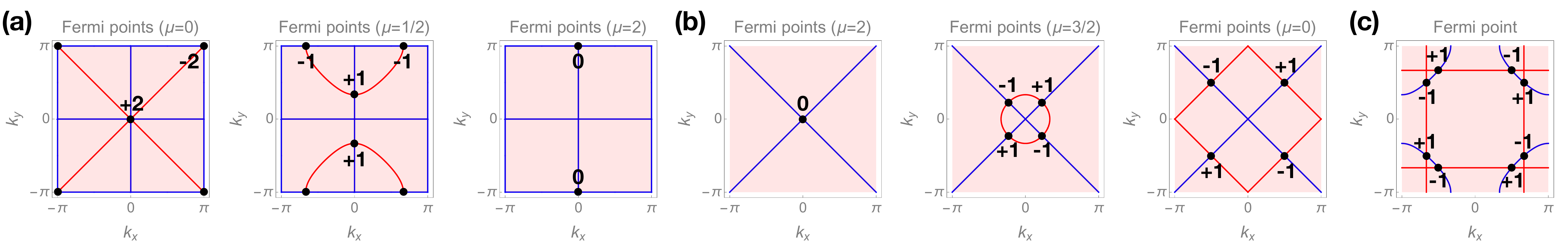}}
	\caption{ FPs with non-zero charges ($\nu_{\rm{abs}}$) are located at the crossings of the red lines (${\rm Re} (\det(H(\bk)-\mu))=0 $) and the blue lines (${\rm Im} (\det(H(\bk)-\mu))=0 $). (a) Two FPs with $\nu_{\rm abs}=\pm 2$ at the rotation centers are split into four FPs with $\nu_{\rm abs}=\pm 1$ in the mirror lines. Finally, these FPs merge to one FP with $\nu_{\rm abs}=0$. (b) A FP with $\nu_{\rm abs}=0$ is split into four FPs with $\nu_{\rm abs}=\pm 1$ in the mirror lines. (c) Eight FPs $\nu_{\rm abs}=\pm 1$ are the general points. }
	\label{FP}
\end{figure*}	

(a) We first consider a one-band model with the Hamiltonian
\bee
H(\bk)  - \mu = \cos k_x - \cos k_y -\mu + i  \sin k_x \sin k_y ,
\ee
where $\mu$ is the chemical potential and imposed to be real to preserve $\bar{M}$ symmetry. 
By checking the symmetries, the Hamiltonian obeys
\bee
C_2(H(k_x,k_y)-\mu)C_2^{-1} = H(-k_x,-k_y)-\mu,
\ee
where $C_2=1$, and
\bee
\bar{M}_x(H(k_x,k_y)-\mu)\bar{M}_x^{-1} = H(-k_x,k_y)-\mu,,
\ee
where $\bar{M}_x=\mathcal{K}^*$. Hence, the system belongs to WG$\#6$ with $C_2$ and $\bar{M}_x$. When $\mu=0$, two FPs are located at $(0,0)$ and $(\pi,\pi)$ separately. By expanding the energy near these points, we have
\begin{align}
E(\bk)=& -(k_x -  ik_y)^2/2 + O(\bk^2) \\
=& -(d k_{x\pi} +i d k_{y\pi})^2/2 + O(d\bk_\pi^2),
\end{align}
where $d k_{x\pi} = k_x - \pi, d k_{y\pi} = k_y - \pi$. Hence, by using the winding number equation (\ref{Fermi winding number}) of the FP, the FP at $(0,0)$ possesses $+2$ charge, while the one at $(\pi,\pi)$ possesses $-2$ charge. The even numbers of the charges at the $C_2$ rotation centers is consistent with the rotation constraint (\ref{FP rotation constraint}).

	Now we adjust the value of the chemical potential to $0<\mu< 2$. The two FPs at $\mu=0$ are split into four FPs located at $(0,\pm \cos^{-1}(1-\mu)),(\pm \cos^{-1}(-1+\mu),\pi)$ as shown in Fig.~\ref{FP}(a). The FPs at $(0,\pm \cos^{-1}(1-\mu))$, which inherit the charge from the FP at $(0,0)$, both have $+1$ winding number. Similarly, the FPs at $(\pm \cos^{-1}(-1+\mu),\pi)$ have $-1$ winding number. The distribution of the four FPs at the mirror lines is consistent with WG$\#6$ with $C_2$ and $\bar{M}_x$ in Table \ref{no-go FPs}. Finally, as $\mu=2$, the four FPs merge at $(0,\pi)$ and the total charge of this single FP is neutralized.

(b) For the second example, we change the one-band model to
\bee
H(\bk) - \mu = \cos k_x + \cos k_y  - \mu + i (\cos k_x - \cos k_y )
\ee
The Hamiltonian with real chemical potential $\mu$ obeys
\begin{align}
\bar{C}_4(H(k_x,k_y)-\mu)\bar{C}_4^{-1} &= H(k_y,-k_x)-\mu, \\
M_x(H(k_x,k_y)-\mu)M_x^{-1} &= H(-k_x,k_y)-\mu,
\end{align}
where $C_4^*=\mathcal{K}^*$ and $M_x=1$. Hence, the system belongs to WG$\#10$ with $\bar{C}_4$ and $M_x$.

	We start with the chemical potential $\mu=2$. The FP satisfying $E(\bK_{\rm{PF}})=\mu$ is located at $(0,0)$. Because this is the only FP in the entire system, to preserve the charge neutralization (\ref{FP neutralization}) the winding number of this FP must be zero. Then as $\mu$ is decreased, the FP at $(0,0)$ is split into four FPs located at $(\pm \cos^{-1} \frac{\mu}{2},\pm \cos^{-1} \frac{\mu}{2})$ and $(\pm \cos^{-1} \frac{\mu}{2},\mp \cos^{-1} \frac{\mu}{2})$ as shown in Fig.~\ref{FP}(b). To identify the charge of the FP, we particularly consider the FP at $(\pi/2,\pi/2)$ for $\mu=0$. The leading order of the energy with respect to the chemical potential can be written as
\bee
E(\bk) - \mu = -(1+i)dk_x - (1-i)dk_y + O(d\bk^2),
\ee
where $dk_{x/y}=k_{x/y}-\pi/2$. By using the definition (\ref{Fermi winding number}) of the winding number, the charge of the FP at $(\pi/2,\pi/2)$ is given by $+1$. The $\bar{C}_4$ rotation and reflection symmetries extend to this FP to the remaining three with the winding number relation (\ref{FP gbar relation}). Hence, the FPs at $(\pm \cos^{-1} \frac{\mu}{2},\pm \cos^{-1} \frac{\mu}{2})$/$(\pm \cos^{-1} \frac{\mu}{2},\mp \cos^{-1} \frac{\mu}{2})$ possess $+1/-1$ charge.

(c) Consider a non-Hermitian Chern insulator and its Hamiltonian $H_C(\bk)$ is written in Eq.~\ref{non-Hermitian Chern insulator}. It has been shown in EP example (c)  that the Chern insulator Hamiltonian preserves $C_4$ rotation symmetry and $M_x$ mirror symmetry with $C_4=e^{i\pi\sigma_z/4}$ and $M_x=\sigma_z$. Even in the presence of the complex chemical potential $\mu$, the symmetries are still preserved. According to Table \ref{no-go FPs}, the minimal number of FPs with charge $\pm 1$ is given by 8.
To show the distribution of the FPs concretely, consider $\gamma_0=0$, $\gamma_z=\sqrt{3}/2$, $M=1/2$, and $\mu=1$. Therefore, the Fermi momentum $\bK_{\rm{FP}}$ is determined by $\det(\mathcal{H}_{\rm C}(\bK_{\rm{PF}})-1)=0$. One of the FPs is located at $\bK^{++}_{\rm{FP}}=(\pi/2,2\pi/3)$. Near this point, we expand the determinant to the leading order
\bee
\det \big( \mathcal{H}_{\rm C}(\bk) - \mu \big ) = -i \frac{\sqrt{3}}{2}dk_{x} + \frac{\sqrt{3}-3i}{2}dk_{y}+O(d\bm{k}^2),
\ee
where $dk_{x}=k_x-\pi/2$ and $dk_{y}=k_y-2\pi/3$. The linear terms determine the winding number $\nu(\bK^{++}_{\rm{FP}})=-1$. The $C_4$ rotation symmetry generates other three FPs with $\nu=-1$ located at $(-\pi/2,-2\pi/3),(\mp 2\pi/3,\pm\pi/2)$. Furthermore, the transpose reflection symmetry extends 4 additional FPs with $\nu=+1$ located at $(\pm\pi/2,\mp2\pi/3)$ and $(\pm2\pi/3,\pm\pi/2)$ as shown in Fig.~\ref{FP}(c).

\section{Applications} \label{application}
	
	The generalized no-go theorem can lead to three useful applications in condensed matter physics --- layer groups, anomalous topological surface, and twisted bilayer physics.

\subsection{Layer groups}

 \begin{table*}[tb!]
\begin{tabular}{|c|c|c|c|c|c|c|c|c|c|c|c|}
 \colrule
LG$\#$ & $1,4,5$ &   $2,3,6,7$ & $8-13,27-36$ & $14-26,37-48$ & 49-52 & 53-64 & 65,74 & 67,69,78 & 68,70,79 & 66,73,75 & 71,72,76,77,80  \\
 \colrule
WG$\#$ &  $1$ & $2$ & $3,4,5$ & $6,7,8,9$ & $10$ & $11,12$ & $13$ & $14$ & $15$ & $16$ & $17$ \\
 \hline
\end{tabular}
\caption{\label{layer group table}%
The table shows the corresponding wallpaper groups from the projection of the 80 layer groups. Each layer follows the no-go theorem of its wallpaper group projection. }
\end{table*}

In chiral-symmetric, space-time-inversion symmetric and non-Hermitian systems, the no-go theorem for the 17 wallpaper groups can be directly extended to the 80 layer groups. Compared with wallpaper groups, layer groups have an additional degree of freedom in $z$ direction without translational symmetry. With this additional direction, some layer groups can preserve $z$-flipping symmetries~\cite{delaFlor:ug5030}, which can be reflection symmetry $M_z$, rotation symmetries ($C_{2x},C_{2y}$), 3D inversion symmetry ($\mathcal{I}$). Each layer group can be formed by one of the wallpaper groups with/without {\it one} $z$-flipping symmetry generator~\cite{2013arXiv1306.1280H}.

The $z$-flipping symmetry operation contains $M_z$ operation. Here we assume $M_z$ commutes with chiral operator ($S$) and  2D space-time inversion operator ($C_{2z}T$). For the discussion associated with the no-go theorem of the layer groups, for any related $z$-flipping symmetry operation, the effect of the $z$-flipping can be neglected, and the remaining operation in the other directions are kept. That is, $M_z\rightarrow$ no change; $C_{2x},C_{2y}\rightarrow M_y,M_x$; $\mathcal{I}\rightarrow C_{2z}$. In this regard, the layer group can be projected to one of the wallpaper groups. Table \ref{layer group table} shows the wallpaper group projections from all the 80 layer groups. The no-go theorem of the layer group follows the corresponding wallpaper group projection; this projection can be used for nodal points protected by chiral symmetry and space-time inversion symmetry as well as non-Hermitian exceptional points and Fermi points.



By treating the layer structure as the 2D lattice, we can project any layer group to one of the wallpaper groups. Its no-go theorem inherits the theorem of the corresponding wallpaper group. Let us use 2 layer groups as examples to demonstrate the 2D projection.

$LG\#48$ has four symmetry generators --- $C_{2z}$ rotation, $M_x$ reflection, $T_{\frac{y}{2}}M_z$ glide reflection, $T_{\frac{x+y}{2}}$ half-lattice-constant translation. Global symmetry operations are usually independent of crystalline symmetry operations. Therefore, we can assume  $M_zT_{\frac{y}{2}}$ and $T_{\frac{x+y}{2}}$ commute with global symmetry operations so that  $M_zT_{\frac{y}{2}}$ and $T_{\frac{x+y}{2}}$ do not affect the constraints of the nodal points. This layer group can be projected into WG$\#9$ with $C_{2z},\ M_x,\ T_{\frac{x+y}{2}}$  and its no-go theorem is based on WG$\#9$.

$LG\#54$ possesses two symmetry generators --- $C_{4z}$ rotation, $T_{\frac{x+y}{2}}C_{2y}$ screw rotation. Although $T_{\frac{x+y}{2}}C_{2y}$ flips the $z$ direction, this operator can be projected to $T_{\frac{x+y}{2}}M_x$. Therefore, the projection group is $WG\#12$.

\subsection{Anomalous topological surface} \label{anomalous surface}

\begin{figure}[b!]
\centerline{\includegraphics[width=0.42\textwidth]{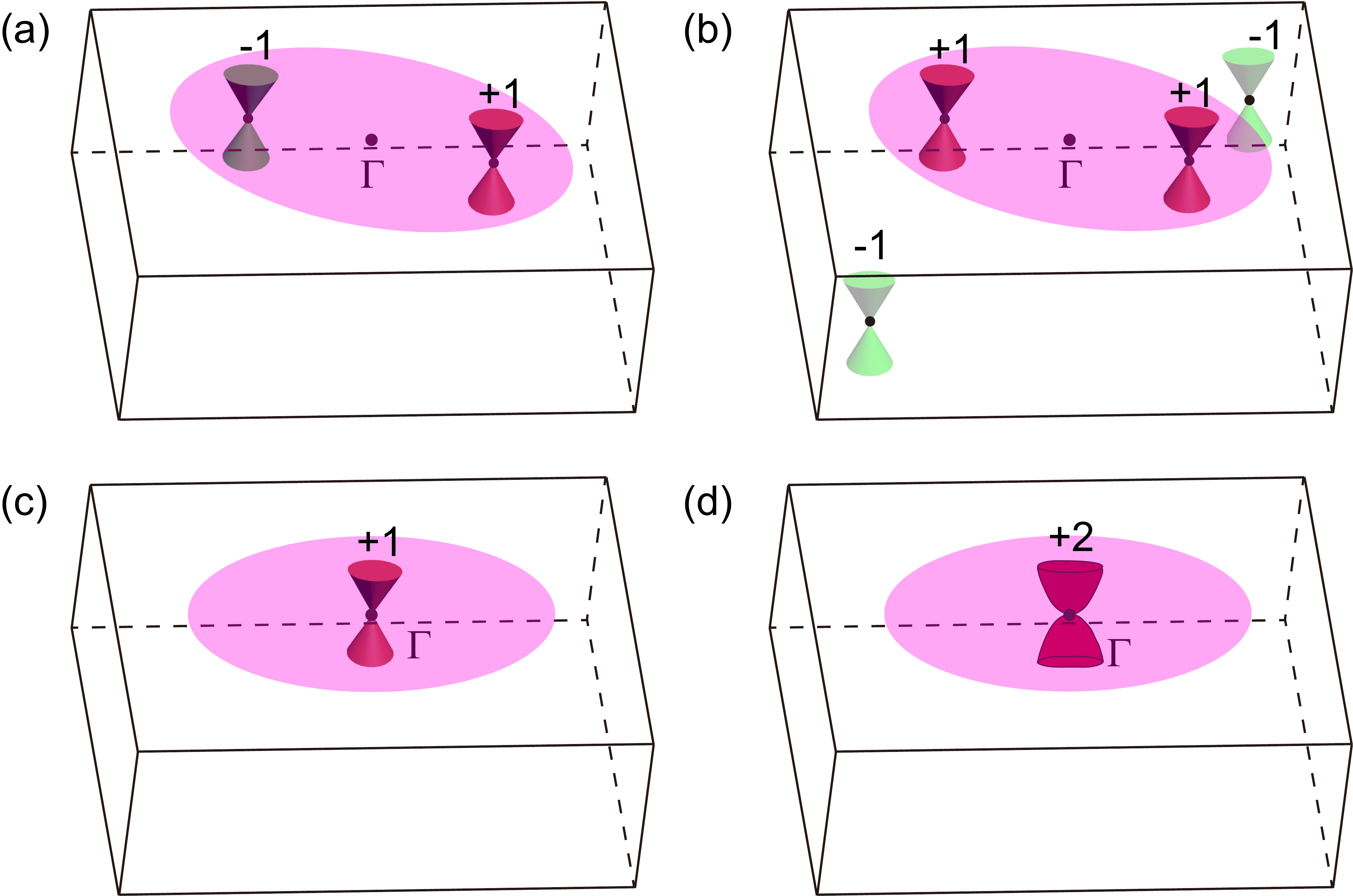}}
\caption{(color online) 
Topological points (nodal points, Fermi points, and exceptional points) are located on the top surface. (a) Because two topological points are connected by $\bar{C}_2$ rotation symmetry, the total charges on the surface are neutralized. (b) Although in the entire 3D system the total charges is neutralized in the presence of the green topological points in the bulk, charge $+2$ on the top surface violates the no-go theorem. (c) A topological point with charge $+1$ is located at a TRI point. (d) A topological point (quadratic-dispersion nodal point, FP, and EP) with charge $+2$ is located at a TRI point. }
\label{anomaly}
\end{figure}

The no-go theorem of the wallpaper groups also serves as clear guidance to search for 3D non-trivial topological phases with anomalous surfaces. This anomalous surface stems from the breaking of the 2D surface no-go theorem. The reason is that the surface BZ (SBZ) can have boundaries and does not cover the entire 2D BZ without boundaries, as shown in Fig.~\ref{anomaly}. Therefore, including all charges in the SBZ, the integral path can not vanish due to the emergence of the SBZ boundaries
\bee
\sum_{\bk_i \in \rm{SBZ}} C(\bk_i) \neq 0.
\ee

 The violation of the charge neutralization can occur in the SBZ and brings anomaly~\cite{PhysRevB.93.075135,RevModPhys.88.035001}. However, symmetries in the wallpaper group might impose additional restrictions to remove the anomaly.
First, in the generalized no-go theorem tables (\ref{no-go away}, \ref{no-go TRS}, \ref{no-go FPs}, \ref{space time inversion table}), ``0" indicates the absence of the topological points for the corresponding WG symmetry classes; hence, the anomalous surface is forbidden so that the bulk is trivial. Second, $\sim$ indicates all nodal points connected by symmetries in the minimal configuration; meanwhile, the area of the SBZ also respects the symmetries. The SBZ either includes or excludes all the nodal points connected by the symmetries, as illustrated in Fig.~\ref{anomaly}(a). Thus, the charges are always neutralized, and the anomalous surface is absent.


The way to realize the anomalous surface is to have all nodal points in the minimal configuration NOT connected by symmetries. In the tables, the non-zero numbers without $\sim$ indicate this case. Although the entire 3D system obeys the 2D no-go theorem, some of the topological points can be outside of the SBZ without breaking any symmetries, as illustrated in Fig.~\ref{anomaly}(b). In the SBZ, the charges of the remaining topological points can violate the neutralization, and the anomalous surface leads to the non-trivial bulk. The 3D bulk topological invariant is given by the sum of the nodal-point charges on the surface
\be
\nu_{\rm{3D}}=\sum_{\bk_i\in{\rm SBZ}} C(\bk_i)
\ee
Let us investigate the non-trivial bulks by studying the violation of our generalized no-go theorem on the surfaces for Hermitian and non-Hermitian systems.

(a) For chiral symmetry, there are five AZ symmetry classes possessing stable nodal points at zero energy. Since the absolute numbers in class BDI and CII are either $0$ or number labeled by $\sim$, the anomalous surface cannot exist in these classes. The trivial 3D topology in these two symmetry classes is consistent with the ten-fold classification~\cite{Kitaev2009,Schnyder2008}. On the other hand, for class AIII, DIII, and CI, to realize the anomalous surface, for a WG preserving rotation symmetry or reflection symmetry, it is required that rotation operator $C_n^+$ commutes with $S$ and reflection operator $M_i^-$ anticommutes with $S$. Otherwise, the crystalline symmetries always keep either the charge-neutralization or the trivialization on the surface. On the anomalous surface, the sum $\nu_{\rm{3D}}$ of the nodal-point charges is equal to the 3D winding number quantized by chiral symmetry~\cite{Ryu2010ten}. Hence, for any WGs with $C_n^+$ symmetry or/and $M_i^-$ symmetry, chiral-symmetric topological insulators/superconductors in class AIII, DIII, or CI are characterized by this $\bZ$ winding number. Particularly, $C_2^+$-inversion-symmetric and $M_i^+$-refection-symmetric phases with the winding number have been studied in the literature~\cite{Sato_Crystalline_PRB14,Chiu_reflection}.


Since nodal points off TRI points are connected by symmetries, to realize the minimal unneutralized charges, the nodal point has to be located at a TRI point, as shown in Fig.~\ref{anomaly}(c). For class AIII and DIII, the charge can be $\pm 1$ so that the $\bZ$ winding number characterizes their bulk topology. On the other hand, since Table \ref{no-go TRS} shows that for class CI the charge at a TRI must be an even number as illustrated in Fig.~\ref{anomaly}(d), the $2\bZ$ winding number for the bulk topology is consistent with the ten-fold classification~\cite{RevModPhys.88.035005}.



(b) For space-time inversion symmetry, Dirac nodes off TRI points for the 10 WGs preserving inversion symmetry are always connected by symmetries. Hence, on the SBZ the total charges are neutralized. On the other hand, while for class AI time-reversal symmetry forbids any Dirac node at any TRI points, for class AII Dirac nodes can be present at TRI points; the SBZ can include only one TRI point without breaking symmetries as shown in Fig.~\ref{anomaly}(c), so the anomalous surface can exist only in class AII. Furthermore, since the Dirac cone is characterized by the $\bZ_2$ invariant (Berry phase), the bulk invariant in class AII with $C_2$ rotation symmetry has $\bZ_2$ invariant, which is consistent with the literature~\cite{Sato_Crystalline_PRB14}.

(c) For non-Hermitian systems, to realize unneutralized exceptional points and Fermi points on the surface, WG generators must be one or two of {\clp $C_n$ and $\bar{M}_x$}, which correspond to the absolute numbers without $\sim$ in Table \ref{no-go FPs}. The reason is that $\bar{C}_{n}$ or/and $M_x$ connect all of the topological points in the minimal configuration or trivialize the points. This is in agreement with that the non-trivial 3D bulk with the non-Hermitian anomalous surface is forbidden by reflection symmetry~\cite{PhysRevLett.126.086401}. Back to the anomalous surface, $C_n$ rotation symmetry leads to a topological point with $n\bZ$ charge at the rotation center or $n$ symmetry-connected topological points with $\bZ$ charges off the rotation center as illustrated in Fig.~\ref{anomaly} (b) and (d). The total charge on the surface is always $n\bZ$, and the 3D topological invariant is a multiple of $n$
\be
\nu_{\rm{3D}}= n j, \ j\in \bZ
\ee
due to $C_n$ rotation symmetry.

The no-go theorem violations of Fermi points and exceptional points bring distinct non-trivial bulks. To keep simplicity, we assume the absence of the skin effect. First, for the anomalous Fermi-point surface, the bulk Hamiltonian $H(\vec{k})$ must have a point gap where unneutralized Fermi points with energy $E$ are located. The topology of the bulk can be described by a 3D $\bZ$ topological invariant~\cite{Denner:2021vb,PhysRevX.9.041015,PhysRevX.8.031079}
\be
W_{3D}=-\int_{\rm{3DBZ}}\frac{d^3\vec{k}}{(2\pi)^3} \epsilon_{ijl} {\rm{Tr}} [Q_i(\vec{k})Q_j(\vec{k})Q_l(\vec{k})]
\ee
where $Q_i(\vec{k})=(H(\vec{k})-E)^{-1}\partial_{k_i}(H(\vec{k})-E)$. The sum of the Fermi-point charges on the anomalous surface corresponds to the integer value of the 3D topological invariant. On the other hand, different from Fermi points at fixed energy, the exceptional points on the surface in the entire energy spectrum are counted to examine the violation of the no-go theorem as an anomaly. The violation leads to the non-trivial topological bulk. The sum of the exceptional-point charges on the anomalous surface corresponds to another $\bZ$ bulk topological invariant, which is unknown in the literature. The hint of the no-go theorem leads to a new direction to study this new non-Hermitian topological phase.

\section{Twisted Bilayer Physics }

	Two graphene-monolayers with a small twisted angle form Moir\'{e} superlattice. At the magic angle, the emergence of the flat bands in the Moir\'{e} BZ has become an important playground to explore strongly correlated physics. In the two monolayers, the Dirac nodes located at $K,\ K'$ evolve to complete flat bands in the entire Moir\'{e} BZ through the interlayer coupling. The presence of the Dirac nodes is one of the key elements for the band flatness. Our generalized no-go theorem can guide us to find other platforms hosting the similar Dirac nodes sorted by symmetries and can extend to square lattice and non-Hermitian platforms.

	As discussed previously in sec.~\ref{graphene section}, graphene without spin-orbital coupling can be simplified to an effective spinless system, which preserves time-reversal symmetry with $T^2=1$ and sublattice (chiral) symmetry, and this monolayer system belongs to WG$\#17$ in class BDI with $C_6^-$ and $M_x^-$. To realize alternative similar twisted bilayer systems, it is worth investigating if other platforms can host only two Dirac nodes at $K,\ K'$. Since the symmetry connection of the two nodes ($\sim$) is not a factor affecting the Dirac node locations, we are looking for these two configurations labeled by $2(K,K')$ and $\tilde{2}(K,K')$. 
Table \ref{no-go away} shows that the same WG in class AIII, BDI and CII also can host the two Dirac nodes at $K,\ K'$ as the minimal configuration. In this regard, the subgroups of these three symmetry groups possessing the identical Dirac physics are listed as WG$\#16$ with $C_6^-$, WG$\#15$ with $C_3^+,\ M_x^-$, WG$\#14$ with $C_3^+,\ M_y^+$, and WG$\#13$ with $C_3^+$ in class AIII, BDI, and CII. Thus, based on those listed symmetry groups, we can search for graphene-like platforms.

	Our generalized no-go theorem also shows that in some symmetry classes the Dirac nodes at $K,\ K'$ must be accompanied by other Dirac nodes due to the charge neutralization. Two Dirac nodes with $+1$ charges located at $K,\ K'$ can be neutralized by two approaches as illustrated in Fig.~\ref{eg TRI} (d) and (b) --- 1.~a quadratic dispersion node with $-2$ charge at $\Gamma$: $4^*(K,K',\Gamma)$, 2.~a Dirac node with $+1$ charge at $\Gamma$ and three Dirac nodes with $-1$ charges at $M_{1/2/3}$ separately: $6(K,K',\Gamma,M_{1/2/3})$. According to Table \ref{no-go away} and Table \ref{no-go TRS}, the corresponding symmetry classes are given by WG$\#17$ with $C_6^+,\ M_x^-$, WG$\#16$ with $C_6^+$, WG$\#15$ with $C_3^+,\ M_x^-$, WG$\#14$ with $C_3^+,\ M_y^-$, and WG$\#13$ in class AIII, DIII, and CI. These symmetry classes provide new directions to search for alternative Dirac platforms.

\begin{figure}[b!]
\centerline{\includegraphics[width=0.48\textwidth]{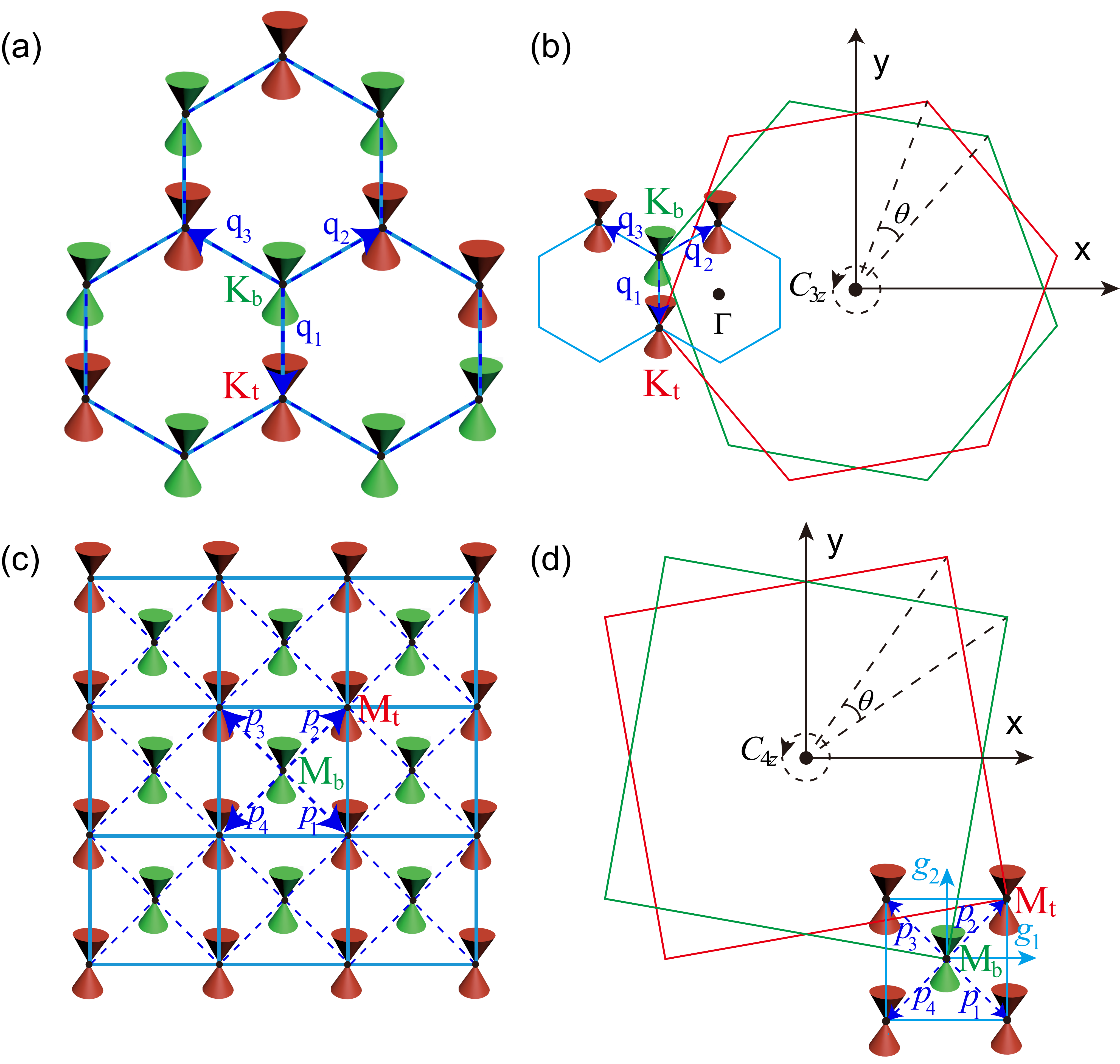}}
\caption{(color online) Momentum-space geometry of twisted bilayers shows the similarity between twisted graphene (hexagonal) and square lattices. While green and red Dirac cones are separately located in the different layers, the blue lines indicate the boundaries of the Moir\'{e} BZs  and the dashed lines indicate the momentum hoppings. (a) The momentum hopping network connecting the Dirac ones captures the the effective physics of the twisted bilayer graphene. (b) The momentum hoppings $\bq_{n}$ in the three directions stem from the momentum difference between the two Dirac cones. (c) The twisted square lattice with Dirac points at $M$ point can also be described by the hopping network. (d) Similarly, the momentum difference between the two Dirac cones leads to the hoppings $\bp_{m}$ in the four directions. 
} 
\label{twist}
\end{figure}

To extend the twisted physics to square lattices, we consider the low-energy effective continuum Hamiltonian of the twisted bilayer graphene~\cite{Bistritzer-pnas11} describes the single-valley physics of the two Dirac cones at $K$ separately in the different layers with the interlayer coupling and can be written in the real space with twisted angle $\theta$ and lattice constant $a_0$~\cite{Ashvin-prl19-flat}
\bee
H_{K}=
\bma -iv_F\bm{\sigma_{\theta/2}\cdot\nabla} & \mathcal{T}_3(\br)  \\
\mathcal{T}_3^\dagger(\br) & -iv_F\bm{\sigma_{-\theta/2}\cdot\nabla}
\ema \label{TBG}
\ee
where $\bm{\sigma_{\pm\theta/2}}=e^{\mp i\theta\sigma_z/4}(\sigma_x,\sigma_y)e^{\pm i\theta\sigma_z/4}$, $\bm{\nabla}=(\partial_x,\partial_y)$, and $\mathcal{T}_3(\br)=\sum_{n=1}^3T_n e^{-i \bq_n\cdot \br}$. The two diagonal terms represent the two Dirac cones located at the different layers and the interlayer coupling $\mathcal{T}_3(\br)$  is composed of three momentum hoppings differed by angle $\psi=2\pi/3$ 
\begin{align}
\bq_{n+1}&=k_\theta (\sin n\psi, - \cos n\psi ), \\
T_{n+1}&=w_{\rm{AA}}\sigma_0 + w_{\rm{AB}} (\sigma_x \cos n \psi + \sigma_y \sin n \psi ),
\end{align}
where $k_\theta =   \frac{8\pi \sin (\theta/2)}{3a_0}$ and $w_{\rm{AA}/\rm{BB}}$ indicates the strength of the AA/AB-basis coupling. Fig.~\ref{twist}(a,b) illustrate the effective physics is captured by the network of the two Dirac cones with the extension of the momentum hoppings. 
The platform for the twisted physics can be naturally extended to a Dirac node at $M$ point in a square lattice with lattice constant $b_0$~\cite{PhysRevB.104.035136,PhysRevResearch.1.033076}. Consider the low-energy physics of the twisted square-lattice bilayer and the Hamiltonian, which shares a similar form of the twisted bilayer graphene (\ref{TBG}), reads
\bee
H_{M}=
\bma -iv_F\bm{\sigma_{\theta/2}\cdot\nabla} & \mathcal{S}_4(\br)  \\
\mathcal{S}_4^\dagger(\br) & -iv_F\bm{\sigma_{-\theta/2}\cdot\nabla}
\ema.
\ee
The physics describes the coupling between the two Dirac nodes at $M$ separately located in the two layers. The interlayer coupling has momentum hoppings in four directions $\mathcal{S}_4(\br)=\sum_{m=1}^4 S_m e^{-i \bp_m\cdot \br}$; Using the form of $\mathcal{T}_3(\br)$ in the graphene,  we can write the details of the hoppings as  
\begin{align}
\bp_{m+1}&=p_\theta(\sin (m\phi+\phi_0),-\cos (m\phi+\phi_0) ), \\
S_{m+1}&= d_0 + d_1 (\sigma_x \cos (m\phi +\phi_0) + \sigma_y  \sin (m\phi +\phi_0)  ),
\end{align}
where $p_\theta=\pi \sin (\theta/2)/b_0$, $\phi=\pi/2$, and $\phi_0=\pi/4$. Fig.~\ref{twist}(c,d) illustrate the expansion of the momentum hoppings form the network of the two Dirac cones from the different layers.   


	
	Now we are looking for symmetry classes hosting a Dirac node at $M$. Due to the no-go theorem, the Dirac node is always accompanied with other nodes. According to Table \ref{no-go away} and Table \ref{no-go TRS}, in class AIII and DIII the minimal configuration $4(\Gamma,M,X,Y)$ can be realized in WG$\#11,12$ with $C_4^+,\ M_x^-$ and WG$\#10$ with $C_4^+$ as new platforms for twisted bilayer physics.

	Graphene hosts Dirac nodes at $K$ and $K'$ points with $\pm 1$ charges quantized by chiral (sublattice) symmetry. It is natural to seek a similar non-Hermitian platform possessing topological points (EPs and FPs) at the same locations. However, since $K$ and $K'$ points are the $C_3$ rotation center, different from the Hermitian nodal phase, the values of the non-Hermitian charges must be a multiple of 3 (\ref{FP rotation constraint},\ref{rotation constraint EP}). On the other hand, for the square BZ, the charge of the topological point at $M$ point, which is the $C_4$ rotation center, must be a multiple of 4. Thus, nodal points for twisted Hermitian and non-Hermitian bilayers are different. The charge of the nodal point at a $C_n$ rotation center in a non-Hermitian layer must be a multiple of $n$, while there is no restriction for Hermitian layers. 
	

	
	








\section{Conclusion} \label{conclusion}

	The stable topological points in 2D lattices arise in chiral-symmetric, space-time-inversion-symmetric, and non-Hermitian systems. In the presence of crystalline symmetries and time-reversal symmetry, the minimal number of the topological points might be more than two beyond the original no-go theorem (the fermion doubling theorem~\cite{NIELSEN1981173,NIELSEN198120,NIELSEN1981219,2019arXiv191202788Y}). We exhaustively study the generalized no-go theorem for the three distinct types of lattice systems and list the minimal numbers of the topological points with the configurations for given wallpaper groups. This theorem provides an alternative scheme to classify the nodal topological materials and includes the nodal topological points located at general points in BZ, which might be overlooked in the symmetry indicator schemes~\cite{Tang:2019aa,Zhang:2019aa,Vergniory:2019aa,2019arXiv190909634O,2019arXiv191011271G}.
	
	This generalized no-go theorem is a unified principle and a fundamental rule for forming Dirac semimetals, nodal time-reversal-symmetric topological superconductors, and non-Hermitian lattices with exceptional points or Fermi points. We start with chiral-symmetric nodal systems for the 17 wallpaper groups. Space-time-inversion-symmetric nodal semimetals and non-Hermitian nodal systems inherit the principle from the chiral-symmetric system with small modifications. The examples in the manuscript show that the topological materials in the literature must obey this ground rule, such as graphene and d-wave superconductors. In addition, the theorem predicts several new Dirac semimetals, such as six Dirac nodes in a hexagonal lattice (\ref{H6}) and a quadratic node with two Dirac nodes (\ref{H4}). Following this guidance, we can reveal and realize new types of novel nodal topological materials with topological nodal points. In particular, these Dirac semimetals in our prediction can potentially possess Majorana bound states with superconductivity~\cite{PhysRevResearch.2.012060}. On the other hand, the non-Hermitian topology of Fermi points and exceptional points is captured by $\bZ$ invariants. Unlike the Hermitian Dirac nodes, the charges of those non-Hermitian topological points at the rotation center, say $C_n$ rotation, must be multiple of $n$. Although $\bZ$ invariants also protect Dirac nodes in chiral-symmetric lattices, the non-Hermitian no-go theorem is distinct from the Hermitian one.

	The generalization of the no-go theorem also leads to several applications in condensed matter physics. The no-go theorem for the 17 wallpaper groups can be directly extended to the 80 layer groups through the projection. Our classification can determine if the no-go theorem on the surface can be violated; since the violation implies non-trivial bulks, this approach indirectly classifies Hermitian and non-Hermitian 3D topological bulk with $\bZ$ and $\bZ_2$ invariants. Twisted bilayer graphene brings fruitfully interesting physics from the coupling of the Dirac cones. New predicted Dirac semimetals from the generalized no-go theorem can be new platforms of twisted bilayers for strong correlations.

	


\section{acknowledgement}

We thank Chen Fang, Akira Furusaki, Guang Bian, and Qiang Zhang for the helpful discussions and comments.

\appendix

\section{Hamiltonian properties} \label{Bloch wave}

In certain basis, each entry of the Hamiltonian can be a single-valued function of momentum $\bk$; this single-valued property is the key to prove the fermion no-go theorem. On the other hand, the symmetry equations for the Hamiltonian through crystalline symmetry play an essential role to determine the minimal number of the topological points in the BZ. To set up these foundations of the generalized no-go theorem, we prove these two features above for 2D lattice systems in this appendix.

We intentionally choose the annihilation operator in the momentum basis in the unit cell convention
\bee
\phi_{\alpha a}(\boldsymbol{k})=\frac{1}{\sqrt{N}}\sum_{\boldsymbol{R}_n}e^{-i\boldsymbol{k} \cdot \boldsymbol{R}_n}\psi_{\alpha }(\boldsymbol{R}_n+\boldsymbol{r}_{a}),
\ee
where $\boldsymbol{R}_n$ is the central position of the $n$-th unit cell, $N$ is the number of the unit cell. Label $a$ indicates different spatial locations of atoms and $\boldsymbol{r}_a$ is the location of the atom, whereas label $\alpha$ indicates non-spatial dependent index, such as spin. In this convention, the phase in the Fourier transformation does not have $\boldsymbol{r}_a$-dependence and the location-dependence of the annihilation operator $\phi_{\alpha a}$ is absorbed in index $a$. By the definition, the annihilation operator has periodicity
\bee
\phi_{\alpha a}(\boldsymbol{k}+ m_1\boldsymbol{G}_1+m_2\boldsymbol{G}_2)=\phi_{\alpha a}(\boldsymbol{k}),
\ee
where $\boldsymbol{G}_i$ is the reciprocal lattice vector and $m_i$ is an integer; therefore, $\phi_{\alpha a}(\boldsymbol{k})$ is single-valued in the BZ.
On the other hand, the Hamiltonian in the form of the second quantization in momentum basis is written as
\bee
\hat{H}=\sum_{\bk }\phi_{\beta b}^\dagger(\boldsymbol{k}) H_{\beta b, \alpha a} (\bk)\phi_{\alpha a}(\boldsymbol{k}). \label{2nd H}
\ee
Thus, each entry $H_{\beta b, \alpha a}(\bk) $ in the free fermion Hamiltonian is a single-valued function of momentum in the BZ.

	Consider crystalline symmetry operator with translation $\tau$ acts on the annihilation operator
\begin{align}
g_\tau & \psi_{\alpha a} (\bR+\br_a)  \nonumber \\
=& \sum_{\beta} W_{\alpha   \beta } \psi_{\beta } (g_\tau (\bR +\br_a)) \nonumber   \\
=& \sum_{\beta} W_{\alpha   \beta } \psi_{\beta } (\bR'^{ab} + \sum_{b} V_{ab} \br_{b}) \label{crystal location}  \\
=& \sum_{\beta,b} W_{\alpha \beta }V_{ab} \psi_{\beta } (g \bR+\br_b+n_1^b g \boldsymbol{a}_1 + n_2^b g\boldsymbol{a}_2 ), \nonumber  
\end{align}
where $W_{\alpha   \beta }$ is a constant transformation unitary matrix for crystalline operation, such as spin rotation and the orbital phase changes.
We note that the crystalline operator also alters the location of the annihilation operator
\begin{align}
g_\tau (\bR +\br_a)=\bR' + \sum_b V_{ab} \br_{b},
\end{align}
where $\bR'$ is the central location of another unit cell. Matrix $V_{ab}$ directly switches the original atom location to another and does not mix other atoms locations. That is, each entry in $V_{ab}$ is either 0 or 1. In this regard, $V_{ab}$ and the summation of $b$ in Eq.~\ref{crystal location} can be moved out of the parenthesis. In general, $\bR'^{ab} \neq g \bR $, where crystalline operator for position vector $g$ acts on $R_n$ without translation and the invariant point of the operation (rotation center or reflection) is in the original point; hence, we have
\bee
\bR'^{ab} = g \bR+ n_1^{ab} g \boldsymbol{a}_1 + n_2^{ab}  g\boldsymbol{a}_2,
\ee
where integer $n_i^{ab} $, which depends on the atom location, indicates the number of the unit cell shift after the crystalline symmetry operation and $\boldsymbol{a}_i$ is the primitive vector obeying
\bee
\bk \cdot (n_1\boldsymbol{a}_1 + n_2\boldsymbol{a}_2)=k_1 n_1 + k_2 n_2,
\ee
where $\bk=(k_1 \boldsymbol{G}_1 + k_2 \boldsymbol{G}_2)/2\pi$.
Since the crystalline symmetry is preserved, $g\boldsymbol{a}_i$ is still the primitive vector; the last two terms in the equation above represent the shift of the unit cell.
	
	We can discuss the $g$ acts on the annihilation operator in the momentum space
\begin{small}	
\begin{align}
g_\tau & \phi_{\alpha a}(\boldsymbol{k}) \nonumber \\
=&\frac{1}{\sqrt{N}}\sum_{\boldsymbol{R}_n}e^{-i\boldsymbol{k} \cdot \boldsymbol{R}_n} g_\tau \psi_{\alpha }(\boldsymbol{R}_n+\boldsymbol{r}_{a}) \nonumber  \\
=&\frac{1}{\sqrt{N}}\sum_{\boldsymbol{R}_n,\beta,b}e^{-i\boldsymbol{k} \cdot \boldsymbol{R}_n} W_{\alpha \beta } V_{ab} \psi_{\beta } ( \bR'^{ab} +\br_b) \nonumber  \\
=&\frac{1}{\sqrt{N}}\sum_{\boldsymbol{R}_n,\beta,b}e^{-ig\boldsymbol{k} \cdot g\boldsymbol{R}_n} W_{\alpha \beta }V_{ab} \psi_{\beta } ( \bR'^{ab} +\br_b) \nonumber  \\
=&\frac{1}{\sqrt{N}}\sum_{\boldsymbol{R}_n,\beta,b}e^{-ig\boldsymbol{k} \cdot (\boldsymbol{R}_n'^{ab} -n_1^{ab} g \boldsymbol{a}_1-n_2^{ab} g\boldsymbol{a}_2 )} W_{\alpha \beta } V_{ab} \psi_{\beta } ( \bR'^{ab} +\br_b) \nonumber  \\
=&\frac{1}{\sqrt{N}}\sum_{\boldsymbol{R}_n,\beta,b}e^{ig\boldsymbol{k} \cdot (n_1^{ab} g \boldsymbol{a}_1+n_2^{ab} g\boldsymbol{a}_2 )}  W_{\alpha \beta }V_{ab} e^{-ig\boldsymbol{k} \cdot \boldsymbol{R}_n'^{ab}  }  \psi_{\beta } ( \bR'^{ab} +\br_b) \nonumber  \\
 =&\sum_{\beta,b}e^{i (k_1n_1^{ab} +k_2n_2^{ab}  )}  W_{\alpha \beta } V_{ab}  \phi_{\beta b}( g\bk)
\end{align}
\end{small}	
To simplify the expression of the crystalline symmetry operation, we can define the single-valued $\bk$-dependent unitary matrix
 \begin{align}
U_{\alpha a, \beta b}^\dagger (\bk) = e^{i (k_1n_1^{ab} +k_2n_2^{ab}  )}  W_{\alpha  \beta } V_{ab}.
\end{align}	
The annihilation operator and creation operator can be rewritten in the economic form
\bee
g_\tau \phi(\bk)= U^\dagger(\bk) \phi (g\bk),\ g_\tau \phi^\dagger(\bk)=  \phi^\dagger (g\bk) U(\bk)
\ee

Due to the symmetry, the second quantization Hamiltonian (\ref{2nd H}) is invariant under the crystalline symmetry operation
\begin{align}
&\sum_{\bk} \phi^\dagger (\bk) H(\bk ) \phi(\bk) = g_\tau   \sum_{\bk} \phi^\dagger (\bk) H(\bk ) \phi(\bk) \nonumber \\
=&   \sum_{\bk} \phi^\dagger (g\bk) U(\bk) H(\bk )U^\dagger(\bk) \phi(g\bk)
\end{align}
Hence, the symmetry equation for the free fermion Hamiltonian is given by
\bee
H(g\bk ) =U(\bk) H(\bk )U^\dagger(\bk).
\ee
We will frequently use this symmetry equation through the manuscript for the generalized no-go theorem.

\section{The winding number connection between $\bK_0$ and $g\bK_0$} \label{winding connection}

We provide the detailed deviations for the relations of the two winding numbers at $\bK_0$ and $g\bK_0$. First, let us show that the definition (\ref{winding number eq}) of the winding number is equivalent to the one using the flattened Hamiltonian. The flattened Hamiltonian
\begin{eqnarray}
Q(\boldsymbol{k})=\sum_{0<E_\alpha}^{N_{+}} | \Psi^{\alpha}(\boldsymbol{k}) \rangle\langle \Psi^{\alpha}(\boldsymbol{k})|-\sum_{0>E_\beta}^{N_{-}}| \Psi^{\beta}(\boldsymbol{k})\rangle\langle \Psi^{\beta}(\boldsymbol{k}) |,
\end{eqnarray}
where $| \Psi^{\gamma}(\boldsymbol{k})\rangle$ is the eigenstate of the $\mathcal{H}({\bk})$ with energy $E_\gamma$. Assuming the absence of zero-energy states, chiral symmetry leads to $N_+=N_-$. Since the flattened Hamiltonian inherits chiral symmetry (\ref{chiral symmetry}) from the original Hamiltonian by choosing a proper basis, chiral symmetry operator can be written as $S=\tau_z\otimes \bI_{n\times n}$ and the flattened Hamiltonian has this off-diagonal form
\bee
Q(\bk)=
\bma
0 & q(\bk) \\
q^\dagger(\bk) & 0
\ema.
\ee
Due to $Q^2=\bI_{2n\times 2n}$, $q(\bk)$ is a unitary matrix and invertible due to the absence of the zero-energy state. Using $q(\bk)$, the winding number is defined in a closed loop integral path in the BZ
\begin{equation}\begin{aligned}
\nu&=\frac{i}{2\pi}\oint d \boldsymbol{k}\cdot Tr[q^{-1}(\bm{k})\partial_{\bm{k}}q(\bm{k})]\\
  &=\frac{i}{2\pi}\oint d \boldsymbol{k}\cdot \partial_{\bm{k}}Tr[\ln q(\bm{k})] \\
  &=\frac{i}{2\pi}\oint d \big( \ln\det[q(\bm{k})] \big )
\end{aligned}\end{equation}
This definition of the winding number is regularly used in the literature~\cite{Schnyder2008,RevModPhys.88.035005}.



To show that $h(\bk)$ and $q(\bk)$ lead to the same winding number, we choose the eigenstates as the basis and then prove that $\mH(\bk)$ commutes with $Q(\bk)$ since
\begin{small}
\begin{eqnarray}
&&\mathcal{H}(\bm{k})Q(\bm{k})
\\
=&&\sum_{0<E_\alpha}^{N_{+}} E_\alpha(\bm{k})| \Psi^{\alpha}(\boldsymbol{k}) \rangle\langle \Psi^{\alpha}(\boldsymbol{k})|- \sum_{0>E_\beta}^{N_{-}}E_\beta(\bm{k})| \Psi^{\beta}(\boldsymbol{k})\rangle\langle \Psi^{\beta}(\boldsymbol{k}) |  \nonumber  \\ 
=&&Q(\bm{k})\mathcal{H}(\bm{k})
\end{eqnarray}
\end{small}
The commutation relation can be rewritten in the proper basis
\begin{eqnarray}
&&0=[\mathcal{H}(\bm{k}),Q(\bm{k})]
\\
&&=\left(\begin{array}{cc}
h(\bm{k})q^\dag(\bm{k})-q(\bm{k})h^\dagger(\bm{k}) & 0   \\
   0 &   h^\dagger(\bm{k})q(\bm{k}) -q^\dag(\bm{k})h(\bm{k})    \\
\end{array}\right).   \nonumber
\end{eqnarray}
One can obtain
\begin{eqnarray}
&&\det[h(\bm{k})q^\dagger(\bm{k})]=\det[{q(\bm{k})}h^\dagger(\bm{k})] \nonumber
\\
&&=\det[(h(\bm{k}){q^\dagger(\bm{k})})^\dagger]=\det[h(\bm{k})q^\dagger(\bm{k})]^*.
\end{eqnarray}
The above equation implies that $\det[h(\bm{k})q^\dag(\bm{k})]$ is a real number so that $\ln \det(h(\bk))+\ln \det(q^*(\bk))=\gamma(\bk)$ is a single-valued function, which leads to $\oint d  \gamma(\bk)=0$. Using the property of the unitary matrix $q(\bk)$, we have $\ln \det (q^*) + \ln \det (q) =0$. Hence, the winding number with the closed integral path $\Gamma(\bK_0)$ can be directly written in form of $h(\bk)$
\begin{equation}\begin{aligned}
\nu(\boldsymbol{K}_{0})&=\frac{i}{2\pi}\oint_{\Gamma\left(\boldsymbol{K}_{0}\right)} d \big ( \ln\det[q(\bm{k})] \big ) \\
&=\frac{i}{2\pi}\oint_{\Gamma\left(\boldsymbol{K}_{0}\right)} d \big ( \ln\det[h(\bm{k})] \big )
\end{aligned}\end{equation}
To connect the relation between $\nu(\bK_0)$ and $\nu(\Gamma\bK_0)$, we first consider $g_{\boldsymbol{\tau}}=\{g|\boldsymbol{\tau}\}$ with spatial translation $\boldsymbol{\tau}$ as a $2D$ crystalline symmetry operator.

 Since the matrix form of the symmetry operator commutes with chiral symmetry operator $S=\tau_z\otimes 1$, the unitary symmetry operator in the single-particle basis (or particle-hole basis for superconductors) is written in a generic form of
\bee
U_g^+(\bk)=
\bma
U_{g1}(\bk) & 0 \\
0 & U_{g2}(\bk)
\ema, \label{Ugplus}
\ee
where $U_{gi}(\bk)$ is a momentum-dependent unitary matrix, which can describe a nonsymmorphic symmetry operation. The crystalline symmetry equation (\ref{crystalline symmetry equations}) connects the Hamiltonian at $\bK_0$ and $g\bK_0$
\begin{align}
\mH(g\bk)=&U_g^+(\bk) \mH (\bk) U_g^{+\dag}(\bk) \\
=&\bma
0 & U_{g1}(\bk)h(\bk)U^\dagger_{g2}(\bk) \\
U_{g2}(\bk)h^\dag(\bk)U^\dagger_{g1}(\bk) & 0
\ema.  \nonumber
\end{align}
We have $h(g\bk)=U_{g1}(\bk)h(\bk)U_{g2}^\dag(\bk)$.
The integral path $\Gamma(g\bK_0)$ is an infinitesimal close loop encircling nodal point $g\boldsymbol{K}_{0}$ and then the winding number at $\Gamma (g\bK_0)$ is given by
\begin{equation}\begin{aligned}
\nu(g\boldsymbol{K}_{0})&=\frac{i}{2\pi}\oint_{\Gamma\left(g\boldsymbol{K}_{0}\right)} d \big ( \ln\det[h(\bm{k})] \big ) \\
&=\frac{i}{2\pi}\oint_{\Gamma\left(\boldsymbol{K}_{0}\right)} \det(g) d \big ( \ln\det[h(g\bm{k})] \big ) \\
&=\frac{i}{2\pi}\oint_{\Gamma\left(\boldsymbol{K}_{0}\right)} \det(g) d \big ( \ln\det[h(\bm{k})] \big ) \\
&=\det(g) \nu(\bK_0)
\end{aligned}\end{equation}
We use the integral $\oint_{\Gamma (K_0)} d\big ( \ln \det U_{gi}(\bk) \big)=0$ since $U_{gi}(\bk)$ does not collect additional $U(1)$ winding phase in the infinitesimal integral loop. The reason is that $\det (U_{g\pm}(\bk))$ does not vanish to create a nodal point due to the nature of unitary matrix.

Secondly, consider time reversal symmetry operator $T^+$ commutes with chiral symmetry operator $S$
\bee
\mH(-\bk) = T \mH(\bk ) T^{-1}= U_T \mH^*(\bk ) U_T^\dagger,
\ee
where $T^+=U_T \mathcal{K}$ and $\mathcal{K}$ is the complex conjugate operator. The commutation relation between $T^+$ and $S$ gives the form of
\bee
T^+=
\bma
U_{T1} & 0 \\
0 & U_{T2}
\ema \mathcal{K},
\ee
where $U_{Ti}$ is momentum-independent since time reversal symmetry operator is non-spatial.
In 2D momentum space, time reversal operation is equivalent to the combination of $C_2$ rotation and complex conjugation. Therefore,
\begin{equation}\begin{aligned}
\nu(-\boldsymbol{K}_{0})
&=\frac{i}{2\pi}\oint_{\Gamma\left(-\boldsymbol{K}_{0}\right)} d \big ( \ln\det[h (\bm{k})] \big ) \\
&=\frac{i}{2\pi}\oint_{\Gamma\left(\boldsymbol{K}_{0}\right)} d \big ( \ln\det[h^* (\bm{k})] \big ) \\
&=\frac{-i}{2\pi} \oint_{\Gamma\left(\boldsymbol{K}_{0}\right)}   d \big ( \ln\det[h (\bm{k})] \big ) \\
&=-\nu(\bK_0)
\end{aligned}\end{equation}
Hence, with the commutation relation between $T^+$ and $S$, time reversal operation flips the sign of the winding number for the nodal point at $-\bK_0$.

On the other hand, when symmetry operator anticommutes with $S$, the relation of the winding numbers between $\bK_0$ and $g\bK_0$ flips the sign. Following the similar derivation, we have crystalline symmetry operator $U_g^-(\bk)$, which anticommutes with $S$, in the form of
\bee
U_g^-(\bk) =
\bma
0 & V_{g1}(\bk) \\
V_{g2}(\bk) & 0
\ema, \label{Ugminus}
\ee
where $V_{gi}$ is a unitary matrix. This symmetry operator links the Hamiltonian at $\bK_0$ and $g\bK_0$
\begin{align}
\mH(g\bk)=&U_g^-(\bk) \mH(\bk) U_g^{-\dag}(\bk) \\
=&\bma
0 & V_{g1}(\bk)h^\dag(\bk)V^\dagger_{g2}(\bk) \\
V_{g2}(\bk)h(\bk)V^\dagger_{g1}(\bk) & 0
\ema.   \nonumber
\end{align}
Using $h(g\bk)=V_{g1}(\bk)h^\dag (\bk)V_{g2}^\dag(\bk)$ and following the similar derivation with $U_g^+(\bk)$, we have this relation of the winding numbers at the two nodal points
\begin{equation}\begin{aligned}
\nu(g\boldsymbol{K}_{0})
&=\frac{i}{2\pi}\oint_{\Gamma\left(\boldsymbol{K}_{0}\right)} \det(g) d \big ( \ln\det[h(g\bm{k})] \big ) \\
&=\frac{i}{2\pi}\oint_{\Gamma\left(\boldsymbol{K}_{0}\right)} \det(g) d \big ( \ln\det[h^\dag (\bm{k})] \big ) \\
&=\frac{-i}{2\pi} \oint_{\Gamma\left(\boldsymbol{K}_{0}\right)} \det(g)  d \big ( \ln\det[h (\bm{k})] \big )  \\
&=-\det(g) \nu(\bK_0)
\end{aligned}\end{equation}

For time-reversal symmetry operator $T^-$ anticommuting with $S$, since the complex conjugate operator in $T^-$ flips the sign of the winding number, the relation between $\bK_0$ and $-\bK_0$ with time-reversal symmetry is given by
\bee
\nu(-\bK_0)=\nu(\bK_0)
\ee
The results of the winding number relations are summarized in Eqs.~\ref{crystalline winding}, \ref{time reversal winding} and Table.~\ref{winding number table}.

\section{The algebra between chiral operator and time-reversal operator} \label{AZ chiral}

We know that $T^2=C^2=\pm 1$ in class BDI, CII and $T^2=-C^2=\pm 1$ in class CI, DIII. Let us consider $T^2=C^2=\pm 1$. We can choose a proper basis and phase so that chiral operator $S$ is hermitian and $S^2=1$. By sandwiching $S^2$ with $T$ and $C$, we have
\bee
TC=TS^2C=T^2 CT C^2=CT.
\ee
This commutation relation leads to $0=T[T,C]=[T,TC]=[T,S]$. That is, class BDI, CII preserve $T^+$ time-reversal symmetry. Following the similar derivation, we have $T^-$ time-reversal symmetry for class CI, DIII.

\section{Neutralized winding number summation} \label{neutralization_proof}
	
\begin{figure}[t!]
\centerline{\includegraphics[width=0.5\textwidth]{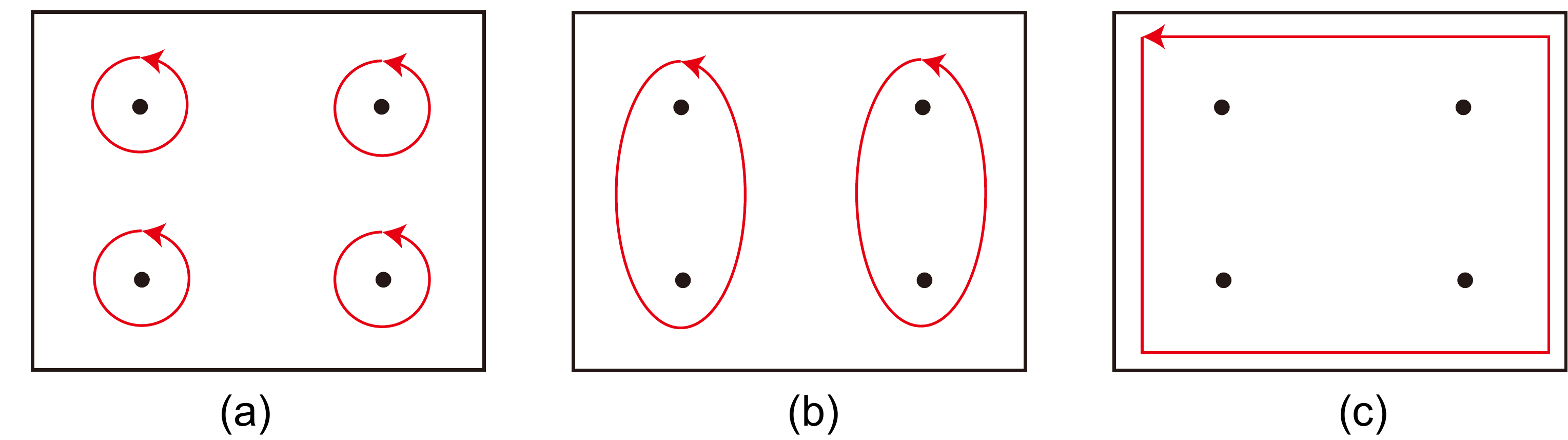}}
\caption{(color online) The integral paths (red line) of the winding numbers for the nodal points (black dots) evolve in the BZ.  The black box represents the entire boundary of the BZ. (a) The integral path of the winding number is an infinitesimal loop encircling a nodal point. (b) The integral path of the multiple loops grows without passing through any nodal point and through the deformation two loops form one big loop. (c) The integral path for all the nodal nodal is equivalent to the integral path along the boundary of the BZ, since through the deformation the path does not pass any nodal points. By using the periodic boundary condition of the BZ, this integral path vanishes. }
 \label{neutralization}
\end{figure}
	
	Each nodal point in the BZ is characterized by winding number. The essence of the no-go theorem is that the summation over winding numbers for all the nodal points must vanish. To prove the neutralization we simply write the definition of the winding number summation
\begin{equation}\begin{aligned}
\sum_i\nu(\boldsymbol{K}_{0}^i)&=\sum_i\frac{i}{2\pi}\oint_{\Gamma\left(\boldsymbol{K}_{0}^i\right)} d \big ( \ln\det[q(\bm{k})] \big ) \\
&=\frac{i}{2\pi}\oint_{\partial \rm{BZ}} d \big ( \ln\det[h(\bm{k})] \big )\\
&=0,
\end{aligned}\end{equation}
where $K_0^i$ represent all of the nodal points ($E=0$) in the BZ. The integral path of the winding number for any nodal point is an infinitesimal loop counterclockwise encircling that nodal point. The integral path for all the nodal points can continuously deform to the boundary of the BZ ($\partial$BZ) as illustrated in Fig.~\ref{neutralization}. Importantly, the integral path never passes any nodal points during the deformation so that the total winding number is unchanged after the deformation; hence, the second line with $\partial$BZ in the equation holds. Since $h(\bm{k})$ can be a single-valued in the entire BZ, by using the periodic boundary, $\partial$BZ vanishes so that the summation must be zero. 



\section{Four Dirac nodes: WG$\#17$ in class DIII}\label{special four}
	
In all the WG symmetry classes, it is the most difficult task to find a tight-binding model for the minimal configuration of a nodal point with $\nu=2$ located at $\Gamma$ and two nodal points with $\nu=-1$ at $K, K'$ in class DIII. In a 2-band model, time reversal symmetry $(T^2=-1)$ limits the winding number of the nodal point at $\Gamma$ to an odd number. Hence, the 2-band model cannot host a nodal point at $\Gamma$ carrying an even winding number. To find the minimal tight-binding model hosting this high-ordered nodal point, we start with a 4-band model consisting of two 6-Dirac-point Hamiltonians $H_6(\bk)$ (\ref{H6}) in class DIII. Since $H_6(\bk)$ possesses 6 Dirac nodes separately located at $\Gamma, M_{1,2,3}, K', K$, two of the twelve Dirac nodes at $\Gamma$ carry $\nu=2$ winding number together, two nodes at $M_{1,2,3}$ carry $\nu=-2$, and two nodes at $K, K'$ carry $\nu=2$. However, the nodal points at $M_{1,2,3}$ should be removed and the signs of the winding numbers at $K, K'$ should be switched. To achieve this, we intensionally add $H_4(\bk)$ (\ref{H4}) from class CI with real coefficient $\alpha$ as a coupling between two $H_6(\bk)$'s. The $4\times 4$ Hamiltonian is written in the form of
\bee
H_{4\times 4}(\bk)=
\bma
0 & h_6(\bk) & 0 & \alpha h_4(\bk) \\
h_6^*(\bk) & 0 & -\alpha h_4^*(\bk) & 0 \\
0 & -\alpha h_4(\bk) & 0 & -h_6(\bk) \\
\alpha h^*_4(\bk) & 0 & -h_6^*(\bk) & 0 \\
\ema. \label{merge H}
\ee
Since time reversal symmetry with $T=i\tau_0 \sigma_y \mathcal{K}$ and particle-hole symmetry $C=\tau_0 \sigma_x \mathcal{K}$ are preserved, the system belongs to class DIII. Hence, chiral symmetry operator is given by $S=\tau_0\sigma_z$. Furthermore, this is WG$\#17$ with $C_6$ rotation and $M_x$ generators since the Hamiltonian obeys $M_xH_{4\times 4}(-k_x, k_y) M_x^{-1}=H_{4\times 4}(k_x,k_y)$ with $M_x=\tau_z \sigma_y$ and $C_6H_{4\times 4}(\frac{k_x+\sqrt{3}k_y}{2},\frac{k_y-\sqrt{3}k_x}{2}) C_6^{-1}=H_{4\times 4}(k_x,k_y)$ with $C_6=\cos \pi/6 \tau_z \sigma_0-i \sin \pi/6 \tau_z \sigma_z$. The algebra between the generators and the chiral symmetry operator is given by $M_x^-$ and $C_6^+$. According to Table \ref{no-go TRS}, this WG symmetry class can host a nodal point at $\Gamma$ with $\nu=2$, two nodal points with $\nu=-1$ at $K, K'$ separately as the minimal configuration.

\begin{figure}[t!]
\centerline{\includegraphics[width=0.5\textwidth]{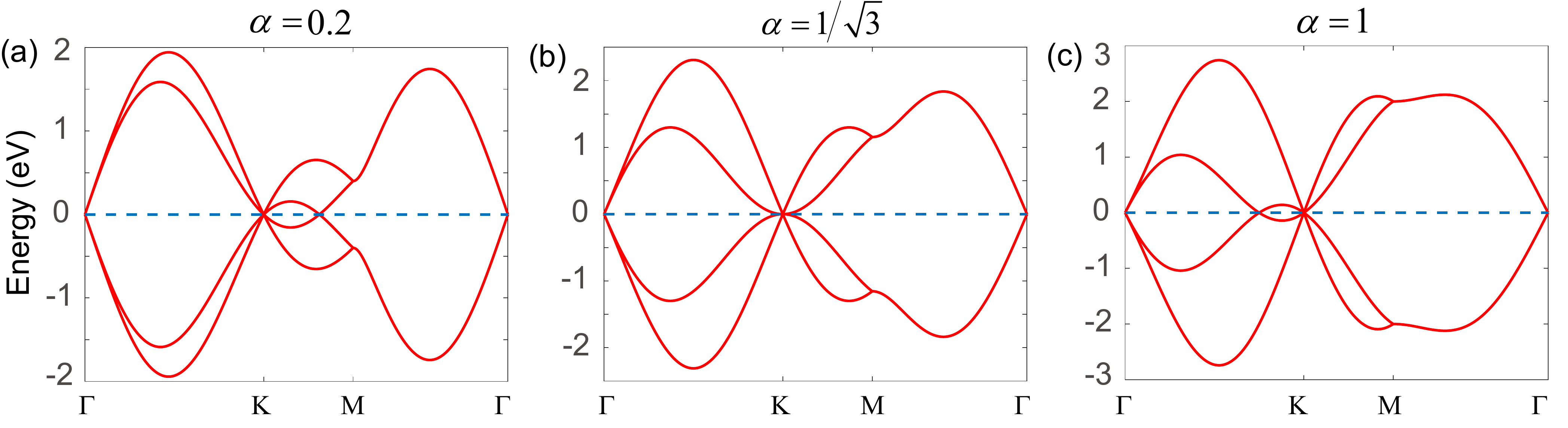}}
\caption{The Dirac points move from $M_{1,2,3}$ as $\alpha$ varies from $0$. (a) The Dirac node moves along $\overline{MK}$ as $0<\alpha<1/\sqrt{3}$. (b) As $\alpha=1/\sqrt{3}$, the moving Dirac node is located at $K$. The total winding number of the 4-fold degenerate nodal point becomes -1. (c) The Dirac point moves away from $K$ toward $\Gamma$. }  
 \label{alpha}
\end{figure}

	To realize the minimal configuration, we begin to increase $\alpha$ from $0$, two Dirac points at each of $M_{1,2,3}$ points move toward $K,\ K'$ along $\overline{MK},\ \overline{MK'}$ separately as shown in Fig.~\ref{alpha}(a). Since $h_4(0)$ vanishes, the node carrying $\nu=2$ at $\Gamma$ is intact. As $\alpha= 1/\sqrt{3}$, the moving Dirac points carrying $\nu=-1$ are exactly located at $K,K'$; Fig.~\ref{alpha}(b) shows the 4-fold degenerate point is located in $K$. Since originally each of $K', K$ has $+2$ winding number and each of the Dirac nodes from $M_{1,2,3}$ carries $\nu=-1$, after merging with the three Dirac points from $M_{1,2,3}$, the 4-fold degenerate nodal point at $K$ or $K'$ carries $\nu=-1$ winding number in total. Thus, the Hamiltonian (\ref{merge H}) with $\alpha= 1/\sqrt{3}$ is the tight-binding model hosting the 4-fold degenerate node at $\Gamma$ with $\nu=2$ and the nodes at $K, K'$ with $\nu=-1$. We note this is the minimal configuration is unstable. That is, as $\alpha> 1/\sqrt{3}$, from $K$ or $K'$ Dirac nodes separately move along $\overline{\Gamma K}$ or $\overline{\Gamma K'}$ toward $\Gamma$ as shown in Fig.~\ref{alpha}(c).

%
%
%
%
%
%
%
%


\section{$C_2T$ symmetry} \label{PT symmetry}

Space-time inversion symmetry ($C_2T$) can protect 2D Dirac cones in the presence of disorders, and quantizes the $\bZ_2$ Berry phase in the 1D integral path. Here, we provide the detailed deviations; the Berry phase with the closed integral path $\Gamma(\boldsymbol{K}_{0})$ at the nodal point $\boldsymbol{K}_{0}$ can be written in Eq.~\ref{Berry}.

Space-time inversion operator $C_2T$ is the combination of a unitary matrix and complex conjugation $C_2T=V_\bk \mathcal{K}$; the unitary matrix $V_{\bk}$ might be $\bk$-dependent and does not have any singularity point in the BZ.
By assuming the absence of degenerate states, the relation of wavefunctions under $C_2T$ symmetry is given by
\bee
\ket{u_n (\bk)}=e^{i\beta_{\bk}^j} C_2T \ket{u_n (\bk)} = e^{i\beta_{\bk}^j} V_\bk \ket{u_n ^*(\bk)}.
\ee
The Berry phase with a non-contractible integral path can be written as
\begin{align}
\gamma(\boldsymbol{K}_{0})=&i \sum_{n\in \rm{occ}} \oint    \bra{u_n (\bk)} \partial_{\bk} \ket{u_n (\bk)} d\bk   \nonumber \\
=& i \sum_{n\in \rm{occ}}   \oint   \bra{u_n^* (\bk)} V_{\bk}^\dagger e^{-i\beta^j_{\bk}} \partial_{\bk} e^{i\beta^j_{\bk}}  V_k \ket{u_n ^*(\bk)} d\bk \nonumber \\
=& - \sum_{n\in \rm{occ}} (\beta^j_{+}-\beta^j_{-})+ i \sum_{n\in \rm{occ}}   \oint     \bra{u_n^* (\bk)} \partial_{\bk}  \ket{u_n ^*(\bk)} d\bk \nonumber  \\
& +i \sum_{n\in \rm{occ}}   \oint \bra{u_n (\bk) }V_k^\dagger (\partial_{\bk} V_k ) \ket{u_n ^*(\bk)} d\bk,   \label{PT1}
\end{align}
where $\beta^j_{\mp}$ represent the phases at the beginning and the end of the integration path respectively; since the integration path is a closed loop, the start point and the end point are identical. Hence, the first summation is $2m\pi $, where $m$ is an integer. Furthermore, due to the absence of any singularity in $V_{\bk}$, the third term vanishes. After using the identity
\bee
 \bra{\phi _{\bk,j}^*} \partial_{\bk}  \ket{\phi _{\bk,j}^*}= \braket{\partial_{\bk} \phi_{\bk,j}}{\phi_{\bk,j}}=- \bra{\phi_{\bk,j}} \partial_{\bk}  \ket{\phi_{\bk,j}}, \label{conjugate}
\ee
we have the quantized Berry phase
\bee
\gamma(\boldsymbol{K}_{0})= \sum_{n\in \rm{occ}}  (\beta^j_{+}-\beta^j_{-})/2  = 0,\ \pi\ (\rm{mod}\ 2\pi)
\ee
The modulo operation of $2\pi$ stems from gauge transform $\ket{u_n ^*(\bk)}\rightarrow e^{il\theta }\ket{u_n ^*(\bk)}$ by using $\boldsymbol{K}_{0}$ as the singularity point for the transformation; Hence, the two Berry phases differed by $2\pi l$ are equivalent.

The crystalline symmetry $\{g|\boldsymbol{\tau}\}$ connects two nodal points $\boldsymbol{K}_{0}$ and $g\boldsymbol{K}_0$ so that the Hamiltonians at the two points obeys
\begin{align}
\mH(g\bk)=&U_g(\bk) \mH (\bk) U_g^{\dag}(\bk),
\end{align}
where the crystalline symmetry operator $U_g(\bk)$ is a momentum-dependent unitary matrix. Therefore, the wavefunctions at the two points are connected by
\bee
\ket{u_n(g\bk)} = U_g(\bk) \ket{u_n(\bk)}.
\ee
With this relation the Berry phases at $g\boldsymbol{K}_0$ point is written in the form of the Berry  phase at $\boldsymbol{K}_0$
\begin{eqnarray}
&&\gamma(g\boldsymbol{K}_{0}) \nonumber
\\
&=&  i\sum_{n\in \rm{occ}} \oint_{\Gamma(g\boldsymbol{K}_{0})}  \langle u_{n}(\boldsymbol{k})|\partial_{\boldsymbol{k}}| u_{n}(\boldsymbol{k})\rangle \cdot \mathrm{d} \boldsymbol{k} \nonumber
\\
&=& i\det(g) \sum_{n\in \rm{occ}} \oint_{\Gamma(\boldsymbol{K}_{0})}  \langle u_{n}(g\boldsymbol{k})|\partial_{g\boldsymbol{k}}| u_{n}(g\boldsymbol{k})\rangle \cdot \mathrm{d} g\boldsymbol{k} \nonumber
\\
&=& i\det(g)\sum_{n\in \rm{occ}} \oint_{\Gamma(\boldsymbol{K}_{0})}  \langle U_g(\boldsymbol{k})u_{n}(\boldsymbol{k})|\partial_{\boldsymbol{k}}| U_g(\boldsymbol{k})u_{n}(\boldsymbol{k})\rangle \cdot \mathrm{d} \boldsymbol{k} \nonumber
\\
&=& i\det(g) \sum_{n\in \rm{occ}} \oint_{\Gamma(\boldsymbol{K}_{0})}\langle u_{n}(\boldsymbol{k})|\partial_{\boldsymbol{k}}|u_{n}(\boldsymbol{k})\rangle \cdot \mathrm{d} \boldsymbol{k} \nonumber
\\
&=&\det(g)\gamma(\boldsymbol{K}_{0})
\end{eqnarray}
Here we use that $U_g(\bk)$ does not have any singularity in the BZ. Since the Berry phase is either $0$ or $\pi$, the relation between $\boldsymbol{K}_{0}$ and $g\boldsymbol{K}_{0}$ can be simplified to
\bee
\gamma(g\boldsymbol{K}_{0})=\gamma(\boldsymbol{K}_{0}).
\ee
Similarly, for time-reversal symmetry, the Berry phases at $\boldsymbol{K}_{0}$ and $-\boldsymbol{K}_{0}$ are related by
\bee
\gamma(\boldsymbol{K}_{0})=\gamma(-\boldsymbol{K}_{0})
\ee
\bibliography{TOPO_v20,NH,aGBZ}
 \bibliographystyle{apsrev4-1}

\end{document}